\documentclass[11pt,a4paper]{article} 
\usepackage{jheppub}
\pdfoutput=1
\usepackage{epsfig}
\usepackage{graphicx}
\usepackage{amsmath}
\usepackage{amsfonts}
\usepackage{amssymb}
\usepackage{slashed}
\usepackage{natbib}
\usepackage{tikz}
\usepackage{bm}
\usepackage{xcolor}
\usepackage{subfigure}
\usepackage{placeins} %For placing float barriers
\usepackage{cleveref} %NB:Must be loaded last.

\usepackage{soul} %For strikethrough \st{text}

\allowdisplaybreaks[1]

\usepackage{tikz-cd}

\newcommand{\ord}{{\cal O}}

\def \tr {\mbox{tr}}

\newcommand{\Li}{{\rm Li_2}}

\newcommand\real[1]{{\mathrm{Re}\left[{#1}\right]}}

\newcommand\disp[1]{{\mathrm{Disp}\left[{#1}\right]}}

\newcommand{\MRK}{{ \mathrm{MRK}}}
\newcommand{\MRKx}{{ \mathrm{MRK}^\prime}}

\newcommand{\zb}{{ \bar{z}}}

\newcommand{\ibio}{{\imath_2 \bar{\imath}_1 }}
\newcommand{\ibi}{{\imath_3 \bar{\imath}_2 }}

\newcommand{\toRegge}{\underset{\mathrm{Regge}}{\to}}
\newcommand{\toMRK}{\underset{\sss{\mathrm{MRK}}}{\to}}
\newcommand{\toNMRK}{\underset{\sss{\mathrm{NMRK}}}{\to}}
\newcommand{\toMRKx}{\underset{\sss{\mathrm{MRK}^\prime}}{\to}}
\newcommand{\toYinf}{\underset{Y \to \infty}{\to}}
\newcommand{\toXinf}{\underset{X \to \infty}{\to}}

\newcommand\sss{\scriptscriptstyle}
\newcommand\as{\alpha_{\mathrm{s}}} 
\newcommand\gs{g_{\mathrm{s}}}

\newcommand\Nf{N_f}

\newcommand\qb{{\bar{q}}}

\def \al #1 {\frac {\as({#1})}{\pi} }
\def \ds #1 {\ooalign{$\hfil/\hfil$\crcr$#1$}}

\makeatletter
\newcommand*{\encircled}[1]{\relax\ifmmode\mathpalette\@encircled@math{#1}\else\@encircled{#1}\fi}
\newcommand*{\@encircled@math}[2]{\@encircled{$\m@th#1#2$}}
\newcommand*{\@encircled}[1]{%
	\tikz[baseline,anchor=base]{\node[draw,circle,outer sep=0pt,inner sep=.2ex] {#1};}}
\makeatother

\newcommand{\cN}{{\mathcal N}}

\newcommand{\cO}{{\mathcal O}}

\newcommand{\cC}{{\mathcal C}}
\newcommand{\Nfour}{\mathcal{N}=4}
\newcommand{\None}{\mathcal{N}=1_\chi}
\newcommand{\NoneX}{\mathcal{N}=1_\chi}
\newcommand{\NoneV}{\mathcal{N}=1_V}

\usepackage{titlesec}
\titleformat{\paragraph}
{\normalfont\normalsize\bfseries}{\theparagraph}{1em}{}
\titlespacing*{\paragraph}
{0pt}{3.25ex plus 1ex minus .2ex}{1.5ex plus .2ex}

\title{One-loop five-parton amplitudes in the NMRK limit}
\author[a,b]{Emmet P. Byrne,}
\affiliation[a]{Higgs Centre for Theoretical Physics, School of Physics and Astronomy, The University of Edinburgh, Edinburgh EH9 3FD, Scotland, UK}
\affiliation[b]{Department of Physics and Astronomy, The University of Manchester, Manchester M13 9PL, UK}
\emailAdd{Emmet.Byrne@manchester.ac.uk}

\abstract{
We analyse the real part of one-loop five-parton amplitudes in the next-to-multi-Regge kinematic (NMRK) limit, to leading power, and to finite order in the dimensional regularisation parameter. 
To leading logarithmic (LL) accuracy, it is known that five-parton amplitudes in this limit are given to all-orders by a single factorised expression, in which the pair of partons which are not well-separated in rapidity are described by a two-parton emission vertex. In this study, we investigate the one-loop amplitudes at next-to-leading logarithmic (NLL) accuracy, and find that is has a more complex structure. In particular, it is found that the purely gluonic amplitudes are compatible with an analogous factorisation of individual colour structures. From the one-loop amplitudes we extract one-loop two-parton emission vertices, which are functions of a subset of the momenta of the amplitude. In the multi-Regge kinematic (MRK) limit, the vertices themselves factorise into the known one-loop single-parton emission vertices and Lipatov vertex, with rapidity dependence governed by the one-loop gluon Regge trajectory, as required by compatibility with the known MRK limit of amplitudes. The one-loop two-parton emission vertices are necessary ingredients for the construction of the next-to-next-to leading order (NNLO) jet impact factors in the BFKL framework. 
}


\begin{document}

\maketitle
\section{Introduction}

The multi-Regge kinematic (MRK) limit of amplitudes, in which final-state particles are strongly ordered in rapidity, has long been a source of insight into the all-orders structure of amplitudes \cite{Lipatov:1976zz,DelDuca:2019tur}. Knowledge of the MRK limit has been used to compute or constrain amplitudes in general kinematics~\cite{DelDuca:2009au,DelDuca:2010zg,Dixon:2014iba,Henn:2016jdu,Caron-Huot:2016owq,Almelid:2017qju,Caron-Huot:2019vjl,Falcioni:2021buo,Caola:2021izf}. Historically, the next-to-multi-Regge kinematic (NMRK) limit (or single-Regge limit), where one of the strong rapidity-ordering requirements is relaxed, has received far less attention than the more tractable MRK limit, with some notable exceptions, including refs.~\cite{White:1973ola,Brower:1974yv}. The NMRK limit is an intriguing theoretical arena, where on the one hand the simplified kinematics make possible a detailed study of high-multiplicity amplitudes, revealing structures which are obscured in general kinematics~\cite{Bartels:2008ce,Byrne:2022wzk}, and on the other hand, the NMRK limit captures more of the complexity and richness of amplitudes than does the MRK limit. The NMRK limit and its MRK boundaries provide an ideal test bed for studying which properties of MRK amplitudes generalise to more generic kinematic spaces. In this work, we find that the known MRK limit of amplitudes motivates a highly convenient framework for organising the kinematic structure of the amplitudes in the NMRK limit. In particular, this organisation makes it trivial to take the further MRK limit of NMRK-factorised expressions.

The behaviour of amplitudes in Regge limits is important for determining the asymptotic high-energy limit of cross sections in QCD \cite{Lipatov:1976zz,Kuraev:1976ge}. The leading-logarithmic (LL) behaviour of amplitudes in QCD is described by the $t$-channel exchange of a so-called Reggeised gluon. In the language of Regge theory (for pedagogical reviews see refs.~\cite{Collins:1977jy,White:2019ggo,Mizera:2023tfe}), the amplitude is governed by a simple Regge pole in the complex angular momentum plane.  
The Balitsky-Fadin-Kuraev-Lipatov (BFKL) equation~\cite{Lipatov:1976zz,Kuraev:1976ge,Kuraev:1977fs,Balitsky:1978ic} sums the LL contributions to QCD scattering processes to all orders in the coupling (for pedagogical introductions see refs.~\cite{DelDuca:1995hf,Forshaw:1997dc,Fadin:1998sh}).
The leading order (LO) kernel of this equation consists of the square of the Lipatov vertex~\cite{Lipatov:1976zz}, which can be obtained by taking the MRK limit of tree-level amplitudes. At next-to-leading logarithmic (NLL) accuracy, the real part of amplitudes are similarly described by a simple Regge pole~\cite{Fadin:2006bj,Fadin:2015zea}, which allows the BFKL approach to be extended to NLL accuracy~\cite{Fadin:1998py,Ciafaloni:1998gs,Kotikov:2000pm,Kotikov:2002ab}. This requires the  kernel to be extended to next-to-leading order (NLO). The NLO kernel may be computed using building blocks which can be obtained by taking the MRK limit of one-loop amplitudes as well as building blocks which can be obtained by taking the NMRK limit of tree-level amplitudes. At next-to-next-to leading-logarithmic (NNLL) accuracy, the real part of the amplitude is described by a Regge cut in addition to a Regge pole \cite{Fadin:2020lam,Caron-Huot:2017fxr,DelDuca:2001gu,Fadin:2017nka}. Recently, schemes have been introduced to disentangle the pole and cut contributions at fixed order \cite{Falcioni:2021dgr,Fadin:2021csi}. In particular, the three-loop Regge trajectory has recently been determined~\cite{DelDuca:2021vjq,Caola:2021izf,Caola:2022dfa}. It is hoped that the BFKL approach may yet be extended to describe the evolution of the Regge-pole contribution. 

As reviewed in ref.~\cite{Byrne:2022wzk}, there has been much progress towards extending the BFKL approach to NNLL accuracy. More recently, knowledge of the one-loop Lipatov vertex has been extended to $\cO(\epsilon^2)$~\cite{Fadin:2023roz}, as required for the NNLO kernel, and the recent calculation of two-loop five-parton amplitudes in QCD~\cite{Agarwal:2023suw,DeLaurentis:2023nss,DeLaurentis:2023izi} has made possible the extraction of the two-loop Lipatov vertex. 
\begin{figure}
    \centering
    \subfigure[]{\includegraphics[width=0.24\textwidth]{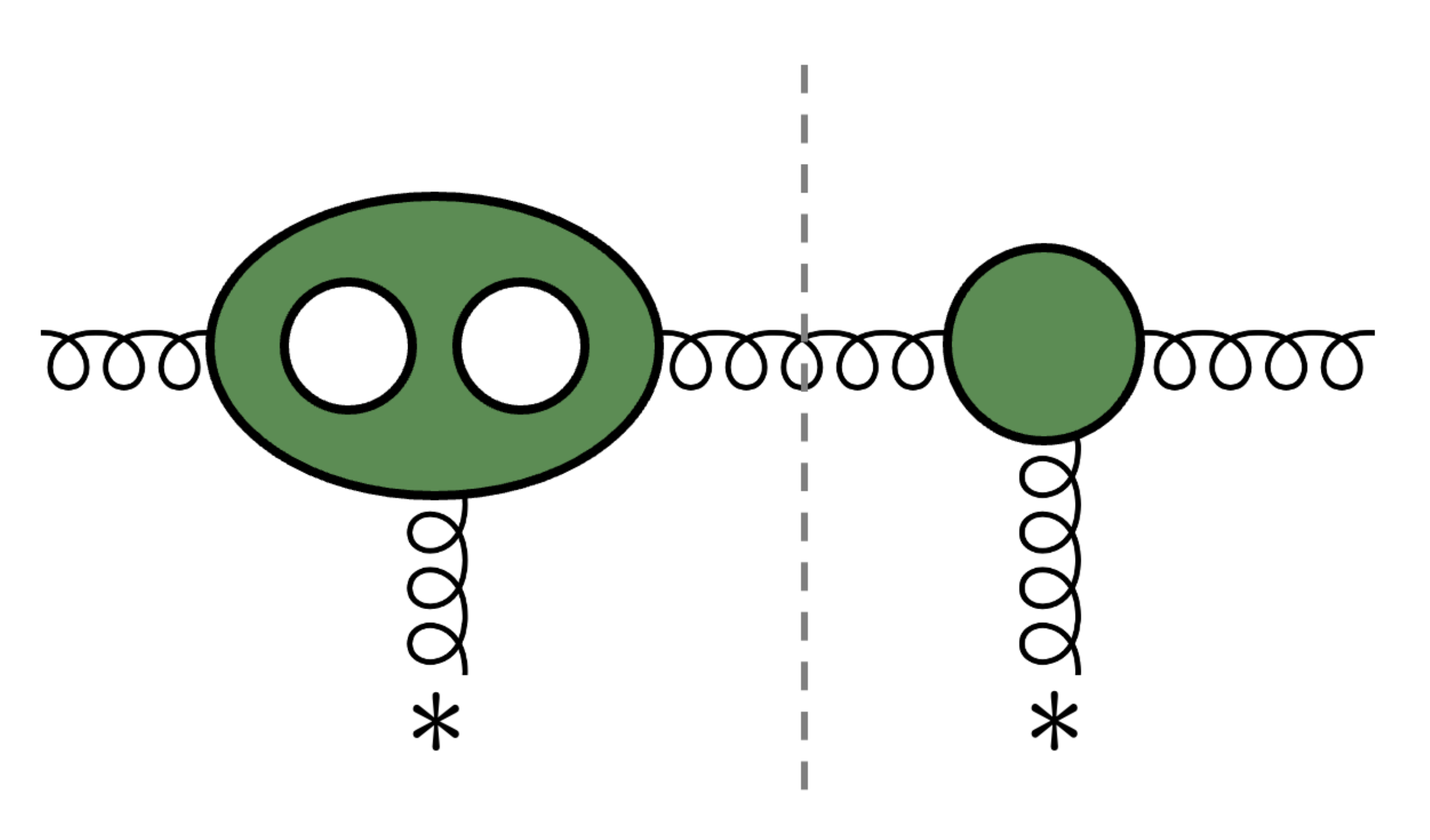}\label{subfig:C2g}} 
    \subfigure[]{\includegraphics[width=0.24\textwidth]{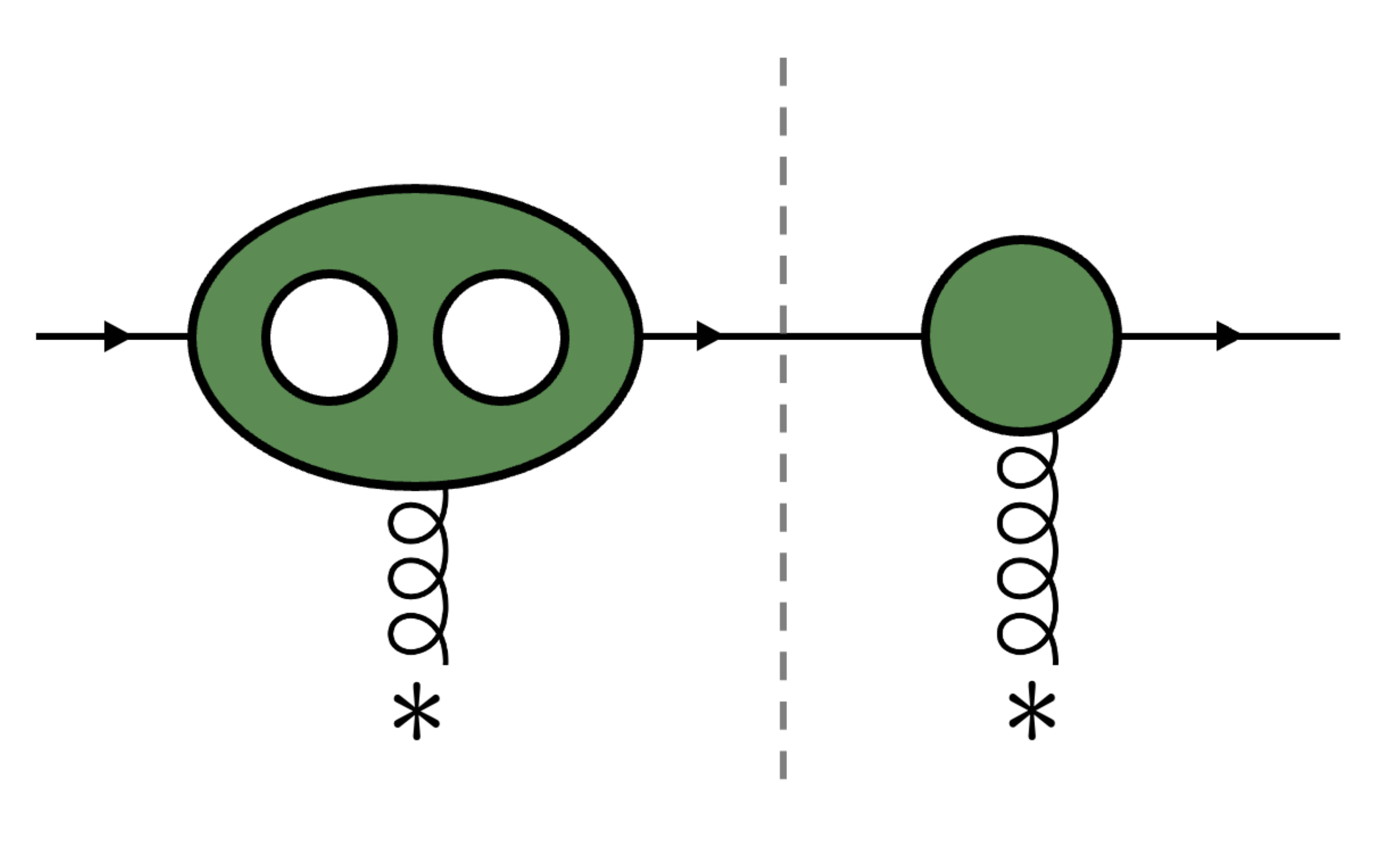}\label{subfig:C2q}} 
    \subfigure[]{\includegraphics[width=0.24\textwidth]{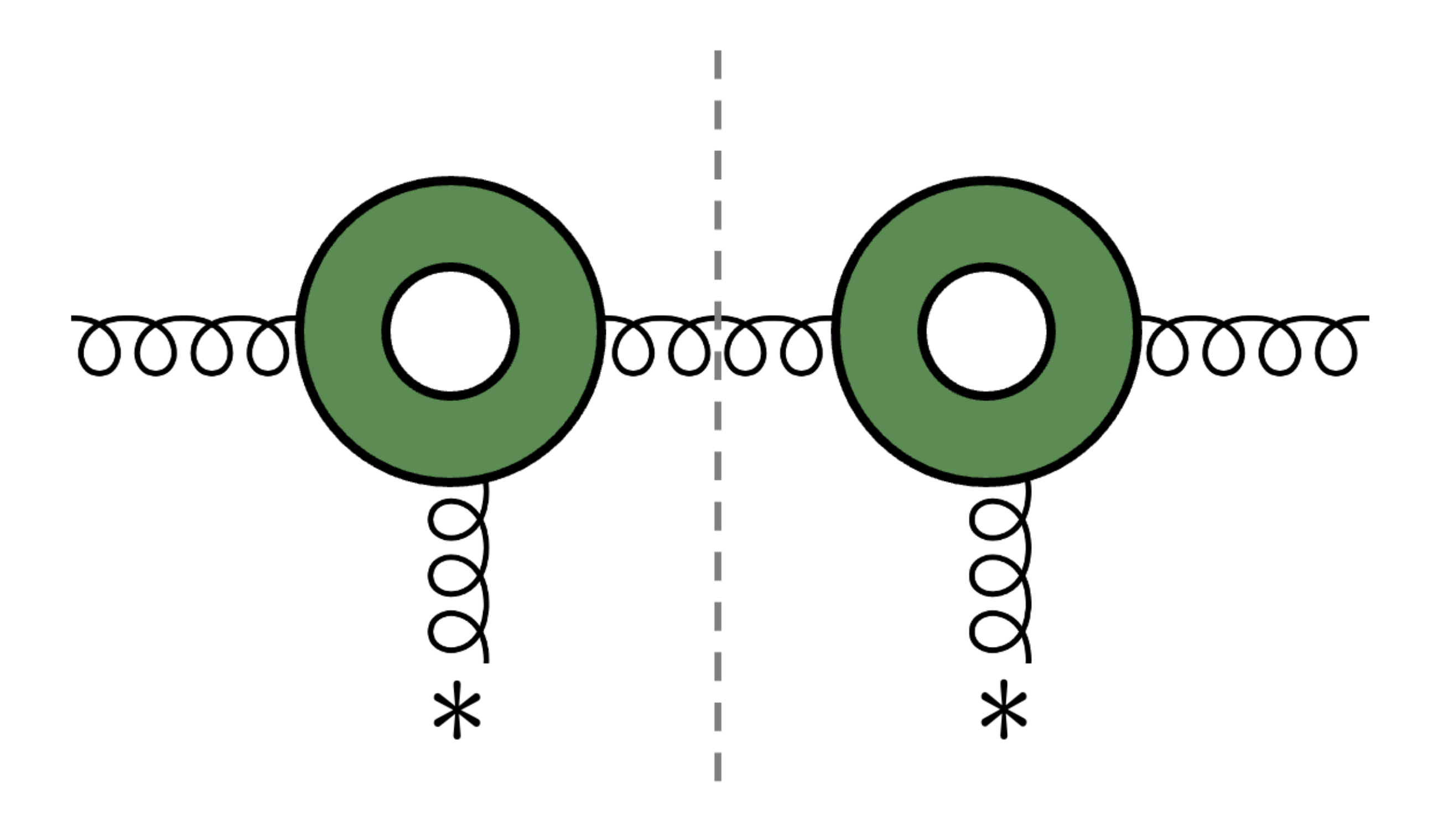}\label{subfig:C1g_C1g}}
    \subfigure[]{\includegraphics[width=0.24\textwidth]{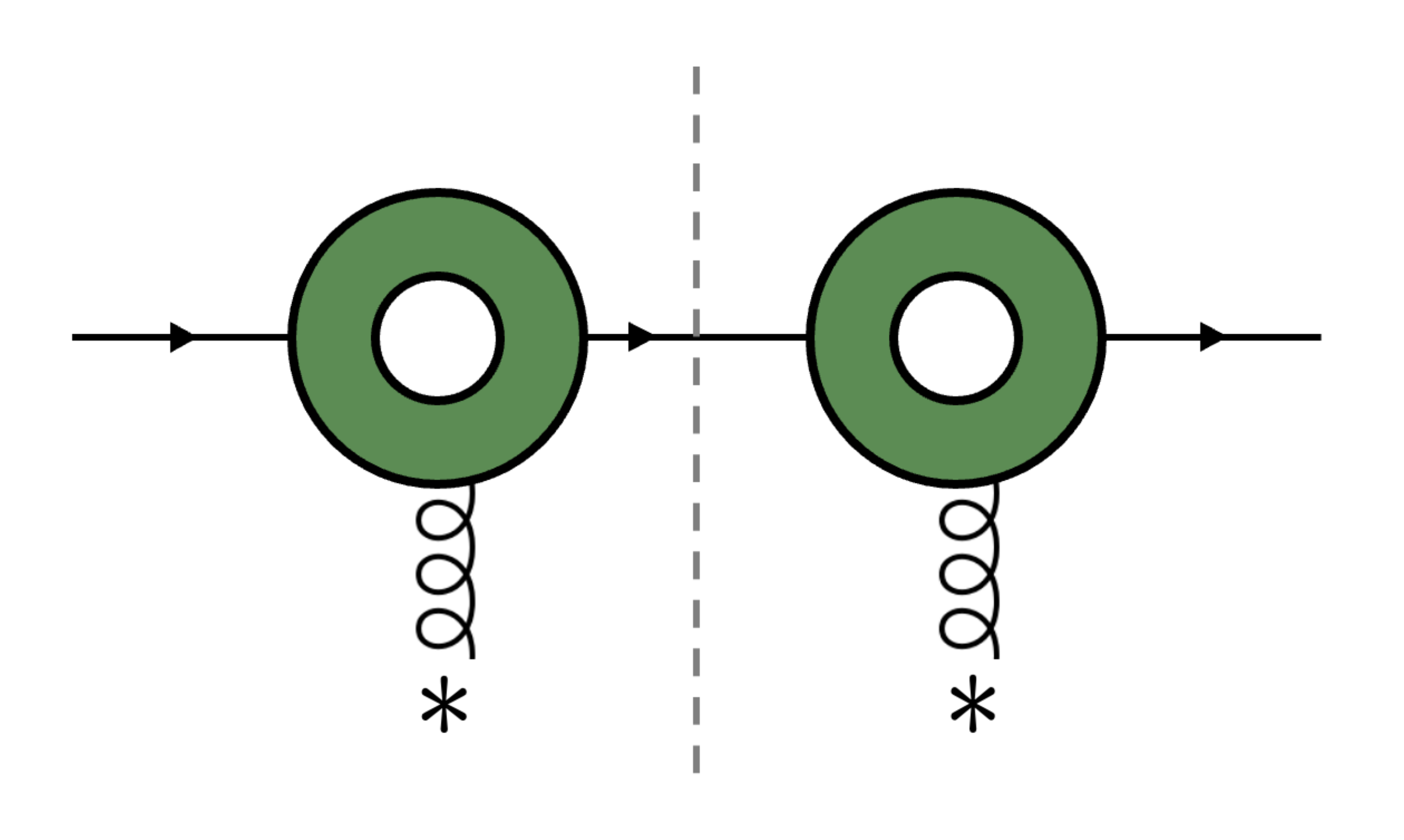}\label{subfig:C1q_C1q}}
    \\
    \subfigure[]{\includegraphics[width=0.24\textwidth]{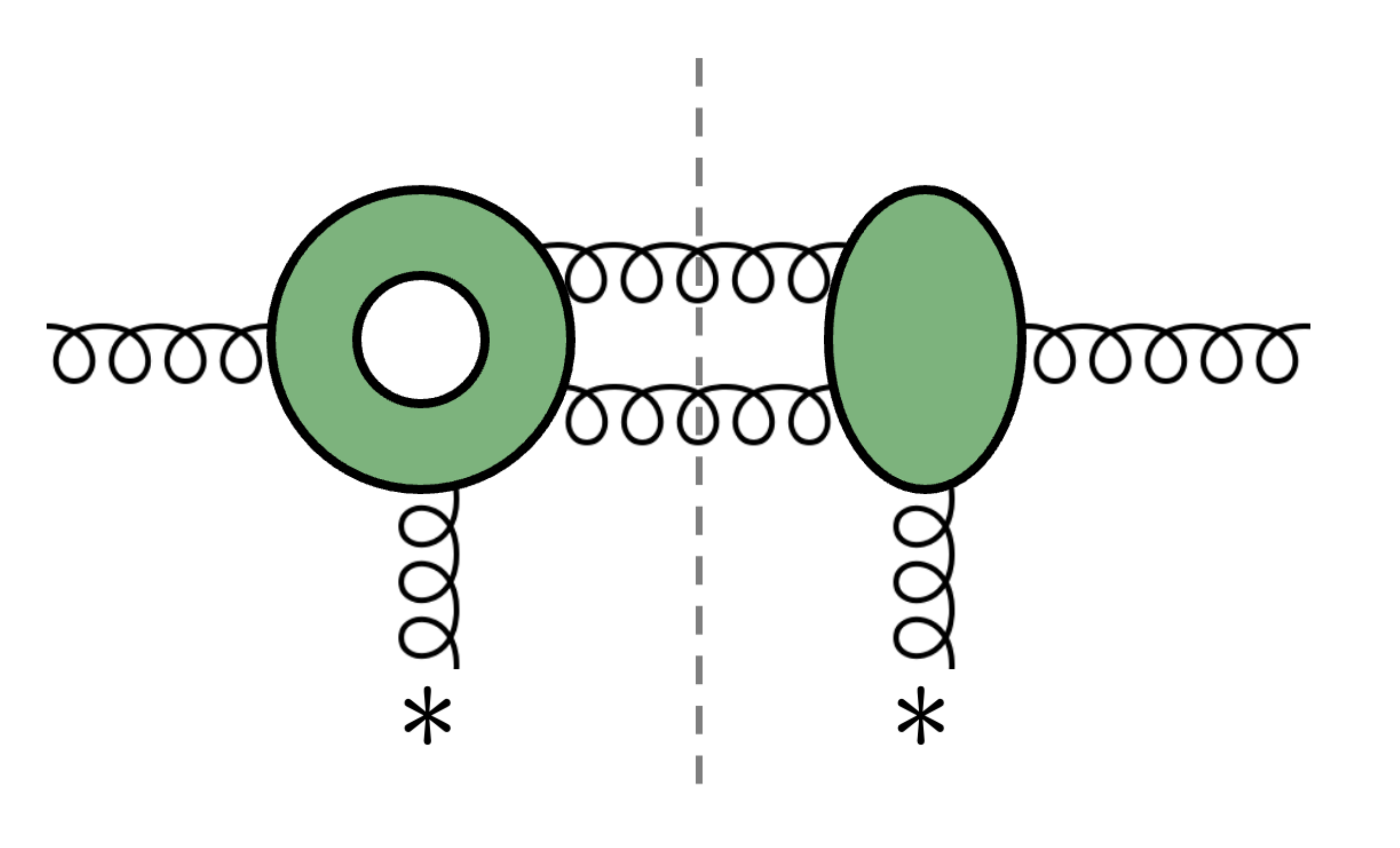}\label{subfig:C1gg}} 
    \subfigure[]{\includegraphics[width=0.24\textwidth]{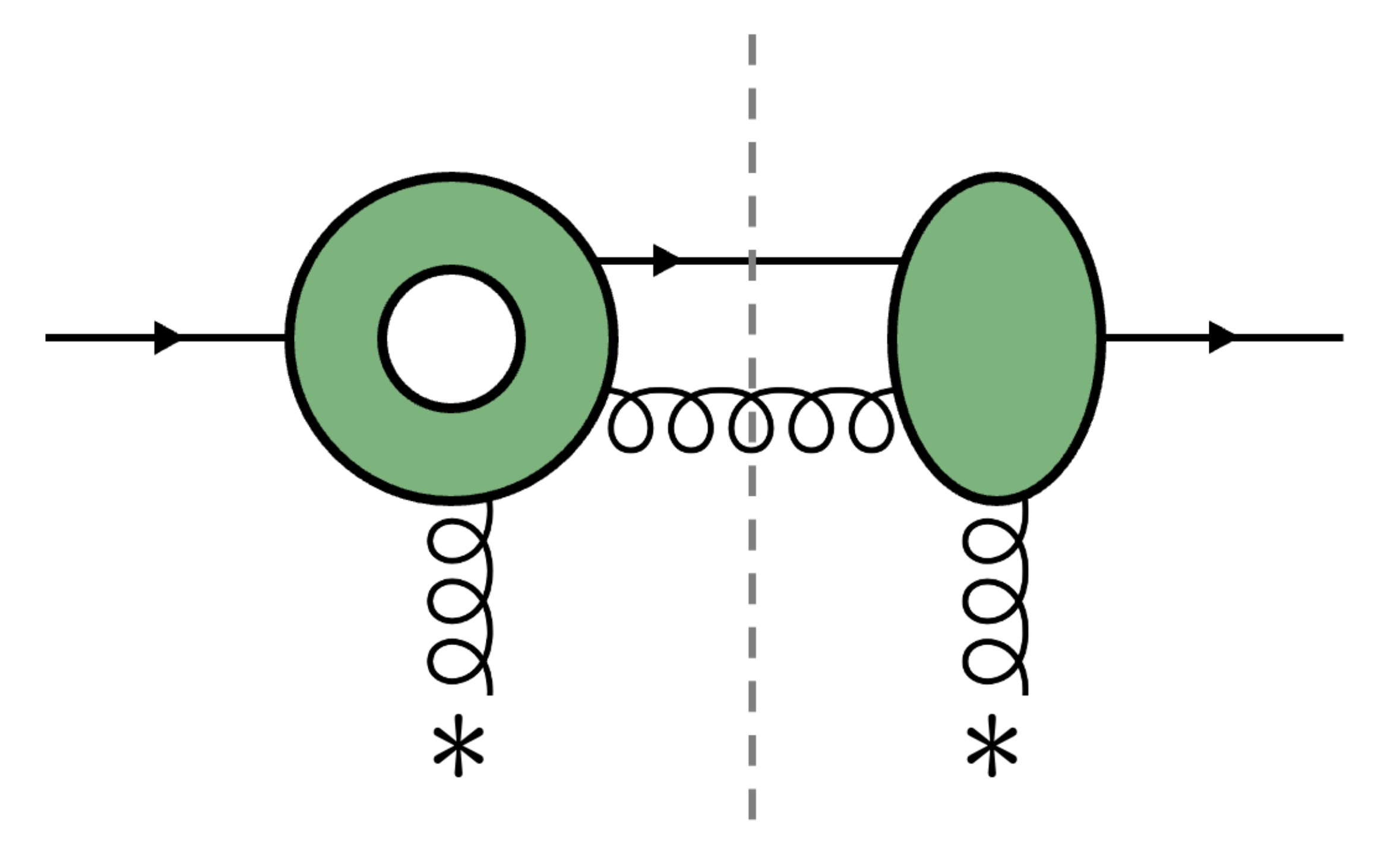}\label{subfig:C1qg}}
    \subfigure[]{\includegraphics[width=0.24\textwidth]{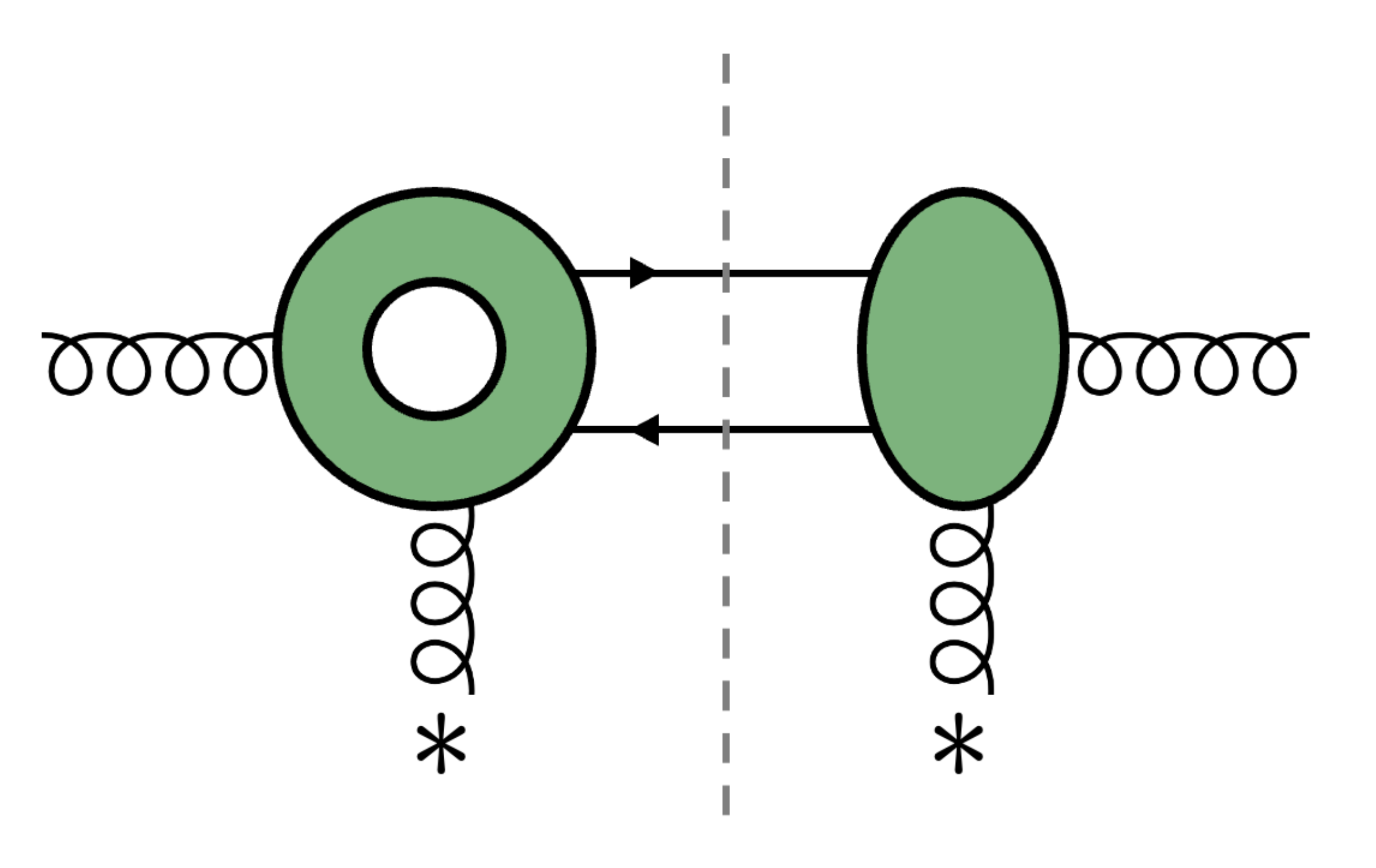}\label{subfig:C1qq}}
    \\
    \subfigure[]{\includegraphics[width=0.24\textwidth]{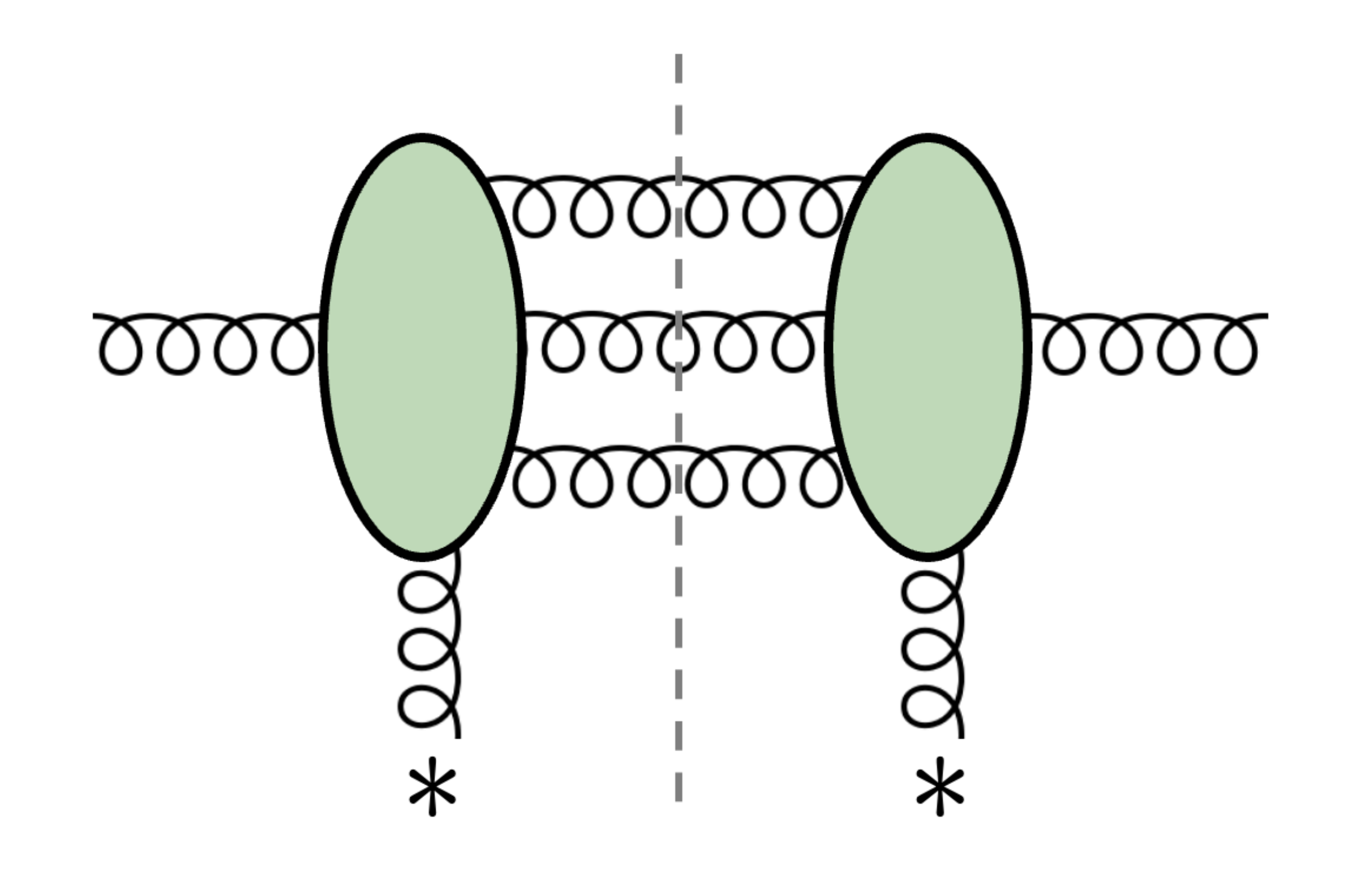}\label{subfig:C0ggg}} 
    \subfigure[]{\includegraphics[width=0.24\textwidth]{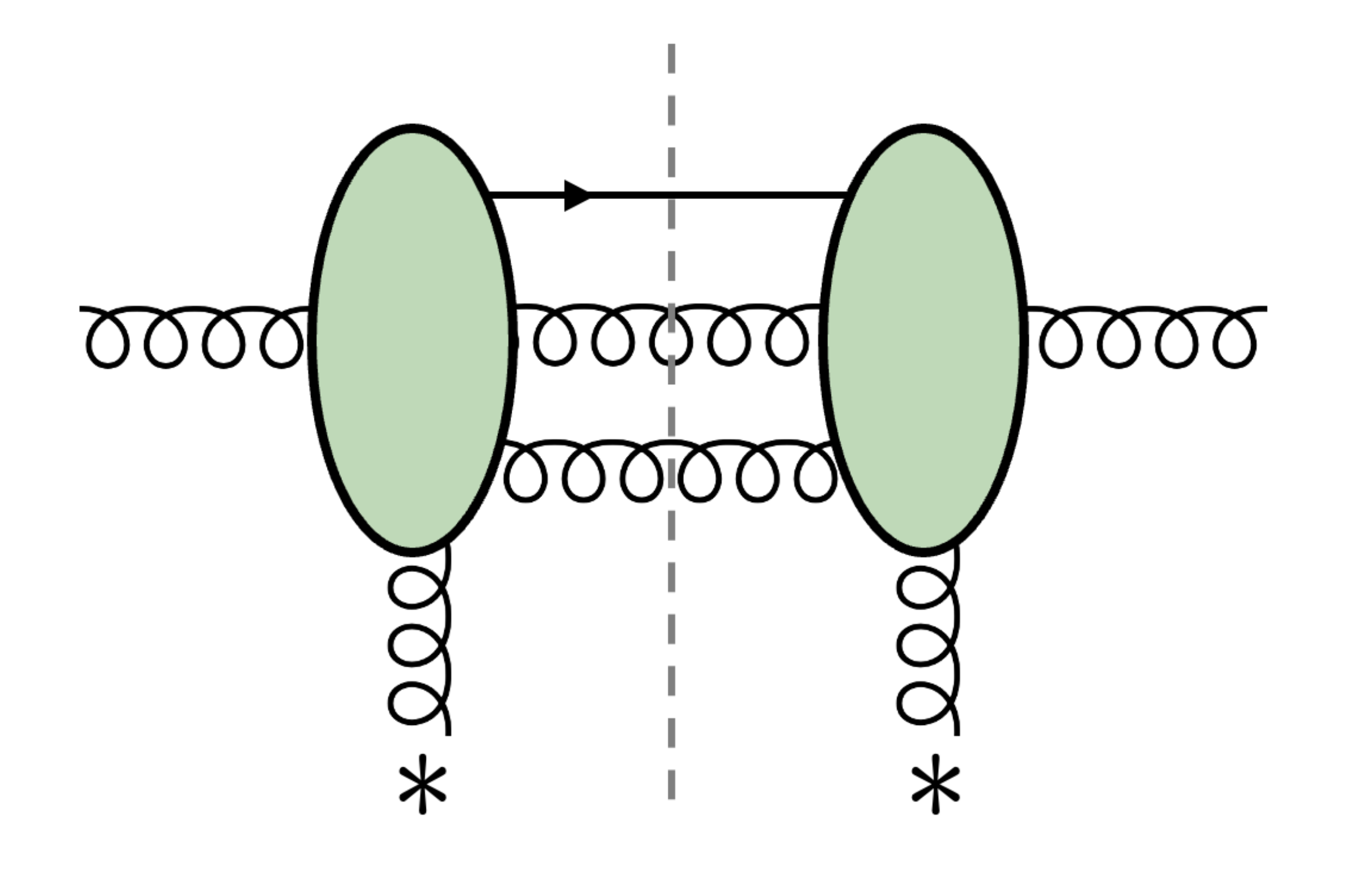}\label{subfig:C0qgg}}
    \subfigure[]{\includegraphics[width=0.24\textwidth]{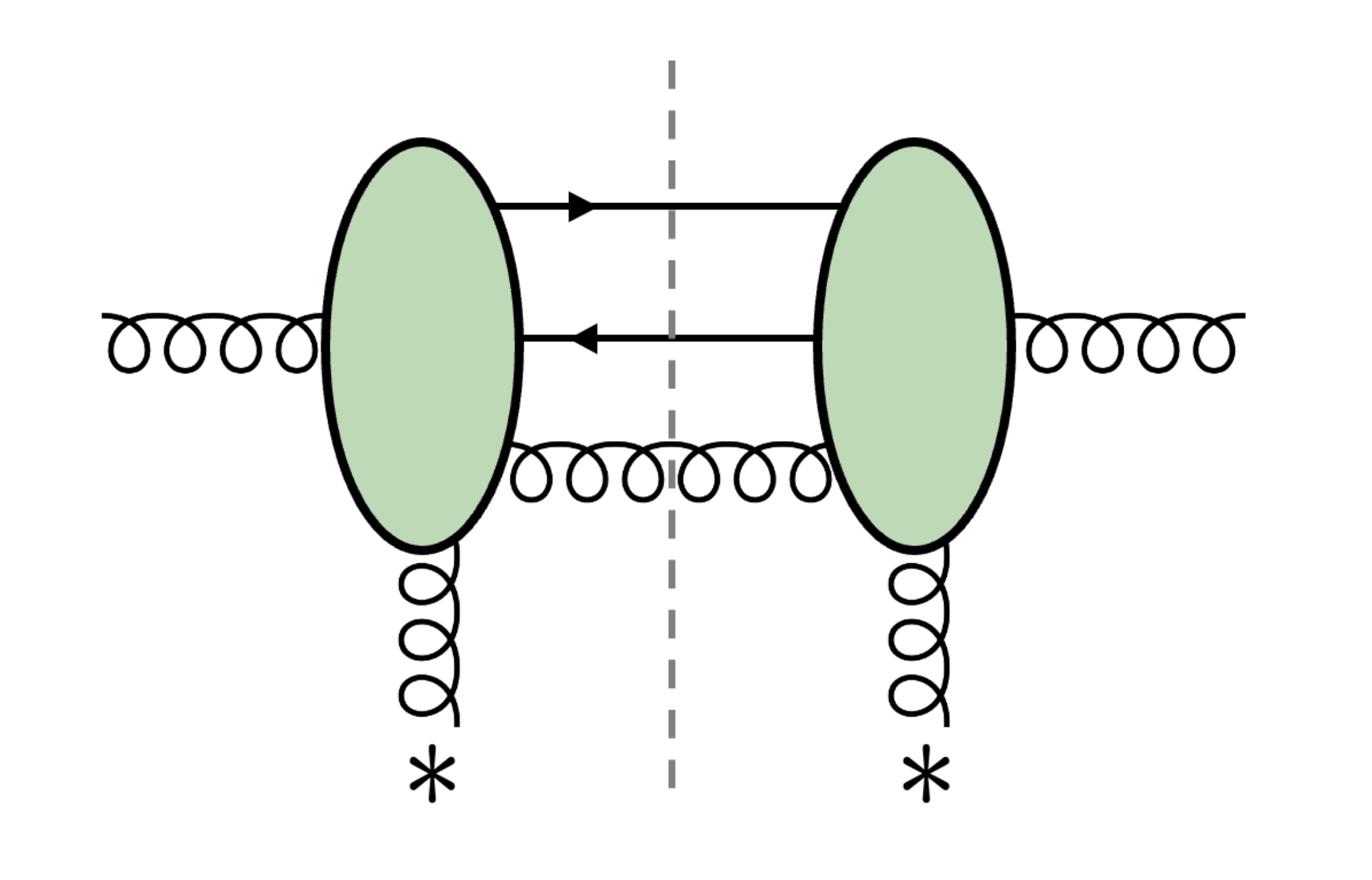}\label{subfig:C0qqg}}
    \caption{Ingredients of the NNLO jet impact factors: (a)-(d) double-virtual corrections; (e)-(g) real-virtual corrections; (h)-(j) double-real corrections. Punctures in blobs denote loop corrections. An asterisk indicates an off-shell gluon. The grey dashed line indicates the final-state cut.}
    \label{fig:IF_NNLO}
\end{figure}

In order to provide NNLL accurate predictions for physical processes it is necessary to also know not only the kernel, but also the relevant impact factors to NNLO accuracy. Let us briefly review the current status of the NNLO impact factors for the production of a jet, which are depicted in \cref{fig:IF_NNLO}.
The double-virtual corrections to the impact factors (\cref{subfig:C2g,subfig:C2q,subfig:C1g_C1g,subfig:C1q_C1q}) require knowledge of the two-loop single-parton peripheral-emission\footnote{The term \emph{impact factor} has been used in the literature to refer to both amplitude-level and cross-section-level quantities. We will use the term impact factor exclusively in the latter context, and refer to the amplitude-level factorised expressions as peripheral-emission vertices, defined in opposition to the central-emission vertices of ref.~\cite{Byrne:2022wzk}. More often, we will refer to the vertices by their explicit parton types, taking all momenta outgoing, e.g. ``the $\qb q g g^*$ vertex". Such vertices are also commonly called parton-parton-parton-Reggeon (PPPR) vertices, but in this work we do not wish to risk implying that the factorisation of amplitudes at one loop implies Reggeisation at higher orders. We use the term \emph{Lipatov vertex} to refer to the $g^*gg^*$ central-emission vertex at any loop order.} vertices~\cite{DelDuca:2014cya} as well as the interference of two one-loop single-parton peripheral-emission vertices~\cite{Fadin:1992rh,Fadin:1992zt,Fadin:1995km,DelDuca:1998kx}. 
The double-real corrections to the impact factors (\cref{subfig:C0ggg,subfig:C0qgg,subfig:C0qqg}) require knowledge of the three-parton emission vertices. These are all known for all partonic channels~\cite{DelDuca:1999iql,Antonov:2004hh,Duhr:2009uxa}. In this paper we obtain some of the missing ingredients for the real-virtual correction to the impact factors (\cref{subfig:C1gg,subfig:C1qg,subfig:C1qq}), namely the one-loop two-parton peripheral-emission vertices for both the $g g g g^*$ and $\qb q g g^*$ case. The one-loop $g g g g^*$ vertex has previously been determined for two colour-orderings and helicity-independence of this vertex was incorrectly assumed~\cite{Canay:2021yvm}. In \cref{sec:colour-5} we clarify the need for determining a third distinct colour ordering of this vertex, and in \cref{sec:2g} we show that two distinct helicity configurations of this vertex are required for the NNLO impact factor.

To regulate the infra-red divergences which will appear when performing the phase space integration of the two-parton emission vertices it will be necessary to know the one-loop two-parton peripheral-emission vertices to order $\epsilon$ in the soft and collinear limits and to order $\epsilon^2$ in the soft-collinear limit. In this paper we focus exclusively on finite ($\cO(\epsilon^0)$) corrections, which can be inferred from available one-loop amplitudes~\cite{Bern:1993mq,Bern:1994fz}. These amplitudes already reveal some interesting structure of amplitudes in the NMRK limit beyond tree-level. We find that, in contrast to the tree-level amplitudes, the one-loop amplitudes are not compatible with an all-orders factorisation of the amplitude. Rather, they are at best compatible with an all-orders factorisation of two different colour structures of the amplitude. 

The one-loop two-parton emission vertices also contain additional colour structures compared to the tree-level vertices. For the $gggg^*$ vertex, this is a totally symmetric colour structure, while for the $\qb q gg^*$ vertex it is a product of delta functions in the adjoint and fundamental representations. In the latter case, the new colour structure survives interference with the tree-level vertex, hence contributes to the NNLO kernel. 

The factorisation-structure and colour-structure of the $\cN{=}4$ part of the one-loop $gggg^*$ is analogous to the structure of the $\cN{=}4$ $g^*ggg^*$ vertex obtained in ref.~\cite{Byrne:2022wzk}. We expect the one-loop $g^*ggg^*$ and $g^* \qb q g^*$ vertices in QCD to have analogous structure to the $gggg^*$ and $\qb q gg^*$ vertices presented in this work.

The structure of this paper is as follows. 
In \cref{sec:Regge} we review the relevant properties of QCD amplitudes in the high-energy limit. In \cref{sec:colour} we study the colour structure of one-loop five-parton amplitudes at leading power in the NMRK limit. In \cref{sec:2g} and \cref{sec:cCgg} we study the respective factorisation properties of colour-ordered and colour-dressed one-loop five-gluon amplitudes in the NMRK limit. In \cref{sec:c_qqbg} and \cref{sec:cC_qqbg} we perform the analogous studies of one-loop five-parton amplitudes with an external quark-antiquark pair.
Finaly, we conclude in \cref{sec:conc}. 

Our approach to obtain new higher-point factorised vertices from kinematic limits of higher-point amplitudes requires us to subtract lower-point vertices derived from lower-point amplitudes. In this paper we make extensive use of some organisational techniques that have been used to study one-loop amplitudes. Such extensive use of these techniques was not necessary to obtain lower-point, colour-dressed vertices in QCD. This means the lower-point vertices we require are not explicit in the published literature. As such, we have included an extensive re-analysis of one-loop $2\to2$ scattering in \cref{sec:rederiving}, using the same organisational principles which we use in the main body of the paper, which provides us with the necessary lower-point expressions.

\section{Review of high-energy limit of amplitudes in QCD}
\label{sec:Regge}
Our conventions for general kinematics are given in \cref{sec:kin_gen} and our notation for kinematics in Regge limits is given in \cref{sec:kin_Regge}.
In this section we briefly review the all-orders structure of amplitudes in the Regge limit, given by \cref{eq:Regge}.
%, where the centre-of-mass energy $s\equiv s_{12}$ is much greater than the momentum transfer $t\equiv s_{23}$,
%\begin{equation}
%    s \gg -t \,.
%\end{equation}
At tree-level, amplitudes are dominated by the $t$-channel exchange of a gluon, and to leading power have the factorised form \cite{Lipatov:1976zz,Kuraev:1976ge}
\begin{align}
\begin{split}
\mathcal{A}_4^{(0)}
(f_2^{\lambda_2}, f_3^{\lambda_3}, f_4^{\lambda_3}, f_1^{\lambda_3}) 
&\underset{\mathrm{Regge}}{\to} 2 s \ \mathcal{C}^{(0)}_{f_2 f_3 g^*}(p_2^{\lambda_2}, p_3^{\lambda_3}, q) \ 
	\frac{1}{t} \ 
\mathcal{C}^{(0)}_{g^* f_4 f_1 }(-q, p_4^{\lambda_4}, p_1^{\lambda_1})\,.
\label{eq:Regge_2}
\end{split}
\end{align}
%where $q= -(p_2 + p_3) = p_4+p_1$ is the momentum flowing in the $t$-channel, i.e. $q^2 = t$. 

\begin{figure}
    \centering
    \subfigure[]{\includegraphics[width=0.255\textwidth]{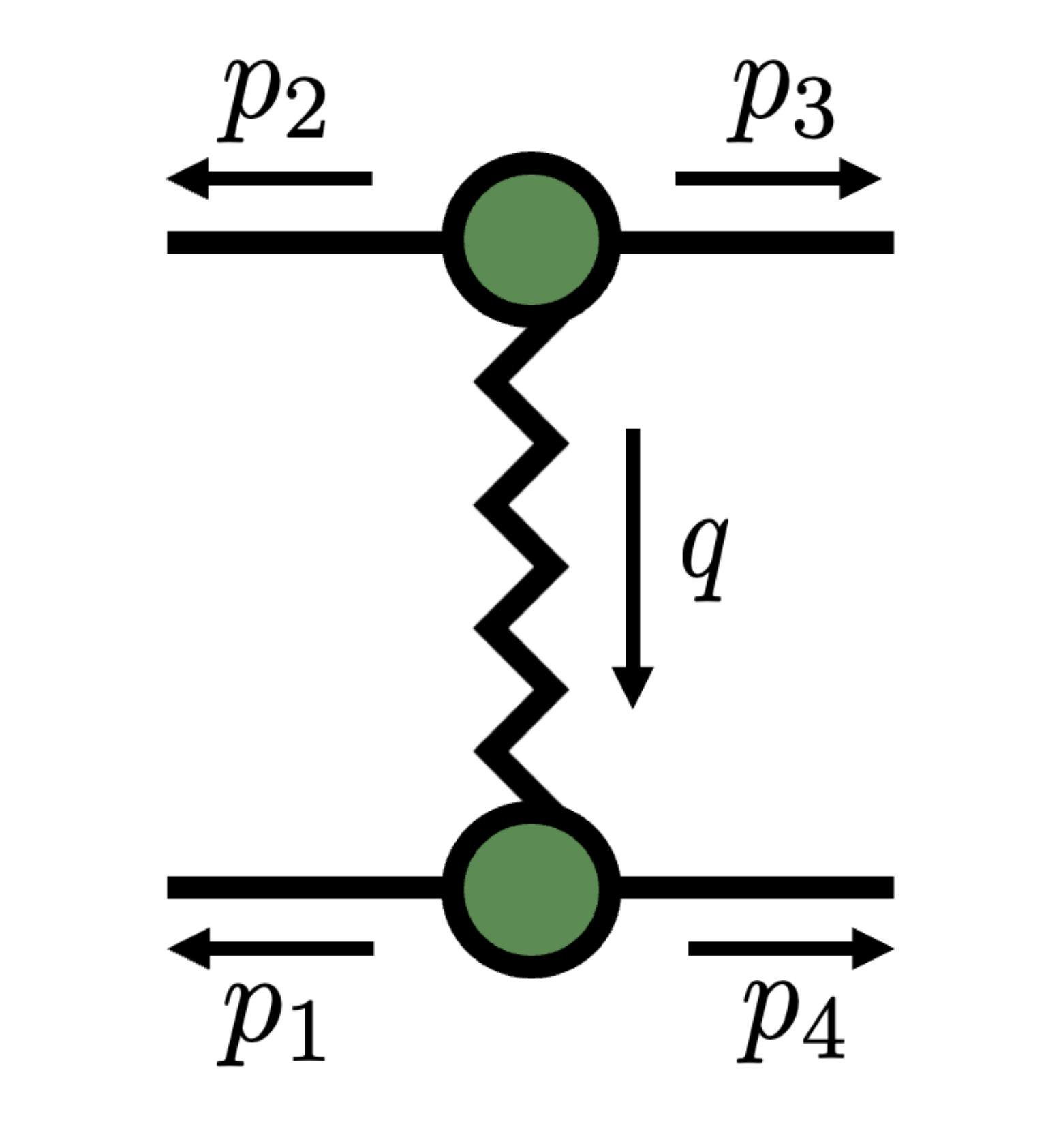}\label{subfig:Fac_Regge}}
    \qquad \qquad
    \subfigure[]{\includegraphics[width=0.2\textwidth]{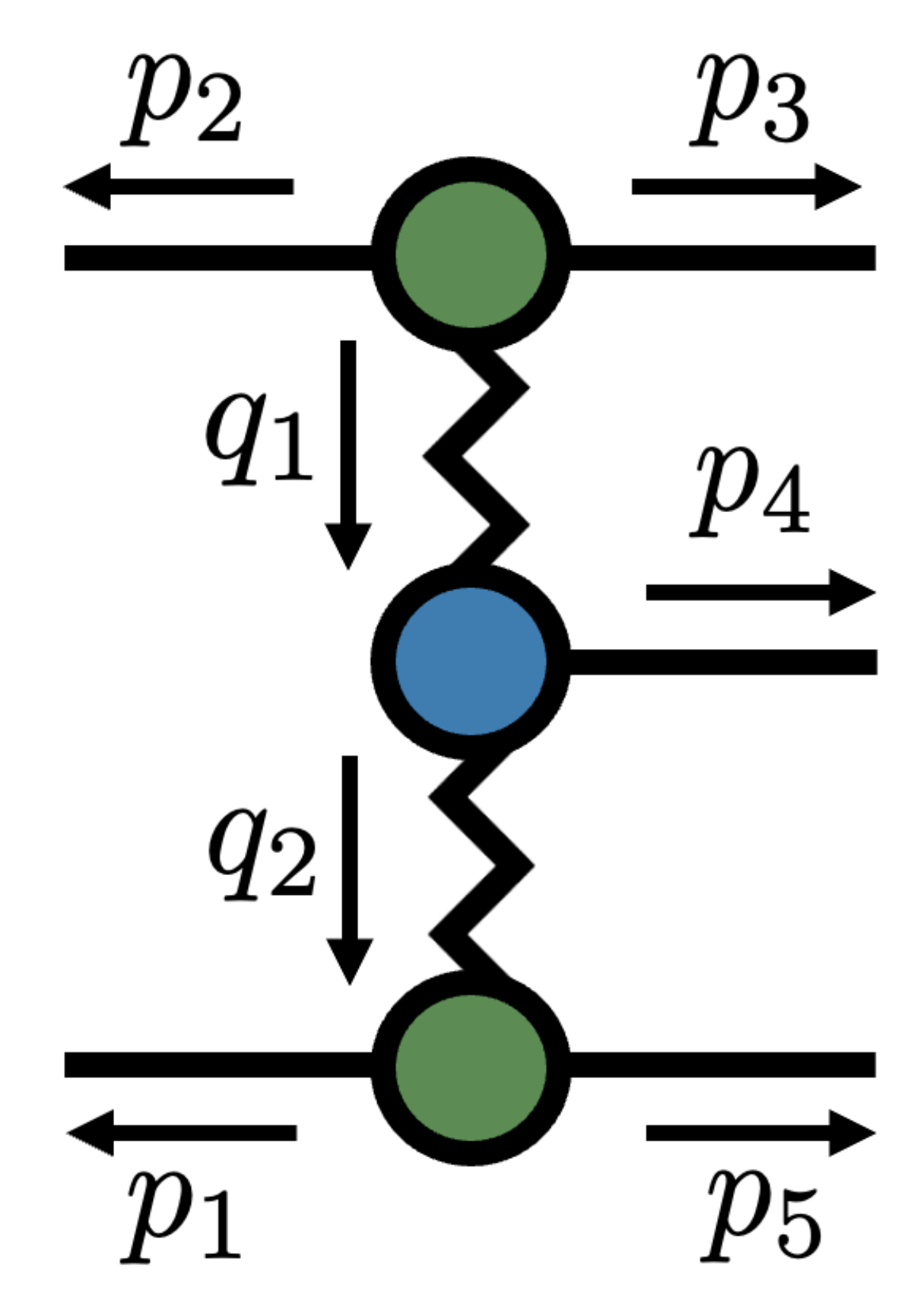}\label{subfig:Fac_MRK}}
    \qquad \qquad
    \subfigure[]{\includegraphics[width=0.23\textwidth]{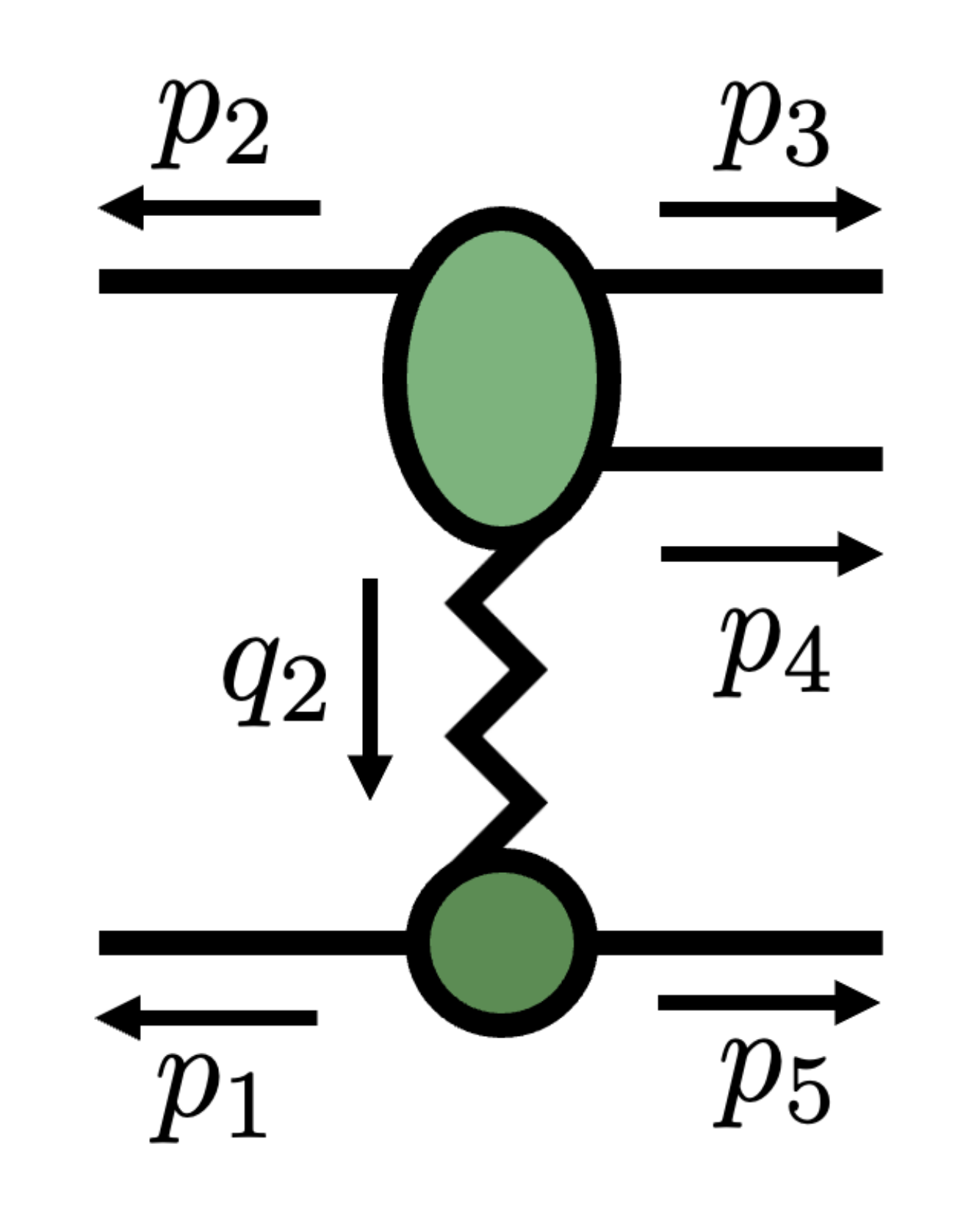}\label{subfig:Fac_NMRK}}
    \caption{Illustrations of the factorised forms of amplitudes discussed in the text: (a) factorisation of the $2 \to 2$ amplitude in the Regge limit; (b) factorisation of the $2 \to 3$ amplitude in the MRK limit; (c) factorisation of the $2 \to 3$ amplitude in the forward NMRK limit. The green and blue blobs denote peripheral- and central-emission vertices respectively, to all orders in the coupling $\gs$. The solid external lines denote the permitted parton types, as given in the main text. The zig-zag lines denote a Reggeised gluon.}
    \label{fig:Fac}
\end{figure}

All the process-dependent information is contained in the vertices $\cC^{(0)}_{f_i f_j g^*}$, which are only non-zero when the colour representation of the external partons $f_i$ and $f_j$ permits the $t$-channel exchange of an anti-symmetric octet:
\begin{subequations}
\label{eq:cC0}
\begin{align}
  \mathcal{C}^{(0)}_{g g g^*}(p_2^{\lambda_2}, p_3^{\lambda_3}, q) &= \gs \, (F^c)_{a_2 a_3} \, C^{(0)}_{g g g^*}(p_2^{\lambda_2}, p_3^{\lambda_3}, q)\,,\label{eq:cC0g}\\
  \mathcal{C}^{(0)}_{\qb q g^*}(p_2^{\lambda_2}, p_3^{\lambda_3}, q) &= \gs \, (T^c)_\ibi \, C^{(0)}_{\qb q g^*}(p_2^{\lambda_2}, p_3^{\lambda_3}, q)\,.
  \label{eq:cC0q}
\end{align}
\end{subequations}
Here $T$ are generators in the fundamental representation of $SU(N_c)$ with normalisation 
$\tr \left(
T^aT^b
\right)
=\Delta_{ab}$\footnote{We acknowledge the use of the Mathematica package ColorMath~\cite{Sjodahl:2012nk} in computing colour factors in this work.}. Note that we use the symbol $\delta_{ij}$ and $\Delta_{ab}$ for the identity in the fundamental and adjoint representations respectively. 
The structure constants, $f$, follow from the commutatation relation
\begin{equation}
[T^a, T^b]_{ij}=if^{abc} \, T^c_{ij}. \label{eq:f}
\end{equation}
We often make use of the notation $(F^b)_{ac}=if^{abc}$ for generators in the adjoint representation, whereby the Jacobi identity is
\begin{equation}
[F^a, F^b]_{d_i d_j}=if^{abc} F^c_{d_i d_j}. \label{eq:Jacobi}
\end{equation}

\Cref{eq:cC0} introduces several conventions we use throughout this paper. In order to emphasise the analogy between amplitudes and factorised vertices we have listed the three outgoing momenta of the vertices. Although this is redundant, as the vertices conserve momentum, this notation will be helpful when keeping track of the colour-orderings of higher-multiplicity vertices. For amplitudes and vertices alike, we use caligraphic type for colour-dressed objects and roman type for colour-stripped objects. In the case of \cref{eq:cC0}, the colour-stripped vertices are simply helicity conserving phases. Using the conventions of \cref{sec:kin_gen}, these phases are \cite{DelDuca:1995zy,DelDuca:1996km}
\begin{subequations}
\begin{align}
    C^{(0)}_{g g g^*}(p_2^{\ominus}, p_3^{\oplus}, q)&=1\,, \qquad
    C^{(0)}_{g^* g g}(-q, p_4^{\oplus}, p_1^{\ominus})=-\frac{q_\perp^*}{q_\perp}\,, \\
    C^{(0)}_{\qb q g^*}(p_2^{\ominus}, p_3^{\oplus}, q)&=1\,, \qquad
    C^{(0)}_{g^* \qb q}(-q, p_4^{\oplus}, p_1^{\ominus})=\left(\frac{q_\perp^*}{q_\perp}\right)^{\frac{1}{2}}\,.
\end{align}
\label{eq:C0}
\end{subequations}
The vertices are antisymmetric under $p_2^{\lambda_2}\leftrightarrow p_3^{\lambda_3}$ or $p_4^{\lambda_4}\leftrightarrow p_1^{\lambda_1}$, and parity conjugation of the vertices is given by complex conjugation. To leading-logarithmic (LL) accuracy in $\frac{s}{t}$, the amplitudes are given to all-orders by the factorised form~\cite{Lipatov:1976zz,Kuraev:1976ge}
\begin{align}
\begin{split}
 \mathcal{A}_4
(f_2^{\lambda_2}, f_3^{\lambda_3}, f_4^{\lambda_3}, f_1^{\lambda_3})
&\toRegge 2 s \ \mathcal{C}^{(0)}_{f_2 f_3 g^*}(p_2^{\lambda_2}, p_3^{\lambda_3}, q) \ 
	\frac{1}{t}\left( \frac{s}{\tau}\right)^{\alpha(t)} \ 
\mathcal{C}^{(0)}_{g^* f_4 f_1 }(-q, p_4^{\lambda_4}, p_1^{\lambda_1})\,,
\label{eq:Regge_LL}
\end{split}
\end{align}
which is illustrated in \cref{subfig:Fac_Regge}. Here $\tau$ is some positive scale of the order $t$. We consider the amplitude to have the expansion in the bare coupling $\gs$,
\begin{equation}
    \mathcal{A}_4 = \mathcal{A}_4^{(0)} + \mathcal{A}_4^{(1)} + \ord(\gs^6) \,,\label{eq:amp}
\end{equation}
where the superscript $(l)$ denotes the loop-order and $\mathcal{A}^{(l)}_4 \sim \ord(\gs^{2(1+l)})$. In \cref{eq:Regge_LL}, the virtual corrections are given to all orders by the exponentiation of the one-loop Regge trajectory~\cite{Lipatov:1976zz,Kuraev:1976ge}
\begin{align}
   \alpha^{(1)}(t) =  c_\Gamma \  \gs^2 \ \frac{2 N_c}{\epsilon}\left(\frac{\mu^2}{-t}\right)^\epsilon \,,\label{eq:trajectory_1}
\end{align}
where $c_\Gamma$ is the ubiquitous loop factor
\begin{equation}
c_\Gamma=\frac{1}{(4 \pi)^{2-\epsilon}}\frac{\Gamma(1+ \epsilon) \Gamma(1-\epsilon)^2}{\Gamma(1-2\epsilon)}\,. \label{eq:cG}
\end{equation}
Beyond LL, it is useful to introduce the concept of signature, or symmetry under crossing $s\leftrightarrow u $. In the Regge limit this is equivalent to $s\leftrightarrow -s $, and signature-even ($[+]$) and signature-odd ($[-]$) ampltiudes are given by
\begin{align}
\begin{split}
    \mathcal{A}_4^{[\pm]}(s,t)
    =\frac{1}{2}\Big(
      &\mathcal{A}_4(s,t)
\pm \mathcal{A}_4(-s,t)
    \Big)\,.
    \end{split}\label{eq:sig}
\end{align}
At NLL, a factorised form analogous to \cref{eq:Regge_LL} only holds for the signature-odd amplitude. The generalisation of \cref{eq:Regge_LL} to NLL accuracy is \cite{Caron-Huot:2017fxr,Fadin:1993wh}
\begin{align}
\begin{split}
\disp{\mathcal{A}^{[-]}_4
(f_2^{\lambda_2}, f_3^{\lambda_3}, f_4^{\lambda_4}, f_1^{\lambda_1})}
&\toRegge 2 s \ \mathcal{C}_{f_2 f_3 g^*}(p_2^{\lambda_2}, p_3^{\lambda_3}, q) \ 
	\mathcal{R}_{g^*}(q; s) \ 
\mathcal{C}_{g^* f_4 f_1 }(-q, p_4^{\lambda_4}, p_1^{\lambda_1})\,.
\label{eq:Regge_NLL}
\end{split}
\end{align}
In this paper we only consider the dispersive part of amplitudes, defined by taking the real part of the transcendental functions. For $2 \to 2$ scattering the Regge cut starts to contribute to the absorptive part of the signature-even amplitude at one-loop, and (beyond the planar limit) to the dispersive part of the signature-odd amplitude at two-loops \cite{Falcioni:2021dgr}. As outlined in the introduction, our interest is in the Regge-pole contribution to the amplitude, so we limit our analysis to the dispersive part of the signature-odd amplitude.

In \cref{eq:Regge_NLL} we have introduced the notation
\begin{align}
   \mathcal{R}_{g^*}(q; s) =\frac{1}{t} \times  \frac{1}{2} \
		\Bigg[
			\bigg(\frac{s}{\tau}\bigg)^{\alpha(t)}+\bigg(\frac{-s}{\tau}\bigg)^{\alpha(t)}
		\Bigg]\,,
\end{align}
for a signature-even Reggeisation factor. This factor is relevant for the signature-odd amplitude due to the overall factor of $s$ in \cref{eq:Regge_NLL}.
%In analogy with \cref{eq:amp} 
We consider the factorised terms in \cref{eq:Regge_NLL} to have an expansion in the coupling. At NLL we need to consider the expansion of the peripheral-emission vertices to one loop,
\begin{align}
   \mathcal{C}_{f_2 f_3 g^*}(p_2^{\lambda_2}, p_3^{\lambda_3}, q)= \mathcal{C}^{(0)}_{f_2 f_3 g^*}(p_2^{\lambda_2}, p_3^{\lambda_3}, q)+\mathcal{C}^{(1)}_{f_2 f_3 g^*}(p_2^{\lambda_2}, p_3^{\lambda_3}, q)+\mathcal{O}\left( \gs^{5}\right)
   \label{eq:cC}
\end{align}
with the (unrenormalised) one-loop correction to \cref{eq:cC0g} given by \cite{Fadin:1992rh,Fadin:1992zt,DelDuca:1998kx}
\begin{equation}
\begin{aligned}
&\mathcal{C}^{(1)}_{g g g^*}(p_2^{\oplus}, p_3^{\ominus}, q)
= \gs^3 \, c_{\Gamma} \, F^c_{a_2 a_3} \, C^{(0)}_{g g g^*}(p_2^{\oplus}, p_3^{\ominus}, q)
\\&\times
\Bigg\{
\left(\frac{\mu^2}{-t}\right)^\epsilon 
    \left[ N_c \left(-\frac{2}{\epsilon^2}-\frac{11}{6 \varepsilon}+\frac{1}{\varepsilon} \log \left(\frac{\tau}{-t}\right)-\frac{32}{9}-\frac{\delta_R}{6}+\frac{\pi^2}{2}\right) 
+N_f\left(\frac{1}{3 \varepsilon}+\frac{5}{9}\right) \right]
%-\frac{\beta_0}{2 \epsilon}
\Bigg\}\,,\label{eq:IF_g}
\end{aligned}
\end{equation}
and the (unrenormalised) one-loop correction to \cref{eq:cC0q} given by \cite{Fadin:1995km,DelDuca:1998kx}
\begin{equation}
\begin{aligned}
&\mathcal{C}^{(1)}_{\qb q g^*}(p_2^{\oplus}, p_3^{\ominus}, q)
= \gs^3 \, c_{\Gamma} \, T^c_\ibi  \, C^{(0)}_{\qb q g^*}(p_2^{\oplus}, p_3^{\ominus}, q)
\\&\times
\Bigg\{
\left( \frac { \mu ^ { 2 } } { - t } \right) ^ { \epsilon } 
\bigg[N_c\left(-\frac{1}{\epsilon^2}+\frac{1}{3 \epsilon}+\frac{1}{\varepsilon} \log \left(\frac{\tau}{-t}\right)+\frac{19}{18}-\frac{\delta_R}{3}+\frac{\pi^2}{2}\right)  -N_f \left(\frac{1}{3 \epsilon}+\frac{5}{9}\right) 
\\
&\qquad +\frac{1}{N_c}\left(\frac{1}{\epsilon^2}+\frac{3}{2 \epsilon}+\frac{7}{2}+\frac{\delta_R}{2}\right) \bigg]
%-\frac{\beta_0}{2 \epsilon}\Bigg\}
.\label{eq:IF_q}
\end{aligned}
\end{equation}
The above vertices are written in terms of the regularisation parameter $\delta_R $, where $\delta_R = 1$ in CDR/HV schemes and  $\delta_R = 0$ in the DR/FDH schemes \cite{Catani:1996pk}. In the following, because we will adopt an organisation of amplitudes into supersymmetric multiplets, we set $\delta_R = 0$.
At one-loop the helicity-violating $ggg^*$ vertex \cite{Fadin:1993qb,DelDuca:1998kx},
\begin{subequations}
\label{eq:IF_pp}
\begin{align}
\begin{split}
\mathcal{C}^{(1)}_{g g g^*}(p_2^{\oplus}, p_3^{\oplus}, q)
&= \gs^3  F^c_{a_2 a_3} (N_c-N_f)
\frac{1}{3 (4 \pi)^2}\left( \frac{-q_\perp^*}{q_\perp}\right)\,,
\label{eq:IF_pp_23}
\end{split}
\\
\begin{split}
\mathcal{C}^{(1)}_{g g g^*}(p_4^{\oplus}, p_1^{\oplus}, q)
&= \gs^3 F^c_{a_4 a_1} (N_c-N_f)
\frac{1}{3 (4 \pi)^2}\,,
\end{split}
\end{align}
\end{subequations}
also contributes at leading-power in the Regge limit. This vertex did not contribute to the NLO impact factors, but the square of \cref{eq:IF_pp} does contribute to the NNLO impact factors.

In \cref{eq:Regge_NLL} we must also consider the expansion of the Regge trajectory to two loops,
\begin{align}
   \alpha(t) =  \alpha^{(1)}(t) +  \alpha^{(2)}(t) +\mathcal{O}\left( \gs^{6}\right)\,, \label{eq:trajectory_expand}
\end{align}
but as the subject of this paper is the analysis of one-loop amplitudes, we will only need \cref{eq:trajectory_1}.
The expansion \cref{eq:trajectory_expand} is equivalent to an expansion of the Reggeised propagator,
\begin{align}
   \mathcal{R}_{g^*}(q; s) = \mathcal{R}_{g^*}^{(0)}(q) + \mathcal{R}_{g^*}^{(1)}(q; s) +\mathcal{O}\left( \gs^{3}\right)\,.
   \label{eq:cR}
\end{align}
As for the vertices, \cref{eq:cC0}, it is useful to consider the kinematic part of \cref{eq:cR}, stripped of colour factors and coupling constants, that is
\begin{align}
   \mathcal{R}^{(0)}_{g^*}(q) &= R^{(0)}_{g^*}(q)\,,\\
   \mathcal{R}^{(1)}_{g^*}(q; s) &=  \gs^2 \, N_c \,  R^{(1)}_{g^*}(q; s)\,,
\end{align}
where for clarity, we list the colour-stripped factors at tree-level and one-loop accuracy:
\begin{align}
   R^{(0)}_{g^*}(q) &= \frac{1}{t}\,,\\
   R^{(1)}_{g^*}(q; s) &= c_\Gamma \frac{1}{t} \frac{2 }{\epsilon}\left(\frac{\mu^2}{t}\right)^\epsilon \log\left( \frac{s}{\tau}\right)\,.
\end{align}
The factorised form of \cref{eq:Regge_NLL} generalises beyond the $2\to2$ case, for example, the MRK limit of the $2\to3$ amplitude (\cref{eq:MRK_5}) is~\cite{Lipatov:1976zz,Kuraev:1976ge}
\begin{align}
\begin{split}
\disp  {\mathcal{A}_5^{[-,-]}
\left(f_2^{\lambda_2}, f_3^{\lambda_3}, g_4^{\lambda_4}, f_5^{\lambda_5}, f_1^{\lambda_1}\right) }
\toMRK 2 s \  &\mathcal{C}_{f_2 f_3 g^*}(p_2^{\lambda_2}, p_3^{\lambda_3}, q_1)
\
\mathcal{R}_{g^*}(q_1; s_{34}) 
\\
\times
	& 
 \mathcal{V}_{g^* g g^*}(-q_1, p_4^{\lambda_4}, q_2) 
 \\
 \times
 & \mathcal{R}_{g^*}(q_2; s_{45})
 \
\mathcal{C}_{g^* f_5 f_1 }(-q_2, p_5^{\lambda_5}, p_1^{\lambda_1})\,,
\label{eq:Regge_5}
\end{split}
\end{align}
which is valid to LL accuracy at amplitude level. Here the amplitude consists of antisymmetric colour representations exchanged in both $t_1$ and $t_2$ channels. As above, we restrict our study to the dispersive part of the ampltiude. The phases of amplitudes in the MRK are properly treated with the analytic representation of amplitudes given in refs.~\cite{Fadin:1993wh,Bartels:1980pe,Bartels:2008ce} but \cref{eq:Regge_5} is sufficient for our present purposes. 

The central-emission vertex has an expansion in the coupling, analogous to \cref{eq:cC}. In the present work we will need to know this vertex to one-loop accuracy,
\begin{align}
\begin{split}
\mathcal{V}_{g^* g g^*}(-q_1, p_4^{\lambda_4}, q_2) = \mathcal{V}_{g^* g g^*}^{(0)}(-q_1, p_4^{\lambda_4}, q_2) + \mathcal{V}_{g^* g g^*}^{(1)}(-q_1, p_4^{\lambda_4}, q_2)+\mathcal{O}\left( \gs^{5}\right)\,.
\end{split}
\end{align}
The leading-order Lipatov vertex is given by
\begin{align}
\begin{split}
\mathcal{V}_{g^* g g^*}^{(0)}(-q_1, p_4^{\lambda_4}, q_2) &= i \, \gs \, f^{c_1  c_2 a_4} \, V_{g^* g g^*}^{(0)}(-q_1, p_4^{\lambda_4}, q_2)\,, \\
\end{split}
\end{align}
with colour-stripped vertex
\begin{align}
\begin{split}
V^{(0)}_{g^* g g^*}(-q_1, p_4^{\lambda_4}, q_2) &= \frac{q_{1\perp}^\ast q_{2\perp}}{p_{4\perp}}\,,\\
\end{split}
\end{align}
and the (unrenormalised) one-loop Lipatov vertex is given by \cite{Fadin:1993wh,Fadin:1994fj,DelDuca:1998cx}
\begin{align}
\begin{split}
&\mathcal{V}_{g^* g g^*}^{(1)}(-q_1, p_4^{\oplus}, q_2) = i \, c_\Gamma \, \gs^3 \, f^{c_1  c_2 a_4} \, V_{g^* g g^*}^{(0)}(-q_1, p_4^{\lambda_4}, q_2) 
\\
& \times
\Bigg\{
    N_c\bigg[
        -\frac{1}{\epsilon^2}\left(\frac{\mu^2}{\left|p_{4 \perp}\right|^2}\right)^\epsilon
        +\frac{1}{\epsilon}\left(\left(\frac{\mu^2}{-t_1}\right)^\epsilon+\left(\frac{\mu^2}{-t_2}\right)^\epsilon\right) \log \left(\frac{\tau}{\left|p_{4 \perp}\right|^2}\right)-\frac{1}{2} \log ^2\left(\frac{t_1}{t_2}\right)+\frac{\pi^2}{3}
    \bigg] \\
    &\qquad  -\frac{\beta_0}{2}\left(t_1+t_2+2 q_{1 \perp} q_{2 \perp}^*\right) \frac{L_0\left(t_1 / t_2\right)}{t_2}
    %-\frac{\beta_0}{2 \epsilon} 
    \\
    & \qquad -\frac{(N_c-N_f)}{3}\left|p_{4 \perp}\right|^2\left[\left(2 t_1 t_2+\left(t_1+t_2+2\left|p_{2 \perp}\right|^2\right) q_{1 \perp} q_{2 \perp}^*\right) \frac{L_2\left(t_1 / t_2\right)}{t_2^3}+\frac{q_{1 \perp} q_{2 \perp}^*}{2 t_1 t_2}\right]
\Bigg\}\,.
\label{eq:Lipatov_1}
\end{split}
\end{align}
The $L_i$ functions are defined in \cref{sec:special}, and the coefficient of the one-loop beta function in QCD is
\begin{equation}
    \beta_0 = \frac{11 N_c - 2 N_f}{3}\,.
\end{equation}

In this paper we are concerned with the factorisation properties of $2 \to 3$ scattering amplitudes in the NMRK limit, \cref{eq:NMRK_5}. The factorised form
\begin{align}
\begin{split}
\disp { \mathcal{A}_5^{[-]}
\left(f_2^{\lambda_2}, f_3^{\lambda_3}, f_4^{\lambda_3}, f_5^{\lambda_5}, f_1^{\lambda_1}\right) }
\toNMRK 2 s \ &\mathcal{C}_{f_2 f_3 f_4 g^*}(p_2^{\lambda_2}, p_3^{\lambda_3}, p_4^{\lambda_4}, q_2) \\
\times &
	\mathcal{R}_{g^*}(q_2; s_{45}) 
 \\
 \times &
\mathcal{C}_{g^* f_5 f_1 }(-q_2, p_5^{\lambda_5}, p_1^{\lambda_1})\,,
\label{eq:Regge_conjecture}
\end{split}
\end{align}
is valid to LL accuracy. Here the superscript $[-]$ now denotes the exchange of anti-symmetric colour representations in just the $t_2$ channel. At LL accuracy, it is sufficient to know the tree-level two-parton peripheral-emission vertices. The pure-gluon vertex has the structure \cite{Fadin:1989kf,DelDuca:1995ki}
\begin{equation}
\mathcal{C}^{(0)}_{g g g g^*}(p_2^{\lambda_2}, p_3^{\lambda_3}, p_4^{\lambda_4}, q_2) =\gs^2 \sum_{\sigma \in S_2} 
  (F^{a_{\sigma_3}} F^{a_{\sigma_4}})_{a_2 c_2} 
  \ C^{(0)}_{g g g g^*}(p_2^{\lambda_2}, p_{\sigma_3}^{\lambda_{\sigma_3}}, p_{\sigma_4}^{\lambda_{\sigma_4}}, q_2)\,,
  \label{eq:cC0gg}
\end{equation}
while the case with an external quark-antiquark pair has the structure
\begin{align}
\begin{split}
  \mathcal{C}^{(0)}_{\qb q g g^*}(p_2^{\lambda_2}, p_3^{\lambda_3}, p_4^{\lambda_4}, q_2) =\gs^2 \Big[
  &(T^{a_{4}} T^{c_2})_{\imath_3 \bar{\imath}_2} 
  \ C^{(0)}_{\qb q g g^*}(p_2^{\lambda_2}, p_{3}^{\lambda_3}, p_{4}^{\lambda_{4}}, q_2) \\ 
 +&(T^{c_2} T^{a_{4}} )_{\imath_3 \bar{\imath}_2} 
  \ C^{(0)}_{\qb q g g^*}(p_{4}^{\lambda_{4}}, p_2^{\lambda_2}, p_{3}^{\lambda_{3}}, q_2) 
  \Big]\,.
  \label{eq:cC0qqb}
\end{split}
\end{align}
A notable feature of \cref{eq:cC0gg,eq:cC0qqb} is that they consist of two colour structures each. The colour-stripped tree-level vertices, $C^{(0)}_{f_2 f_3 f_4 g^*}$, are reviewed in
\cref{sec:C0234}. In ref.~\cite{DelDuca:1995ki,DelDuca:1996km}, they were obtained at fixed helicity from colour-ordered amplitudes. The colour-ordered amplitudes themselves obeyed a colour-stripped analogue of the Regge pole factorisation in \cref{eq:Regge_conjecture},
\begin{align}
\begin{split}
 A^{(0)[-]}_4
\left(f_2, f_3, f_4, f_5, f_1\right)
\toNMRK 2 s \ & C^{(0)}_{f_2 f_3 f_4 g^*}(p_2^{\lambda_2}, p_3^{\lambda_3}, p_4^{\lambda_4}, q_2) \\
	 \times & R_{g^*}^{(0)}(q_2) \\
  \times &
C_{g^* f_5 f_1 }^{(0)}(-q_2, p_5^{\lambda_5}, p_1^{\lambda_1})\,.
\label{eq:Regge_CO_LO}
\end{split}
\end{align}
Motivated by ref.~\cite{Byrne:2022wzk}, it is interesting to ask whether this factorisation holds beyond tree level, that is, whether a colour-ordered version of \cref{eq:Regge_conjecture},
\begin{align}
\begin{split}
\mathrm{Disp} \left[ A_5^{[-]}
\left(f_2^{\lambda_2}, f_3^{\lambda_3}, f_4^{\lambda_3}, f_5^{\lambda_5}, f_1^{\lambda_1}\right) \right]
\toNMRK 2 s \ & C_{f_2 f_3 f_4 g^*}(p_2^{\lambda_2}, p_3^{\lambda_3}, p_4^{\lambda_4}, q_2) \\
	 \times & R_{g^*}(q_2; s_{45}) \\
  \times &
C_{g^* f_5 f_1 }(-q_2, p_5^{\lambda_5}, p_1^{\lambda_1})\,,
\label{eq:Regge_CO}
\end{split}
\end{align}
holds to LL or NLL accuracy, even at one loop. The amplitude on the left-hand side of \cref{eq:Regge_CO} represents the expansion
\begin{align}
A_n = \gs^{n-2} \ A_n^{(0)} + \gs^{n} \ A_n^{(1)} + \cO (\gs^{n+2})\,,
\end{align}
and the factorised terms on the right-hand side of \cref{eq:Regge_CO} represent analogous expansions of the colour-stripped Reggeised propagator and emission vertices,
\begin{subequations}
\label{eq:C}
\begin{align}
R_{g^*} &=  R_{g^*}^{(0)}  \quad \quad \quad \, + \gs^2 \ R_{g^*}^{(1)} \quad \quad + \cO (\gs^{4})\,,\\
C_{g^* f_5 f_1}&= \gs \ C^{(0)}_{g^* f_5 f_1}  \ \  \, +\gs^3 \ C^{(1)}_{g^* f_5 f_1} \ \  + \cO (\gs^{5})\,,\\
C_{f_2 f_3 f_4 g^*}&= \gs^2 \ C^{(0)}_{f_2 f_3 f_4 g^*}+\gs^4 \ C^{(1)}_{f_2 f_3 f_4 g^*} + \cO (\gs^{6})\,. 
\end{align}
\end{subequations}
For colour-stripped amplitudes we define signature in the $t_2$ channel by the crossing of momenta and helicity, $p_5^{\lambda_5}\leftrightarrow p_1^{\lambda_1}$, that is,
\begin{align}
\begin{split}
    &A_5^{[\pm]}(2^{\lambda_2},3^{\lambda_3},4^{\lambda_4},5^{\lambda_5},1^{\lambda_1})
    \\
    &=\frac{1}{2}\left(
A_5(2^{\lambda_2},3^{\lambda_3},4^{\lambda_4},5^{\lambda_5},1^{\lambda_1})
\pm A_5(2^{\lambda_2},3^{\lambda_3},4^{\lambda_4},1^{\lambda_1},5^{\lambda_5})
    \right)\,.
    \label{eq:sig_CO}
    \end{split}
\end{align}

In this paper our primary investigation is whether \cref{eq:Regge_conjecture} holds to NLL accuracy at one loop, which would allow us to define a one-loop correction to the two-parton emission vertex of the form
$$\mathcal{C}^{(1)}_{f_2 f_3 f_4 g^*}(p_2^{\lambda_2}, p_3^{\lambda_3}, p_4^{\lambda_4}, q_2).$$
We impose the requirement that this vertex factorises in the further MRK and $\MRKx$ limits according to \cref{eq:Regge_5}, i.e.
\begin{align}
    \mathcal{C}_{f_2 f_3 f_4 g^*}(p_2^{\lambda_2}, p_3^{\lambda_3}, p_4^{\lambda_4}, q_2)&\underset{y_3 \gg y_4}{\to} 
    \mathcal{C}_{f_2 f_3 g^*}(p_2^{\lambda_2}, p_3^{\lambda_3}, q_1)
    \mathcal{R}_{g^*}(q_1; s_{34})
    \mathcal{V}_{g^* g g^*}(-q_1, p_4^{\lambda_4}, q_2) 
    \,,\\
    \mathcal{C}_{f_2 f_3 f_4 g^*}(p_2^{\lambda_2}, p_3^{\lambda_3}, p_4^{\lambda_4}, q_2)&\underset{y_4 \gg y_3}{\to} 
    \mathcal{C}_{f_2 f_4 g^*}(p_2^{\lambda_2}, p_4^{\lambda_4}, q_1^\prime)
    \mathcal{R}_{g^*}(q_1^\prime; s_{43})
    \mathcal{V}_{g^* g g^*}(-q_1^\prime, p_3^{\lambda_3}, q_2)\,,
\end{align}
whenever the partonic identities $f_i$ permit an antisymmetric octet exchange in the $t$- or $t^\prime$- channel respectively. 
In \cref{sec:cCgg,sec:cC_qqbg}, we will find that \cref{eq:Regge_conjecture} holds for the $\bar{q}qgg^*$ case, but it fails in the $gggg^*$ case. The pure-gluon case is the only partonic channel which has leading-power behaviour in both MRK and $\MRKx$ limits, and we find that as a consequence, \cref{eq:Regge_conjecture} must be modified to include at least two such factorised terms, as in \cref{eq:5g_permitted}. In the $\bar{q}qgg^*$ case, the amplitude is power suppressed in one or both of the MRK and $\MRKx$ limits, depending on whether a fermion or gluon is physically incoming. This ultimately enables us to define a colour dressed vertex, \cref{eq:cC1q}, for which the expansion of \cref{eq:Regge_conjecture} is correct up to $\cO(g_s^5)$.

For the $gggg^*$ case, the breakdown of \cref{eq:Regge_conjecture} at one loop and NLL follows from the breakdown of \cref{eq:Regge_CO} at one loop and NLL, for a specific colour ordering (namely, where the physically incoming parton is colour adjacent to both peripherally-emitted partons). For the other one-loop colour ordered amplitudes, \cref{eq:Regge_CO} does hold to NLL, and proves to be a useful framework to organise amplitudes, even in general kinematics.

Of course, the all-orders structure of amplitudes cannot be determined from an analysis of one-loop amplitudes. Vital clues about the all-orders structure of the five-parton ampliudes in the NMRK limit can be obtained from the two-loop five parton amplitudes, which have recently been computed \cite{Agarwal:2023suw,DeLaurentis:2023nss,DeLaurentis:2023izi}. It is hoped that the current paper will provide a clear account of what we can already learn from the available one-loop amplitudes, and which will aid these future studies.
\section{Colour structure of one-loop five-parton amplitudes in the NMRK limit}
\label{sec:colour}
One of the most interesting aspects of Regge limits of amplitudes is the interplay of colour and kinematic structure. In this section we study the colour structure of one-loop five-parton amplitudes in the NMRK limit. We will utilise the one-loop DDM colour decompositions \cite{DelDuca:1999rs} for one-loop five-point amplitudes which we review in \cref{sec:colour_n}. In that section, we introduce the notation we will use for partial and primitive amplitudes. In \cref{sec:colour-5}, we study how the colour structure of one-loop five-parton amplitudes simplifies in the NMRK limit, utilising only minimal knowledge of the kinematic parts of the amplitudes (namely, whether a given partial amplitude is power suppressed in this limit). Building on ref.~\cite{Byrne:2022wzk}, in \cref{sec:simple} we move to particularly convenient bases to study one-loop amplitudes in the NMRK limit. The detailed study of the purely kinematic partial amplitudes is performed in \cref{sec:2g} (for $ggggg$ amplitudes) and \cref{sec:c_qqbg} (for $q \qb ggg$ amplitudes). In \cref{sec:cCgg} and \cref{sec:cC_qqbg} we bring together our colour and kinematic results to discuss the respective colour-dressed one-loop amplitudes and their factorisation properties in the NMRK limit.

\subsection{Review of colour structure of one-loop amplitudes}
\label{sec:colour_n}
Beyond tree level it is useful to make a distinction between partial and primitive amplitudes. By primitive amplitude we mean a gauge invariant subset of colour-ordered Feynman diagrams. By partial amplitude we mean a kinematic coefficient of a colour structure where the colour ordering is not necessarily fixed. 

For the one-loop amplitudes that we analyse in this paper we will use the DDM colour bases introduced in ref.~\cite{DelDuca:1999rs}. For $n$-gluon amplitudes the relevant colour decomposition is 
\begin{align}
\begin{split}
\mathcal{A}_{n}^{(1)}(g_1,\ldots, g_n)&= \gs^n  \sum_{\sigma \in S_{n-1} / R} \tr 
\left(F^{a_{\sigma_1}} \cdots F^{a_{\sigma_n}}\right) A^{(1,  \, g)}_{n}\left({\sigma_1}^{\lambda_{\sigma_1}}, \ldots, {\sigma_n}^{\lambda_{\sigma_n}}\right) \\
 &+ N_f \ \gs^n  \sum_{\sigma \in S_n } \tr \left(T^{a_{\sigma_1}} \cdots T^{a_{\sigma_n}}\right) 
 A^{(1, \, f)}_{n}\left(\sigma_1^{\lambda_{\sigma_1}}, \ldots, {\sigma_n}^{\lambda_{\sigma_n}}\right)\,,
 \label{eq:DDM_ng}
 \end{split}
\end{align}
where $S_{n-1}=S_n/Z_n$ is the group of non-cyclic permutations and $R$ is the reflection operation $R(1, 2, \ldots, n-1, n)=R(n, n-1 , \ldots, 2,1)$. Here, $A_n^{(1,\ m)}$ are the one-loop primitive amplitudes with $n$ external partons, and particle content $m$ circulating in the loop. We use the notation $g$ or $f$ to indicate a gluon or fermion is circulating in the loop. We will later also consider the possibility of a complex scalar, $m{=}s$, circulating in the loop. We note the $n$-gluon partial amplitudes obey the reflection identity
\begin{equation}
    A_n^{(1, \, m)}\left(1, 2, \ldots, n-1, n\right) = (-1)^n A_n^{(1, \, m)}\left(n ,n-1 , \ldots, 2,1\right) \,.\label{eq:reflection}
\end{equation}
For amplitudes with $(n-2)$ gluons and a quark-antiquark pair the analogous colour decomposition is~\cite{DelDuca:1999rs}
\begin{align}
\begin{split}
\mathcal{A}_{n}^{(1)}\left(\bar{q}_1, q_2, g_3, \ldots, g_n\right)
= & \gs^n
\Bigg[
\sum_{p=2}^n 
\sum_{\sigma \in S_{n-2}}
\left(T^{c_2} T^{a_{\sigma_3}} \cdots T^{a_{\sigma_p}} T^{c_1}\right)_\ibio
\left(F^{a_{\sigma_{p+1}}} \cdots F^{a_{\sigma_n}}\right)_{c_1 c_2} \\
& 
\times A^{(1, \, g_R)}_{n} 
\left(1_{\qb}^{\lambda_1}, {\sigma_{p+1}}^{\lambda_{\sigma_{p+1}}}, \ldots, {\sigma_{n}}^{\lambda_{\sigma_{n}}}, 2_q^{\lambda_2} , {\sigma_3}^{\lambda_3}, \ldots, {\sigma_p}^{\lambda_p}\right)
\\
& \mkern-180mu
+ 
\frac{\Nf}{N_c} \sum_{j=1}^{n-1} \sum_{\sigma \in S_{n-2} / S_{n ; j}} \operatorname{Gr}_{n ; j}^{(\bar{q} q)}\left(\sigma_3, \ldots, \sigma_n\right) 
%A_{\qb q g\cdots g ; j}^{(1, \ q)}
A_{n; j}^{(1, \, f)}
\left(1_\qb^{\lambda_1}, 2_q^{\lambda_2} ; {\sigma_3}^{\lambda_3}, \ldots, {\sigma_n}^{\lambda_n}\right)\Bigg]\,.
\label{eq:DDM_qqbn}
\end{split}
\raisetag{2\normalbaselineskip}
\end{align}
For primitive amplitudes with an external quark-antiquark pair, diagrams without closed fermion or scalar loops are assigned to $m{=}g$ because these diagrams mix under gauge transformations with diagrams with closed gluon loops. 
Following ref.~\cite{Bern:1994fz}, we separate primitive amplitudes with an external quark-antiquark pair into individually gauge invariant terms,
\begin{equation}
A^{(1, \, m)}_{n} = A^{(1, \, m_R)}_{n} +A^{(1, \, m_L)}_{n} \,.
\end{equation}
The label $R$ or $L$ denotes the gauge invariant subset of colour-ordered Feynman diagrams in which the fermion line passes to the right or left of the closed loop (with the diagrammatic convention that colour ordered partons are labeled clockwise). The $L$ and $R$ primitive amplitudes are related by a reflection identity analogous to \cref{eq:reflection}:
\begin{equation}
A_n^{(1, \, m_R)}\left(1_{\bar{q}}, 3,4, \ldots, 2_q, \ldots, n-1, n\right)=(-1)^n A_n^{(1, \, m_L)}\left(1_{\bar{q}}, n, n-1, \ldots, 2_q, \ldots, 4,3\right)\,.
\label{eq:reflection_q}
\end{equation}
Note that for $m {\in} \{f,s\}$ the following primitive amplitudes vanish:
\begin{subequations}
\label{eq:vanish}
\begin{align}
    A_n^{(1,\, m_R)}(1_\qb, 2_q, 3, 4, \ldots, n )&= A_n^{(1,\, m_L)}(1_\qb, n, \ldots, 4, 3, 2_q)=0\,,
    \label{eq:tadpole}
    \\
    A_n^{(1,\, m_R)}(1_\qb, 3, 2_q, 4, \ldots, n )&= A_n^{(1,\, m_L)}(1_\qb, n, \ldots, 4, 2_q, 3)=0\,.
    \label{eq:bubble}
\end{align}
\end{subequations}
These identities can be understood by considering the Feynman diagrams which contribute to the respective amplitudes. The amplitudes in \cref{eq:tadpole} and \cref{eq:bubble} permit only tadpole and bubble diagrams respectively, which vanish in dimensional regularisation. We emphasise that \cref{eq:vanish} does not hold for $m{=}g$ because, as stated previously, the external fermion line may form part of the gluon loop. 

The pure-gluon decomposition, \cref{eq:DDM_ng}, is written solely in terms of primitive amplitudes. This is not the case for one-loop amplitudes with an external quark-antiquark pair, \cref{eq:DDM_qqbn}, where partial amplitudes, not primitive amplitudes, multiply the 
colour structures $\operatorname{Gr}^{(q \qb)}_{n;j}$. These colour structures are defined by
\begin{align}
\begin{split}
\operatorname{Gr}_{n ; 1}^{(\bar{q} q)}(3, \ldots, n) & =N_c\left(T^{a_3} \cdots T^{a_n}\right)_\ibio \,, \\
\operatorname{Gr}_{n ; 2}^{(\bar{q} q)}(3 ; 4, \ldots, n) & =0 \,,\\
\operatorname{Gr}_{n ; j}^{(\bar{q} q)}(3, \ldots, j+1 ; j+2, \ldots, n) & =\tr\left(T^{a_3} \cdots T^{a_{j+1}}\right)\left(T^{a_{j+2}} \cdots T^{a_n}\right)_\ibio, \quad 3 \leq j \leq n-2, \\
\operatorname{Gr}_{n ; n-1}^{(\bar{q} q)}(3, \ldots, n) & =\tr\left(T^{a_3} \cdots T^{a_n}\right) \delta_\ibio\,,
\end{split}
\raisetag{1\normalbaselineskip}
\end{align}
and $S_{n;j}=Z_{j-1}$ is the subgroup of $S_{n-2}$ that leaves $\operatorname{Gr}^{(q \qb)}_{n;j}$ invariant. The partial amplitudes which multiply these colour structures can be written as a sum of the simpler primitive amplitudes~\cite{Bern:1994fz},
\begin{subequations}
\begin{align}
A_{n ; 1}^{(1, \, f)}\left(1_{\bar{q}}, 2_q, 3, \ldots, n\right) & =A_n^{(1, \, f_L)}\left(1_{\bar{q}}, 2_q, 3, \ldots, n\right), \label{eq:partial_1}\\
A_{n ; j>1}^{(1, \, f)}\left(1_{\bar{q}}, 2_q ; 3, \ldots, j{+}1 ; j{+}2, \ldots, n\right) & =(-1)^j \sum_{\sigma \in \operatorname{COP}\{\alpha\}\{\beta\}} A_n^{(1, \, f_R)}\left({\sigma_1}_{\bar{q}}, {\sigma_2}_q, \sigma_3, \ldots, \sigma_n\right),
\label{eq:partial_j}
\end{align}
\end{subequations}
with $\{\alpha \}=\{j{+}1, j, \ldots, 4,3\},\{ \beta \}=\{1,2, j{+}2, j{+}3, \ldots, n{-}1, n\}$, and where $\operatorname{COP}\{\alpha\}\{\beta\}$ is the set of all permutations of $\{1, \ldots, n\}$ with leg 1 held fixed that preserve the cyclic ordering of the elements $\alpha_i$ within $\{\alpha \}$ and the elements $\beta_i$ within $\{\beta \}$, while allowing for all possible relative orderings of the $\alpha_i$ with respect to the $\{\beta \}$ \cite{Bern:1993mq}.

This concludes our review of the colour structure of one-loop amplitudes. In the next section we show how the colour structure of one-loop five-parton amplitudes simplifies in the NMRK limit.

\subsection{Colour structure of one-loop five-parton amplitudes in a peripheral NMRK limit}
\label{sec:colour-5}
In \cref{sec:colour-4} we review the colour structure of one-loop four-parton amplitudes in the Regge limit. In this section we perform a similar analysis of one-loop five-parton amplitudes in the forward NMRK limit, \cref{eq:NMRK_5}.
Let us first consider the five-gluon amplitude for the physical process 
$$g(p_1) g(p_2) \to g(p_3) g(p_4) g(p_5),$$
for which we need the $n=5$ case of \cref{eq:DDM_ng}, 
\begin{align}
\begin{split}
\mathcal{A}_{5}^{(1)}(g_1,g_2,g_3,g_4,g_5)&= \gs^5  \sum_{\sigma \in S_{4} / \mathcal{R}} \tr 
\left(F^{a_{\sigma_1}} \cdots F^{a_{\sigma_5}}\right) A^{(1, \ g)}_{5}\left({\sigma_1}, \ldots, {\sigma_5}\right) \\
 &+ N_f \ \gs^5  \sum_{\sigma \in S_5 } \tr \left(T^{a_{\sigma_1}} \cdots T^{a_{\sigma_5}}\right) 
 A^{(1, \ f)}_{5}\left(\sigma_1, \ldots, {\sigma_5}\right)\,.
 \label{eq:DDM_5g}
 \end{split}
\end{align}
%we repeat the steps of \cref{sec:colour-4}. 
We then make use of the fact that the only colour orderings which contribute to leading power are those that have gluons 1 and 5 cyclically adjacent. We have explicitly checked that all other colour orderings are power suppressed, for all helicity configurations. We therefore obtain
\begin{align}
\begin{split}
\mathcal{A}_{5}^{(1)}(g_1,g_2,g_3,g_4,g_5)\toNMRK \gs^5  
\sum_{\sigma \in Z_3 } 
\Big[
\tr 
\left(F^{a_{\sigma_2}} F^{a_{\sigma_3}} F^{a_{\sigma_4}} F^{a_5} F^{a_1}\right) 
&A^{(1, \ g)}_{5}\left({\sigma_2}, {\sigma_3},{\sigma_4}, 5,1\right)
\\
+
\tr 
\left(F^{a_{\sigma_2}} F^{a_{\sigma_3}} F^{a_{\sigma_4}} F^{a_1} F^{a_5}\right) 
&A^{(1, \ g)}_{5}\left({\sigma_2}, {\sigma_3},{\sigma_4}, 1,5\right)
\\
+ N_f 
\big(\tr 
\left(T^{a_{\sigma_2}} T^{a_{\sigma_3}} T^{a_{\sigma_4}} T^{a_{5}} T^{a_{1}}\right) 
-
\tr \left(T^{a_{\sigma_4}} T^{a_{\sigma_3}} T^{a_{\sigma_2}} T^{a_{1}} T^{a_{5}}\right) 
\big)
& A^{(1, \ f)}_{5}\left({\sigma_2}, {\sigma_3},{\sigma_4}, 5,1\right)
\\
+ N_f 
\big(\tr \left(T^{a_{\sigma_2}} T^{a_{\sigma_3}} T^{a_{\sigma_4}} T^{a_{1}} T^{a_{5}}\right) 
-
\tr \left(T^{a_{\sigma_4}} T^{a_{\sigma_3}} T^{a_{\sigma_2}} T^{a_{5}} T^{a_{1}}\right) 
\big)
& A^{(1, \ f)}_{5}\left({\sigma_2}, {\sigma_3},{\sigma_4}, 1,5\right)
\Big]
\,,
\label{eq:DDM_5gNMRK}
\end{split}
\end{align}
where the sum is over cyclic permutations of the set $\{2,3,4\}$.
The terms in \cref{eq:DDM_5gNMRK} are pairwise related by the interchange of gluons 1 and 5, which naturally suggests that we consider amplitudes of definite signature in the $t_2=s_{51}$ channel.
For the signature-odd amplitude, we obtain
\begin{align}
\begin{split}
&\mathcal{A}_{5}^{(1)[-]}(g_1,g_2,g_3,g_4,g_5)\toNMRK \gs^5  \left(-F^{d}_{a_5 a_1}\right)
\\
&\times
\sum_{\sigma \in Z_3 } 
\Bigg[
\tr 
\big(F^{a_{\sigma_2}} F^{a_{\sigma_3}} F^{a_{\sigma_4}} F^{d}\big) 
A^{(1, \ g)[-]}_{5}\left({\sigma_2}, {\sigma_3},{\sigma_4}, 5,1\right)
\\
& \quad + N_f 
\Big(
    \tr 
    \big(T^{a_{\sigma_2}} T^{a_{\sigma_3}} T^{a_{\sigma_4}} T^{d}
    \big) 
    +
    \tr 
    \big(T^{a_{\sigma_4}} T^{a_{\sigma_3}} T^{a_{\sigma_2}} T^{d}
    \big) 
\Big)
 A^{(1, \ f)[-]}_{5}\left({\sigma_2}, {\sigma_3},{\sigma_4}, 5,1\right)
\Bigg]
\,,
\label{eq:DDM_5gNMRK_odd}
\end{split}
\end{align}
where we have used the structure constant definitions and Jacobi identity (\cref{eq:Jacobi,eq:f}) to simplify the traces
\begin{subequations}
\begin{align}
\tr\big(\cdots [F^a, F^b ] \cdots \big)
&=\tr\big( \cdots F^c\cdots \big)F^c_{ba} \,,
\\
\tr\big(\cdots [T^a, T^b ] \cdots \big)
&
=\tr\big( \cdots T^c\cdots \big)F^c_{ba} \,.
\end{align}
\end{subequations}
We note that an overall factor $F^d_{a_5 a_1}$ factorises from the colour structures in \cref{eq:DDM_5gNMRK_odd}, and furthermore, that the remaining colour structures are exactly that of the one-loop four-gluon amplitude, \cref{eq:DDM_4g}, with $a_1 \to d$. %No such simplification occurs in the signature-even case. 

We now turn our attention to the one-loop amplitude with an external quark-antiquark pair. We consider the physical process
$$q(p_1) g(p_2) \to q(p_3) g(p_4) g(p_5)$$
as an example, but the following analysis applies to the other physical processes obtained by crossing particles with momenta in the set $\{p_2, p_3, p_4\}$, e.g.,
$gg \to q \qb g.$
We start from the $n=5$ case of \cref{eq:DDM_qqbn}:
\begin{align}
\begin{split}
\mathcal{A}_5^{(1)}
\left(\bar{q}_2, q_3, g_4, g_5, g_1\right)
=  g^5
\Bigg\{
    \sum_{\sigma \in S_3}  
    \bigg[
        &\left(T^{c_2} T^{c_1}\right)_\ibi
        \left(F^{a_{\sigma_4}} F^{a_{\sigma_5}} F^{a_{\sigma_1}}\right)_{c_1 c_2} 
        A_{5}^{(1, \ g_R)}
        \left(2_{\bar{q}}, \sigma_4, \sigma_5, \sigma_1, 3_q \right) 
        \\
         +&\left(T^{c_2} T^{a_{\sigma_4}} T^{c_1}\right)_\ibi
         \left(F^{a_{\sigma_5}} F^{a_{\sigma_1}}\right)_{c_1 c_2} 
         A_5^{(1, \ g_R)}
         \left(2_{\bar{q}}, \sigma_5, \sigma_1, 3_q, \sigma_4\right) 
         \\
         +&\left(T^{c_2} T^{a_{\sigma_4}} T^{a_{\sigma_5}} T^{c_1}\right)_\ibi
         \left(F^{d_{\sigma_1}}\right)_{c_1 c_2} 
         A_5^{(1, \ g_R)}\left(2_{\bar{q}}, \sigma_1, 3_q, \sigma_4, \sigma_5\right) \\
         +&\left(T^{c_2} T^{a_{\sigma_4}} T^{a_{\sigma_5}} T^{a_{\sigma_1}} T^{c_1}\right)_\ibi 
         \delta_{c_1 c_2} 
         A_5^{(1, \ g_R)}
         \left(2_{\bar{q}}, 3_q, \sigma_4, \sigma_5, \sigma_1\right)
     \bigg] 
     \\
    + 
    \frac{N_f}{N_c}
    \bigg[
        \sum_{\sigma \in S_{3}} 
        &N_c
        (T^{a_{\sigma_4}}T^{a_{\sigma_5}}T^{a_{\sigma_1}})_\ibi
        A_{5 ; 1}^{(1, \ f)}
        \left(2_{\bar{q}}, 3_q ; \sigma_4, \sigma_5, \sigma_1 \right)
        \\
        +\sum_{\sigma \in S_{3}/Z_2} 
        &\tr \left(T^{a_{\sigma_4}}T^{a_{\sigma_5}}\right)
        (T^{\sigma_1})_\ibi
        A_{5 ; 3}^{(1, \ f)}
        \left(2_{\bar{q}}, 3_q ; \sigma_4, \sigma_5, \sigma_1 \right)
        \\
        +\sum_{\sigma \in S_{3}/Z_3} &\tr(T^{a_{\sigma_4}}T^{a_{\sigma_5}}T^{a_{\sigma_1}})
        \delta_\ibi
        A_{5 ; 4}^{(1, \ f)}
        \left(2_{\bar{q}}, 3_q ; \sigma_4, \sigma_5, \sigma_1 \right)
    \bigg]
\Bigg\}
\end{split}
\end{align}
Again we find that the only leading-power primitive amplitudes are those with gluons 1 and 5 colour-adjacent. As for the pure-gluon case, the signature-odd amplitude simplifies considerably in the NMRK limit:
\begin{align}
\begin{split}
&\mathcal{A}_5^{(1)[-]}
\left(\bar{q}_2, q_3, g_4, g_5, g_1\right)
\toNMRK  g^5 \left( -F^d_{a_5 a_1}\right)
\\
& \hspace{4 em} \times
\Bigg\{
        \big(T^{c_2} T^{c_1}\big)_\ibi
        \big(F^{a_4} F^{d}\big)_{c_1 c_2} 
        \,
        A_{5}^{(1, \, g_R)[-]}
        \left(2_{\bar{q}}, 4,5,1, 3_q \right) 
        \\
        & \hspace{5 em}
        +\big(T^{c_2} T^{c_1}\big)_\ibi
        \big(F^{d} F^{a_4}\big)_{c_1 c_2} 
        \,
        A_{5}^{(1, \, g_R)[-]}
        \left(2_{\bar{q}}, 5,1,4, 3_q \right) 
        \\& \hspace{5 em}
         +\big(T^{c_2} T^{a_4} T^{c_1}\big)_\ibi
         \big(F^{d}\big)_{c_1 c_2} 
         \,
         A_{5}^{(1, \, g_R)[-]}
         \left(2_{\bar{q}}, 5,1, 3_q, 4\right) 
         \\
        & \hspace{5 em}
         +\big(T^{c_2} T^{d} T^{c_1}\big)_\ibi
         \big(F^{a_4}\big)_{c_1 c_2} 
         \,
         A_{5}^{(1, \, g_R)[-]}
         \left(2_{\bar{q}}, 4, 3_q, 5,1\right) 
         \\& \hspace{5 em}
         +\left(T^{c_2} T^{a_4} T^{d} T^{c_1}\right)_\ibi 
         \delta_{c_1 c_2} 
         \,
         A_{5}^{(1, \, g_R)[-]}
         \left(2_{\bar{q}}, 3_q, 4,5,1\right)
        \\& \hspace{5 em}
         +\left(T^{c_2} T^{d} T^{a_4} T^{c_1}\right)_\ibi 
         \delta_{c_1 c_2} 
         A_{5}^{(1, \, g_R)[-]}
         \left(2_{\bar{q}}, 3_q, 5,1,4\right)
     \\
    &\hspace{5 em}
    + 
    \frac{N_f}{N_c}
    \bigg[
        N_c
        (T^{a_4}T^{d})_\ibi
        A_{5 ; 1}^{(1, \, f)[-]}
        \left(2_{\bar{q}}, 3_q ; 4,5,1 \right)
        \\
        &\hspace{7 em}
        +
        N_c
        (T^{d}T^{a_4})_\ibi
        A_{5 ; 1}^{(1, \, f)[-]}
        \left(2_{\bar{q}}, 3_q ; 5,1,4 \right)
        \\&
        \hspace{7 em}
        +\tr(T^{a_4}T^{d})
        \delta_\ibi
        A_{5 ; 4}^{(1, \, f)[-]}
        \left(2_{\bar{q}}, 3_q ; 4,5,1 \right)
    \bigg]
\Bigg\}
\label{eq:DDM_qqb_NMRK_odd}
\end{split}
\end{align}
As for the pure-gluon case, we see that an overall colour factor of $F^d_{a_5 a_1}$ factorises from all terms, that is to say, of the antisymmetric representations, only $\mathbf{8}_a$ contributes to leading power. We note that the remaining colour structures in the braces of \cref{eq:DDM_qqb_NMRK_odd} are equal to the colour structures of the one-loop four-parton amplitude \cref{eq:DDM_4_q} with the replacement of $a_1 \to d$. 

We now write the partial amplitudes in the final two lines of \cref{eq:DDM_5gNMRK_odd} in terms of primitive amplitudes. From \cref{eq:partial_1} we simply obtain
\begin{align}
A_{5 ; 1}^{(1, \, f)[-]}
\left(2_{\bar{q}}, 3_q ; 4,5,1 \right)
=
&A_{5}^{(1, \, f_L)[-]}
\left(2_{\bar{q}}, 3_q , 4,5,1 \right)\,,
\\
A_{5 ; 1}^{(1, \, f)[-]}
\left(2_{\bar{q}}, 3_q ; 5,1,4 \right)
=
&A_{5}^{(1, \, f_L)[-]}
\left(2_{\bar{q}}, 3_q , 5,1,4 \right)\,,
\end{align}
even in general kinematics. For the remaining partial amplitude, the leading-power terms of \cref{eq:partial_j} are
\begin{align}
\begin{split}
A_{5 ; 4}^{(1, \, f)[-]}
\left(2_{\bar{q}}, 3_q ; 4,5,1 \right)
\toNMRK
&A_{5}^{(1, \, f_R)[-]}
\left(2_{\bar{q}},1,5,3_q,4 \right)
\\
+&A_{5}^{(1, \, f_R)[-]}
\left(2_{\bar{q}},1,5,4,3_q \right)
\\
+&A_{5}^{(1, \, f_R)[-]}
\left(2_{\bar{q}},4,1,5,3_q \right)\,. \label{eq:A51}
\end{split}
\end{align}
This partial amplitude has the same colour structure as the four-parton partial amplitude \cref{eq:A43q}. In the four-parton case, this partial amplitude vanishes due to a combination of \cref{eq:tadpole} and \cref{eq:furry}. If we were to take seriously the analogy of the antisymmetrised pair of momenta $p_1$ and $p_5$ with an effective of-shell gluon in the NMRK limit, we would expect that the first term in \cref{eq:A51} vanishes due to an analogue of \cref{eq:tadpole}, while the second and third terms cancel due to Furry's theorem (which in QED is valid for on-shell and off-shell photons). In \cref{sec:c_qqbg} will study the one-loop five-parton primitive amplitudes in the NRMK limit and we will see that this is indeed the case. 
\subsection{Simplified colour bases for one-loop amplitudes in the NMRK limit}
\label{sec:simple}
While the DDM bases are very useful for the gauge-invariant organisation of the kinematics, they do not make manifest the factorisation of amplitudes in MRK limits. Furthermore, while the basis in \cref{eq:DDM_qqb_NMRK_odd} provides a useful correspondence between colour structure and spin structure, it will be convenient to move to a smaller basis. In this section we rewrite \cref{eq:DDM_5gNMRK_odd} and \cref{eq:DDM_qqb_NMRK_odd} in a minimal basis. 
As in ref.~\cite{Byrne:2022wzk}, we choose a basis where two elements are the colour-structures of the tree-level vertices \cref{eq:cC0gg,eq:cC0qqb}.
This choice is particularly useful for comparing the NLO and NNLO contributions to the BFKL kernel. 
For the pure-gluon case we obtain
\begin{align}
\begin{split}
  &\mathcal{A}^{(1)[-]}_{5}(2,3,4,5,1) =\gs^5 (-F^d_{a_5 a_1})
  \\
  &\times 
  \Bigg\{
  \sum_{\sigma \in S_2} 
  (F^{a_{\sigma_3}} F^{a_{\sigma_4}})_{a_2 c_2} 
  \\
  &\hspace{2 em} \times
  \ \bigg[ N_c \, 
    \left( \frac{2}{3}A_{5}^{(1,g)[-]}(2,\sigma_3,\sigma_4,5,1)-\frac{1}{3}A_{5}^{(1,g)[-]}(\sigma_3,2,\sigma_4,5,1)-\frac{1}{3}A_{5}^{(1,g)[-]}(\sigma_3,\sigma_4,2,5,1)\right)
  \\
  & \hspace{3 em}
  +N_f \, \left( \frac{2}{3}A_{5}^{(1,f)[-]}(2,\sigma_3,\sigma_4,5,1)-\frac{1}{3}A_{5}^{(1,f)[-]}(\sigma_3,2,\sigma_4,5,1)-\frac{1}{3}A_{5}^{(1,f)[-]}(\sigma_3,\sigma_4,2,5,1)\right)
  \bigg]
  \\
  &\qquad +
  d_A^{a_2 a_3 a_4 c_2}\left( A_{5}^{(1,g)[-]}(2,3,4,5,1)+A_{5}^{(1,g)[-]}(3,2,4,5,1)+A_{5}^{(1,g)[-]}(3,4,2,5,1)\right)
  \\
  &\qquad +
  2 \, d_F^{a_2 a_3 a_4 c_2}\left( A_{5}^{(1,f)[-]}(2,3,4,5,1)+A_{5}^{(1,f)[-]}(3,2,4,5,1)+A_{5}^{(1,f)[-]}(3,4,2,5,1)\right)
  \Bigg\}
  \label{eq:A1_g_simple}
\end{split}
\end{align}
The new colour structures at one-loop are the totally symmetrised traces in the adjoint and fundamental representations,
\begin{align}
    d_A^{a_1 a_2 a_3 a_4}&=\frac{1}{4 !} \sum_{\sigma \in S_4} \operatorname{tr}\left(F^{a_{\sigma_1}} F^{a_{\sigma_2}} F^{a_{\sigma_3}} F^{a_{\sigma_4}}\right)\,,\\
    d_F^{a_1 a_2 a_3 a_4}&=\frac{1}{4 !} \sum_{\sigma \in S_4} \operatorname{tr}\left(T^{a_{\sigma_1}} T^{a_{\sigma_2}} T^{a_{\sigma_3}} T^{a_{\sigma_4}}\right)\,.
\end{align}
% Use Tikz?
%\feynmandiagram [horizontal=a to b] {
%  i1 -- [gluon] a -- [gluon] i2,
%  a -- [gluon] b,
%  f1 -- [gluon] b -- [gluon] f2,
%};
Noting the correspondence between colour-representations and loop content $m$, we define the following two linear combinations of primitive amplitudes:
\begin{align}
     A_{FF}^{(1,\, m)}(2,3,4,5,1)&=\frac{2}{3}A_{5}^{(1,\, m)}(2,3,4,5,1)-\frac{1}{3}A_{5}^{(1,\, m)}(2,4,3,5,1)-\frac{1}{3}A_{5}^{(1,\, m)}(4,2,3,5,1)\,,
     \label{eq:AFF}
     \\
    A_{d}^{(1, \, m)}(2,3,4,5,1)&=A_{5}^{(1,\, m)}(2,3,4,5,1)+A_{5}^{(1,\, m)}(2,4,3,5,1)+A_{5}^{(1,\, m)}(4,2,3,5,1)\,,
     \label{eq:Ad}
\end{align}
where we have made use of the reflection identity \cref{eq:reflection} and the antisymmetry of the signature odd amplitudes under permuting gluons 5 and 1.

Similarly, for the quark-antiquark case we find
\begin{align}
\begin{split}
&\mathcal{A}_5^{(1 )[-]}
\left(\bar{q}_2, q_3, g_4, g_5, g_1\right)
\to  g^5 \left( -F^d_{51}\right)
\\
&\times
\Bigg\{
        \big(T^{d} T^{a_4}\big)_\ibi
			\bigg(
N_c A_{5}^{(1, \ g_L)[-]}\left(2_{\bar{q}}, 3_q ,5,1,4  \right) 
-\frac{1}{N_c} A_{5}^{(1, \ g_R)[-]}\left(2_{\bar{q}}, 3_q ,5,1,4  \right) 
\\
&\hspace{24 em}+N_f A_{5}^{(1, \ f_L)[-]}\left(2_{\bar{q}}, 3_q ,5,1,4  \right)
			\bigg)
\\
& \hspace{1 em} +         \big(T^{a_4} T^{d}\big)_\ibi
			\bigg(
N_c A_{5}^{(1, \ g_L)[-]}\left(2_{\bar{q}}, 3_q ,4,5,1  \right) 
-\frac{1}{N_c} A_{5}^{(1, \ g_R)[-]}\left(2_{\bar{q}}, 3_q ,4,5,1 \right) 
\\
&\hspace{24 em}
+N_f A_{5}^{(1, \ f_L)[-]}\left(2_{\bar{q}}, 3_q ,4,5,1 \right)
			\bigg)
        \\
    & \hspace{1 em}  + \delta_\ibi \Delta_{d a_4} \bigg(
    A_{5}^{(1, \ g_L)[-]}\left(2_{\bar{q}}, 3_q ,4,5,1  \right)
   +A_{5}^{(1, \ g_R)[-]}\left(2_{\bar{q}}, 3_q, 4 ,5,1 \right)
   \\
& \hspace{5.5 em}
+A_{5}^{(1, \ g_L)[-]}\left(4, 2_{\bar{q}}, 3_q ,5,1  \right) 
   +A_{5}^{(1, \ g_R)[-]}\left(4, 2_{\bar{q}}, 3_q ,5,1 \right) 
   \\
   & \hspace{5.5 em}
   +A_{5}^{(1, \ g_L)[-]}\left(2_{\bar{q}},4, 3_q ,5,1  \right) 
   +A_{5}^{(1, \ g_R)[-]}\left(2_{\bar{q}},4, 3_q ,5,1  \right) 
\bigg)
\Bigg\}\,.
\label{eq:A1_q_simple}
\end{split}
\end{align}
We see the kinematic coefficient of the $\delta \Delta$ colour structure can be conveniently written in terms of 
\begin{align}
    A_{\delta \Delta}^{(1, \, m)}(2_\qb,3_q,4,5,1)&=A_{5}^{(1,\, m)}(2_\qb,3_q,4,5,1)+A_{5}^{(1,\, m)}(2_\qb,4,3_q,5,1)+A_{5}^{(1,\, m)}(4,2_\qb,3_q,5,1)\,.
    \label{eq:A_delta}
\end{align}
We note that the totally symmetric colour-structures $d_A$ and $d_F$ will not contribute to the NNLO impact factors, as they vanish when interfered with the tree-level colour structures. This is not the case for the $\delta \Delta$ colour structures.

Our goal is to study the factorisation properties of \cref{eq:AFF,eq:Ad} for $m{\in}\{g,f\}$ and \cref{eq:A_delta} for $m{\in}\{g_L,g_R\}$. In the next section we first study the factorisation of the constituent primitive amplitudes of the former.
\section{One-loop primitive vertices for the peripheral emission of two gluons}
\label{sec:2g}
In this section we investigate the NMRK limit of one-loop, colour-ordered, five-gluon amplitudes. Such amplitudes can be organised along the lines of colour ordering, helicity configuration, and loop content. We have organised this section firstly in terms of loop content, that is, by what particle type, or particle multiplet, is circulating in the loop. Following the organisation of ref.~\cite{Bern:1993mq}, we investigate the $\Nfour$ amplitudes in \cref{sec:gN4}, the $\NoneX$ amplitudes in \cref{sec:gN1X} and the amplitudes with a complex scalar circulating in the loop in \cref{sec:gs}. Other multiplets (not least the physically relevant gluon and fermion contributions) can be obtained through supersymmetric rearrangements \cite{Bern:1993mq,Bern:1994zx,Bern:1994cg}, which we now briefly review. 

We can decompose one-loop pure-gluon primitive amplitudes in supersymmetric theories into a sum of primitive amplitudes with a spin-1 boson, spin-$\frac{1}{2}$ fermion or complex scalar circulating in the loop. In this paper we consider the following supersymmetric multiplets:
\begin{subequations}
\label{eq:susy}
\begin{align}
    A_{n}^{(1, \, \Nfour)} &= A_{n}^{(1, \, g)} + 4 \ A_{n}^{(1, \, f)}+ 3 \ A_{n}^{(1, \, s)}\,,\\
    A_{n}^{(1, \, \NoneV)} &= A_{n}^{(1, \, g)}+ A_{n}^{(1, \, f)}\,,\\
    A_{n}^{(1, \, \NoneX)} &= A_{n}^{(1, \, f)}+ A_{n}^{(1, \, s)}\,,
\end{align}
\end{subequations}
where the subscripts $V$ and $\chi$ denote vector and chiral multiplets respectively. We can then write the primitive amplitudes required for QCD in terms of supersymmetric amplitudes,
\begin{subequations}
\label{eq:gf-susy}
\begin{align}
    A_{n}^{(1, \ g)} &= A_{n}^{(1, \ \Nfour)}
    - 4 \ A_{n}^{(1, \ \NoneX)}
    + A_{n}^{(1, \ s)} \label{eq:Ag_decomp}\\
    A_{n}^{(1, \ f)} &= A_{n}^{(1, \ \NoneX)}
    - A_{n}^{(1, \ s)}\,.\label{eq:Af_decomp}
\end{align}
\end{subequations}
The main benefit of these decompositions is that amplitudes in supersymmetric theories are cut-constructable, that is, they can be computed entirely from generalised unitarity cuts~\cite{Bern:1993mq,Bern:1994zx,Bern:1994cg}. The remaining complex scalar term in \cref{eq:Ag_decomp} has a rational term which (in $D=4-2 \epsilon$) cannot be computed from taking discontinuities. It is nevertheless simpler to compute than the gluon- or quark-loop amplitudes, because the complex scalar does not propagate spin information around the loop.

After the primary separation of \cref{sec:gN4,sec:gN1X,sec:gs} by loop content, these sections are further subdivided by helicity configuration. Colour-ordered amplitudes with differing helicity configuration typically have qualitatively different structure. 
%(An extreme example of this is provided by a comparison of the amplitudes in \cref{sec:div_4} with \cref{sec:fin_4}.)
%In Regge limits, the colour-ordered building blocks generically inherit some structure of the parent amplitudes. 
For our purposes, there are two separate points worth emphasising at this stage. Firstly, a given two-gluon peripheral emission vertex, say
$$C_{gggg^*}^{(1, \, m)}(p_2^{\ominus},p_3^{\oplus},p_4^{\oplus},q_2)\,$$
can be obtained by taking the NMRK limit of four distinct five-gluon amplitudes,
$$
A_5^{(1, \, m)}(p_2^{\ominus},p_3^{\oplus},p_4^{\oplus},p_5^\oplus,p_1^\ominus)
\,,
\qquad
A_5^{(1, \, m)}(p_2^{\ominus},p_3^{\oplus},p_4^{\oplus},p_1^\oplus,p_5^\ominus)\,,
$$
$$
A_5^{(1, \, m)}(p_2^{\ominus},p_3^{\oplus},p_4^{\oplus},p_5^\ominus,p_1^\oplus)
\,,
\qquad
A_5^{(1, \, m)}(p_2^{\ominus},p_3^{\oplus},p_4^{\oplus},p_1^\ominus,p_5^\oplus)\,.
$$
The former two amplitudes, being split-helicity configurations, have a simpler analytic structure than the latter two amplitudes. In, the following, we choose to present the extraction of the peripheral emission vertex from the simplest starting amplitude (although in practice we have checked each helicity configuration to verify the universality of the factorisation).
%$$
%C_{gggg^*}^{(1, \, m)}(p_2^{\ominus},p_3^{\oplus},p_4^{\oplus},q_2) \qquad \mathrm{and} \qquad C_{gggg^*}^{(1, \, m)}(p_2^{\oplus},p_3^{\ominus},p_4^{\oplus},q_2)$$
%$$ C_{gggg^*}^{(1, \, m)}(p_2^{\oplus},p_3^{\ominus},p_4^{\oplus},q_2)$$
Secondly, in close analogue to amplitudes, the vertices themselves have qualitatively different structure based on the helicity configuration of the \emph{vertex}.
We need to consider three distinct helicity configurations, 
$$ C_{gggg^*}^{(1, \, m)}(p_2^{\ominus},p_3^{\oplus},p_4^{\oplus},q_2)\,, \qquad 
C_{gggg^*}^{(1, \, m)}(p_2^{\oplus},p_3^{\ominus},p_4^{\oplus},q_2)\,, \qquad 
C_{gggg^*}^{(1, \, m)}(p_2^{\oplus},p_3^{\oplus},p_4^{\oplus},q_2)\,.
$$
The other helicity configurations can be obtained by complex conjugation and by the reflection property of the vertices which is inherited from \cref{eq:reflection}. We note that a supersymmetric ward identity guarantees the all-plus and single-minus amplitudes vanish to all loop orders in supersymmetric theories. We therefore only need to consider the all-plus vertex for amplitudes with a circulating complex scalar.

Our final subdivision of vertices is by colour ordering. For a given helicity configuration we need to consider the three colour orderings which appear in \cref{eq:DDM_5gNMRK_odd},
$$ 
C_{gggg^*}^{(1, \, m)}(p_2,p_3,p_4,q_2)\,, \qquad
C_{gggg^*}^{(1, \, m)}(p_3,p_4,p_2,q_2)\,, \qquad
C_{gggg^*}^{(1, \, m)}(p_4,p_3,p_2,q_2)
$$
With a notable exception to be discussed in section \cref{sec:N4_mhv}, these colour orderings can be obtained by application of the kinematic maps in \cref{sec:min}. 

Having discussed the organisational structure of this section, we begin with the analysis of amplitudes in $\Nfour$.

\subsection{\texorpdfstring{$\Nfour$}{N=4} contribution}
\label{sec:gN4}
One-loop amplitudes in $\Nfour$ consist of the tree-level amplitude times a purely transcendental function. It is therefore convenient to introduce a normalised one-loop amplitude, $a_n^{(1, m)}$, defined by
\begin{equation}
A^{(1, \, m)}_{n}(1, \ldots, n)= c_\Gamma  \ A_{n}^{(0)}(1, \ldots, n) \  a_{n}^{(1, \, m)}(1, \ldots, n).
\label{eq:mhv_amplitude}
\end{equation}
Throughout this paper, we will use an analogous notation for factorised vertices, where lower-case symbols are used for quantities which have been normalised by the tree-level quantity, and $c_\Gamma$.
For example, we write the one-loop vertices for a peripheral and central emission of a gluon as
\label{eq:bb_notation}
\begin{align}
C^{(1, \, m)}_{g g g^*}(p_2^{\lambda_2}, p_3^{-\lambda_2}, q_1)&=
c_\Gamma \, C_{g g g^*}^{(0)}(p_2^{\lambda_2}, p_3^{-\lambda_2}, q_1) 
\
c_{g g g^*}^{(1, \, m)}(p_2^{\lambda_2}, p_3^{-\lambda_2}, q_1)\,,\\
V^{(1, \, m)}_{g^* g g^*}(q_1, p_4^{\lambda_4}, q_2)&=c_\Gamma \,   V_{g^* g g^*}^{(0)}(-q_1, p_4^{\lambda_4}, q_2)
\
v_{g^* g g^*}^{(1,  \, m)}(-q_1, p_4^{\lambda_4}, q_2)\,,
\end{align}
respectively, and likewise we consider the normalised one-loop Reggeisation factor 
\begin{align}
R^{(1, \, m)}_{g^*}(q;s) &= c_\Gamma \, R^{(0)}_{g^*}(q) 
\
r^{(1, \, m)}_{g^*}(q;s) \,.
\end{align}

In $\Nfour$, the one-loop corrections to the tree-level amplitudes are helicity independent, and given by a simple transcendental function
\cite{Bern:1993mq}
\begin{align}
\begin{split}
a_5^{(1, \ \Nfour)}(1,2,3,4,5)&= -\frac{1}{\epsilon^2} \sum_{i=1}^5\left(\frac{\mu^2}{-s_{i, i+1}}\right)^\epsilon
+\frac{5}{6} \pi^2-\frac{\delta_R}{3}
\\
&+\sum_{i=1}^5 \log \left(\frac{-s_{i, i+1}}{-s_{i+1, i+2}}\right) \log \left(\frac{-s_{i+2, i-2}}{-s_{i-2, i-1}}\right)
+\cO(\epsilon)\,.
\label{eq:VN4_5}
\end{split}
\end{align}
In this paper we only work to $\cO(\epsilon^0)$, and in the following, dependence on higher orders in $\epsilon$ will be left implicit. We emphasise that the all-plus and single-minus amplitudes are zero in supersymmetric theories, so we only need to extract the single-minus vertex. For consistent organisation of this section we treat these identical helicity configurations under the following subsection.
\subsubsection{ 
\texorpdfstring{$ \{ g^\ominus, g^\oplus, g^\oplus, g^*\} $}{(g-,g+,g+,g*)},  \texorpdfstring{$\{ g^\oplus, g^\ominus, g^\oplus, g^*\} $}{(g+,g-,g+,g*)} and  
\texorpdfstring{$\{ g^\oplus, g^\oplus, g^\ominus, g^*\} $}{(g+,g+,g-,g*)}}
\label{sec:N4_mhv}
As alluded to in the introduction to this section, when no large logarithms are present in the amplitude, obtaining vertices with different cyclic orderings is a simple matter of applying the kinematic maps in \cref{sec:min}. This is not the case for one-loop amplitudes in $\Nfour$, and we treat the following two colour orderings separately.

\subsubsection*{Cyclic ordering $\bm{\{p_2,p_3,p_4,q_2\}}$}
%\subsubsection{$(p_2,p_3,p_4,q_2)$}

As in the four-gluon case discussed in \cref{sec:A4loop_Regge}, the function \cref{eq:VN4_5} admits an \emph{exact} factorisation into one-loop MRK building blocks~\cite{Bartels:2008ce}. We thus write the real part of the one-loop correction in the physical region of \cref{eq:outgoing} as
\begin{align}
\begin{split}
 \real{a_5^{(1, \, \Nfour)}(2,3,4,5,1)}=&c_{g g g^*}^{(1, \ \Nfour)}(p_2,p_3,q_1)  +r_{g^*}^{(1, \, \Nfour)}(q_1;s_{34})\\
 +&v^{(1, \, \Nfour)}(t_1,\frac{s_{34}s_{45}}{s_{345}},t_2)\\
 +&r_{g^*}^{(1, \, \Nfour)}(q_2;s_{45})+c_{g^* g g}^{(1, \, \Nfour)}(-q_2,p_5,p_1)
 \,, \label{eq:aN4_5_exact}
 \end{split}
\end{align}
where the colour-stripped one-loop corrections to the peripheral-emission vertex and Reggeised gluon propagator are given in \cref{eq:c3_N4} and \cref{eq:r} respectively. We have also used the special function \cite{Byrne:2022wzk}\begin{align}
\begin{split}
 v^{(1, \ \Nfour)}(t_1,\eta , t_2)
 =&- \frac{1}{\epsilon^2}\, \left(\frac{\mu^2}{ \eta}\right)^\epsilon+ \frac{\pi^2}{3}- \frac{1}{2} \log^2\left( \frac{t_1}{t_2} \right) 
		\\&+ \frac{1}{\epsilon}\, \left[ \left(\frac{\mu^2}{-t_1}\right)^\epsilon + \left(\frac{\mu^2}{-t_2}\right)^\epsilon \right]
		\log\left( \frac{\tau}{\eta} \right)\,.
  \label{eq:vN4}
 \end{split}
\end{align}
As indicated by the lack of subscripts, and the different treatment of the arguments of this function, we treat this function differently to the other building blocks hitherto encountered. We will not use this function as a factorised building block in its own right, but rather to organise the one-loop corrections to amplitudes and multi-gluon emission vertices such that their Regge limits are transparent. 
For example, from \cref{eq:aN4_5_exact} the peripheral-emission vertex for two gluons can be immediately idendified as
\begin{align}
\begin{split}
 c_{g g g g^*}^{(1, \, \Nfour)}(p_2,p_3,p_4,q_2)=&c_{g g g^*}^{(1, \, \Nfour)}(p_2,p_3,q_1)  
 +r_{g^*}^{(1, \, \Nfour)}(q_1;s_{34})\\
 +&v^{(1, \, \Nfour)}(t_1,\frac{s_{34} \, p_4^+}{(p_3^++p_4^+)},t_2)\,,
 \label{eq:234_N4}
 \end{split}
\end{align}
where we needed only to take the NMRK limit of the $\eta$ variable 
\begin{equation}
\frac{s_{34}s_{45}}{s_{345}}\toNMRK \frac{s_{34} \, p_4^+}{(p_3^++p_4^+)}
\end{equation}
to obtain a function that depends on the momenta local in rapidity (that is, on $\{p_2,p_3,p_4\}$).
\\ \\
In turn, this vertex has a trivial decomposition in the MRK limit
\begin{align}
\begin{split}
 c_{g g g g^*}^{(1, \, \Nfour)}(p_2,p_3,p_4,q_2) \toMRK &c_{g g g^*}^{(1, \, \Nfour)}(p_2,p_3,q_1) +r_{g^*}^{(1, \, \Nfour)}(q_1;s_{34})\\
 +&v_{g^* g g^*}^{(1, \, \Nfour)}(-q_1,p_4,q_2)
 \,,
 \end{split}
\end{align}
where the one-loop Lipatov vertex in $\Nfour$ (c.f. the weight-two terms in \cref{eq:Lipatov_1}) is conveniently written in terms of the special function \cref{eq:vN4},
\begin{align}
v^{(1, \, \Nfour)}_{g^* g g^*}(-q_1,p_4,q_2)=v^{(1, \, \Nfour)}(-|q_{1 \perp}|^2,|p_{4 \perp}|^2,-|q_{2 \perp}|^2)
 \,.
\end{align}
As mentioned in the introduction to this section, it is generally just a matter of applying the kinematic maps in \cref{sec:min} to obtain the vertices for other cyclic orderings of the momenta. However, the presence of large logarithms in $\Nfour$ ampltiudes introduces some subtleties which are discussed in the following section.

\subsubsection*{Cyclic ordering $\bm{\{p_3,p_2,p_4,q_2\}}$}
%\subsubsection{$(p_2,p_3,p_4,q_2)$}
%\subsubsection{$(p_3,p_2,p_4,q_2)$}

The exact decomposition \cref{eq:aN4_5_exact} holds for any cyclic permutation of the indices, so it is tempting therefore to write the amplitude with colour ordering $\{3,2,4,5,1\}$ as
\begin{align}
\begin{split}
 \real{a_5^{(1, \, \Nfour)}(3,2,4,5,1)}=&c_{g g g^*}^{(1, \, \Nfour)}(p_3,p_2,q_1)  +r_{g^*}^{(1, \, \Nfour)}(q_1;s_{24})\\
 +&v^{(1, \, \Nfour)}(t_1,\frac{s_{24}s_{45}}{s_{245}},t_2)\\
 +&r_{g^*}^{(1, \, \Nfour)}(q_2;s_{45})+c_{g g g^*}^{(1, \, \Nfour)}(p_5,p_1,-q_2)
 \,. \label{eq:a5_N4_B}
 \end{split}
\end{align}
While this is correct, \cref{eq:a5_N4_B} suggests we should take the peripheral-emission vertex for this colour ordering to be 
\begin{align}
\begin{split}
%\stackrel{?}{=}
 c_{g g g g^*}^{(1, \, \Nfour)}(p_3,p_2,p_4,q_2)=&c_{g g g^*}^{(1, \, \Nfour)}(p_3,p_2,q_1)  
 +r_{g^*}^{(1, \, \Nfour)}(q_1;s_{24})\\
 +&v^{(1, \, \Nfour)}(t_1,\frac{s_{24}\, p_4^+}{(-p_2^++p_4^+)},t_2)\,,
 \label{eq:postulate}
 \end{split}
\end{align}
that is, the vertex \cref{eq:234_N4}, under $P_{324}$. This vertex has the desired MRK limit,
\begin{align}
\begin{split}
 c_{g g g g^*}^{(1, \ \Nfour)}(p_3,p_2,p_4,q_2)\toMRK&c_{g g g^*}^{(1, \ \Nfour)}(p_3,p_2,q_1)  
 +r_{g^*}^{(1, \ \Nfour)}(q_1;p_3^+ p_4^-)\\
 +&v_{g^* g g^*}^{(1, \ \Nfour)}(q_1,p_4,q_2)\,.
 \end{split}
\end{align}
However, this does not obtain the correct $\MRKx$ limit. Indeed, we after somewhat perverse manipulations, we can write the $\MRKx$ limit of \cref{eq:a5_N4_B} in terms of the expected $\MRKx$ building blocks plus an unwanted large logarithmic term:
\begin{align}
\begin{split}
 c_{g g g g^*}^{(1, \ \Nfour)}(p_3,p_2,p_4,q_2)\toMRKx &c_{g g g^*}^{(1, \ \Nfour)}(p_2,p_4,q_1^\prime)  
 +r_{g^*}^{(1, \ \Nfour)}(q_1^\prime; p_3^+ p_4^-)\\
 +&v_{g^* g g^*}^{(1, \ \Nfour)}(-q_1^\prime, p_3,q_2)+\frac{2}{\epsilon}\left(\frac{\mu^2}{|q_{2\perp}|^2}\right)^\epsilon \log\left(\frac{p_3^+}{p_4^+}\right)\,.
 \end{split}
\end{align}
We therefore cannot take our vertex to be \cref{eq:postulate}. 
Rather than starting from \cref{eq:a5_N4_B}, we could instead exploit the reflection identity \cref{eq:reflection} to write the \emph{same} one-loop amplitude in the form \footnote{As in ref.~\cite{Byrne:2022wzk}, the signs of the arguments of the functions are fixed by the requirement that each building block is real.}
\begin{align}
\begin{split}
 \real{a_5^{(1, \ \Nfour)}(4,2,3,1,5)}=&c_{g g g^*}^{(1, \ \Nfour)}(p_4,p_2,q_1^\prime)  +r_{g^*}^{(1, \ \Nfour)}(q_1^\prime; s_{43})\\
 +&v^{(1, \ \Nfour)}(q_1^\prime,\frac{s_{43} \, s_{35}}{s_{435}},q_2)\\
 +&r_{g^*}^{(1, \ \Nfour)}(q_2;-s_{31})+c_{g g g^*}^{(1, \ \Nfour)}(p_5,p_1,-q_2)
 \,. \label{eq:a5_N4_B_R}
 \end{split}
\end{align}
This alternative rewriting of the one-loop amplitude suggests we should take the impact factor to be 
\begin{align}
\begin{split}
c_{g g g g^*}^{(1, \ \Nfour)}(p_4,p_2,p_3,q_2)=
&c_{g g g^*}^{(1, \ \Nfour)}(p_4,p_2,q_1^\prime)  +r_{g^*}^{(1, \ \Nfour)}(q_1^\prime; s_{43})\\
 +&v^{(1, \ \Nfour)}(q_1^\prime,\frac{s_{43} \, p_3^+}{(p_3^++p_4^+)},q_2)\,,
 \end{split}
\end{align}
which correctly describes the known $\MRKx$ limit, but not the MRK limit.

At this point we recall the NMRK limit of the tree-level amplitude of the colour-ordering in question,
\begin{align}
\begin{split}
   A^{(0)}_5(3,2,4,5,1) \toNMRK 
   s \,
   C^{(0)}_{gggg^*}(p_3, p_2, p_4, q_2)
    \frac{1}{t_2}
    C^{(0)}_{g^*gg}(-q_2, p_5, p_1)\,.
\end{split}
\end{align}
We now use the fact that the tree-level colour-ordered vertex inherits a $U(1)$ decoupling identity from its parent amplitudes (see for example \cref{eq:U1} and \cref{eq:U1_v}): 
\begin{equation}
C^{(0)}_{gggg^*}(p_3, p_2, p_4, q_2)=
    -C^{(0)}_{gggg^*}(p_2, p_3, p_4, q_2) 
    -C^{(0)}_{gggg^*}(p_2, p_4, p_3, q_2)\,.\label{eq:U1_2}
\end{equation}
The first and second term vanish in the $\MRKx$ and $\MRK$ limits respectively. This observation gives us a way to utilise the two-fold nature of the one-loop amplitude discussed above. We use \cref{eq:U1_2} and \cref{eq:reflection} to write the full colour-ordered amplitude as 
\begin{align}
\begin{split}
& \frac{1}{2s}\real{A_5^{(1, \ \Nfour)}(3,2,4,5,1)}
\\
&\toNMRK
-c_\Gamma
C^{(0)}_{gggg^*}(p_2, p_3, p_4, q_2) 
\frac{1}{t_2}
C^{(0)}_{g^*gg}(-q_2, p_5, p_1)
\real{a_5^{(1, \ \Nfour)}(3,2,4,5,1)}
\\
&
\phantom{\toNMRK}
-c_\Gamma
C^{(0)}_{gggg^*}(p_2, p_4, p_3, q_2) 
\frac{1}{t_2}
C^{(0)}_{g^*gg}(-q_2, p_5, p_1)
\real{a_5^{(1, \ \Nfour)}(4,2,3,5,1)}
 \,.
 \end{split}
\end{align}
We then use the decomposition \cref{eq:a5_N4_B} for the first line and the decomposition \cref{eq:a5_N4_B_R} for the second line. Explicitly,
\begin{align}
\begin{split}
&\frac{1}{2s}\real{A_5^{(1, \ \Nfour)}(3,2,4,5,1)}
\toNMRK
-c_\Gamma\sum_{\sigma \in S_2}
C^{(0)}_{gggg^*}(p_2, p_{\sigma_3}, p_{\sigma_4}, q_2) 
\\
%&\times 
%\Bigg\{
%c_{g g g^*}^{(1, \ \Nfour)}(p_2,p_{\sigma_3},-%(p_2{+}p_{\sigma_3}))+r_{g^*}^{(1, \ \Nfour)}(-(p_2{+}p_{\sigma_3}); s_{2 \sigma_4})
%+v^{(1, \ \Nfour)}(s_{2 \sigma_3},\frac{s_{34} \,p_{\sigma_3}^+}{(p_{3}^++p_{4}^+)},s_{51})
%\\
%&\qquad +r_{g^*}^{(1, \ \Nfour)}(q_2;p_{\sigma_4}^+ p_5^-)+c_{g g g^*}^{(1, \ \Nfour)}(p_5,p_1,-q_2)
%\Bigg\}
&\times 
\Bigg\{
c_{g g g g^*}^{(1, \ \Nfour)}(p_{\sigma_3},p_2,p_{\sigma_4},q_2)+r_{g^*}^{(1, \ \Nfour)}(q_2;p_{\sigma_4}^+ p_5^-)+c_{g g g^*}^{(1, \ \Nfour)}(p_5,p_1,-q_2)
\Bigg\}
\\
&\frac{1}{t_2}
C^{(0)}_{g^*gg}(-q_2, p_5, p_1)
 \,.
 \label{eq:a5_N4_B_p}
 \end{split}
\end{align}
In this way, the decomposition of the loop-correction which trivialises the $\MRK$ or $\MRKx$ limits only has support in those limits. The requirement of yielding the correct $\MRK$ and $\MRKx$ limits has forced us to write \cref{eq:a5_N4_B_p} as a sum over two permutations, which will carry over to the colour-dressed amplitude, \cref{eq:A1_g_simple_3}.
\subsection{\texorpdfstring{$\NoneX$}{N=1X} contribution}
\label{sec:gN1X}
The analysis of one-loop five-gluon amplitudes in $\None$ are in one respect simpler than the amplitudes in $\Nfour$, due to the lack of any large logarithms in the NMRK or MRK limit. Therefore, the complications discussed in \cref{sec:N4_mhv} do not arise.
However, the amplitudes are not helicity independent. We need to consider two distinct helicity configurations of the starting amplitudes, and two distinct helicity configurations of the vertices. For the convenience of the reader we list the amplitudes here, which have been taken from refs.~\cite{Bern:1993mq,Canay:2021yvm}. Following ref.~\cite{Bern:1994fz,Bern:1993mq}, we write the amplitudes in terms of a divergent piece, $V$, (which contains purely transcendental functions of the momenta) and a further, finite term $G$, (which generically contains both rational functions and products of rational and transcendental functions of the momenta) \footnote{We use the conventional notation $V$ for the divergent, purely transcendental part of the one-loop correction. We hope this does not cause confusion with the central-emission vertices, which will always include subscript flavour indices.},
\begin{equation}
    a^{(1, \, m)}=V^{(1, \, m)} + G^{(1, \, m)}\,.\label{eq:PG}
\end{equation}
The divergent parts of the amplitude are given by
\begin{subequations}
\begin{align}
    &V^{(1, \, \None)}(p_1^\ominus,p_2^\ominus,p_3^\oplus, p_4^\oplus, p_5^\oplus)=\frac{1}{ \epsilon}+\frac{1}{2}\left[\log \left(\frac{\mu^2}{-s_{23}}\right)+\log \left(\frac{\mu^2}{-s_{51}}\right)\right]+2\,, 
    \label{eq:VN1j2}
    \\
    &V^{(1, \, \None)}(p_1^\ominus,p_2^\oplus,p_3^\ominus, p_4^\oplus, p_5^\oplus)=\frac{1}{ \epsilon}+\frac{1}{2}\left[\log \left(\frac{\mu^2}{-s_{34}}\right)+\log \left(\frac{\mu^2}{-s_{51}}\right)\right]+2 \,,
    \label{eq:VN1j3}
\end{align}
\end{subequations}
and the finite remainders are given by
\begin{subequations}
\begin{align}
    G^{(1, \, \None)}(p_1^\ominus,p_2^\ominus,p_3^\oplus, p_4^\oplus, p_5^\oplus)&= -
 \frac{I_{1234}+I_{1245}}{2 s_{12} s_{51}} L_0\left(\frac{-s_{23}}{-s_{51}}\right)\,
 \label{eq:GN1j2}
    \\
    \begin{split}
    G^{(1, \, \None)}(p_1^\ominus,p_2^\oplus,p_3^\ominus, p_4^\oplus, p_5^\oplus)&=
   \frac{I_{1325}+I_{1342}}{2 s_{13} s_{51}} L_0\left(\frac{-s_{34}}{-s_{51}}\right)
   \\
    & \hspace{-6 em}
    -\frac{I_{1324} I_{1342}}{s_{13}^2 s_{51}^2} 
    L s_1\left(\frac{-s_{23}}{-s_{51}}, \frac{-s_{34}}{-s_{51}}\right)
    -
    \frac{I_{1325} I_{1352}}{s_{13}^2 s_{34}^2} 
    L s_1\left(\frac{-s_{12}}{-s_{34}}, \frac{-s_{51}}{-s_{34}}\right)\,.
    \label{eq:GN1j3}
    \end{split}
\end{align}
\end{subequations}
We will now extract the two distinct helicity configurations for the $gggg^*$ vertex from these amplitudes, as outlined in the introduction.

\subsubsection{ \texorpdfstring{$\{ g^\ominus, g^\oplus, g^\oplus, g^*\} $}{(g-,g+,g+,g*)}}\label{sec:N1_mpp}
%\subsubsection{$(p_2^\ominus, p_3^\oplus, p_4^\oplus, q_2)$} \label{sec:N1_2m3p4p}
As discussed in the introduction to this section, it is possible to obtain the vertex
$$c^{(1, \, \None)}_{gggg^*}(p_2^\ominus, p_3^\oplus, p_4^\oplus, q_2)$$ 
from the split-helicity amplitude 
$$a^{(1, \, \None)}_{5g}
(p_2^\ominus, p_3^\oplus, p_4^\oplus, p_5^\oplus, p_1^\ominus).$$
Inspecting \cref{eq:VN1j2}, we note that, even in general kinematics, the real part of this term is nothing other than the sum of two peripheral-emission vertices, \cref{eq:c3_N1}, that is,
\begin{align}
\mathrm{Re}\left[ V^{(1, \, \None)}(p_1^\ominus,p_2^\ominus,p_3^\oplus, p_4^\oplus, p_5^\oplus) \right]=
c^{(1, \ \None)}_{g g g^* }(p_2,p_3,q_1)+c^{(1, \ \None)}_{g g g^* }(p_5, p_1, -q_2)\,.
\end{align}
As an aside, we note this simple separation of the peripheral emission vertices holds also for the one-loop $n$-gluon split-helicity amplitude~\cite{Bern:1994cg}, so the one-loop $n{-}4$-gluon all-plus central-emission vertex is a simple, finite function that can be readily extracted from the $n$-gluon amplitude. 

Returning to the present five-gluon case, we have only to take the forward NMRK limit of  \cref{eq:GN1j2}. We first write it in terms of the minimal set of variables defined in \cref{sec:min},
\begin{align}
\begin{split}
	G^{(1, \, \None)}(p_1^\ominus,p_2^\ominus,p_3^\oplus, p_4^\oplus, p_5^\oplus)&=\frac{1}{2}\left(
\frac{X+Y |z|^2+X Y (z-\zb+1)}{X+|z|^2+X Y |z-1|^2}
\right)L_0\left(\frac{-s_{23}}{-s_{51}}\right)\,.
 \label{eq:GN1}
	\end{split}
\end{align}
It is now straightforward to take various Regge limits.
In particular, we can obtain the contribution to the forward-emission vertex by taking the $Y \to \infty$ (forward NMRK) limit,
\begin{align}
\begin{split}
G^{(1, \, \None)}(p_1^\ominus,p_2^\ominus,p_3^\oplus, p_4^\oplus, p_5^\oplus) &\toYinf
\frac{1}{2}\left(
\frac{
|z|^2+X(z-\zb+1)
}{X|z-1|^2}
\right)L_0 \left(\frac{-s_{23}}{-s_{51}}\right)\,.
\end{split}
\end{align}
One check of this result is to take the further $X \to \infty$ (MRK) limit,
\begin{align}
\begin{split}
G^{(1, \, \None)}(p_1^\ominus,p_2^\ominus,p_3^\oplus, p_4^\oplus, p_5^\oplus) &\toMRK
\frac{1}{2}\left(
\frac{
z-\zb+1
}{|z-1|^2}
\right)L_0 \left(\frac{-s_{23}}{-s_{51}}\right)\,,
\end{split} 
\end{align}
This is the one-loop correction to the Lipatov vertex in $\NoneX$, which can be inferred from e.g. ref.~\cite{DelDuca:1998cx},
\begin{align}
\begin{split}
v_{g^* g g^*}^{(1, \  \NoneX)}(-q_1, p_4, q_2)
&=
\frac{1}{2}
		\left(|q_{1\perp}|^2+|q_{2\perp}|^2-2q_{1\perp}q_{2\perp}^*\right)
		\frac{L_0\left(\frac{|q_{1\perp}|^2}{|q_{2\perp}|^2}\right)}{|q_{2\perp}|^2}
\\
&=
\frac{1}{2}\left(
\frac{
z-\zb+1
}{|z-1|^2}
\right)L_0 \left(\frac{|q_{1\perp}|^2}{|q_{2\perp}|^2}\right)\,.
\label{eq:N1X_Lipatov}
\end{split} 
\end{align}
Collecting the two contributions to the peripheral-emission vertex, we obtain 
\begin{align}
\begin{split}
c^{(1, \, \None)}_{gggg^*}(p_2^\ominus,p_3^\oplus, p_4^\oplus, q_2) = c^{(1, \ \None)}_{g g g^* }(p_2,p_3,q_1)+
\frac{1}{2}\left(
\frac{
|z|^2+X(z-\zb+1)
}{X|z-1|^2}
\right)L_0 \left(\frac{-t_1}{-t_2}\right)\,,
\label{eq:cN1_2m3p4p}
\end{split} 
\end{align}
which has the transparent MRK limit
\begin{align}
\begin{split}
c^{(1, \, \None)}_{gggg^*}(p_2^\ominus,p_3^\oplus, p_4^\oplus, q_2) \toXinf c^{(1, \ \None)}_{g g g^* }(p_2,p_3,q_1)+
v_{g^* g g^*}^{(1, \  \NoneX)}(q_1, p_4, q_2)\,.
\label{eq:cN1_2m3p4p_X}
\end{split} 
\end{align}
The other colour orderings can be obtained using the kinematic maps derived in \cref{sec:min}. For example, applying the map $P_{243}$ given in \cref{eq:P243} we obtain
\begin{align}
\begin{split}
c^{(1, \, \None)}_{gggg^*}(p_2^\ominus,p_4^\oplus, p_3^\oplus, q_2) = c^{(1, \ \None)}_{g g g^* }(p_2,p_4,q_1)+
\frac{1}{2}\left(
\frac{
\zb +z(\zb-1)+X
}{|z-1|^2}
\right)L_0 \left(\frac{-t_1^\prime}{-t_2}\right)\,,
\end{split} 
\end{align}
while applying the map $P_{324}$ given in \cref{eq:P324} to the vertex \cref{eq:cN1_2m3p4p} we obtain
\begin{align}
\begin{split}
c^{(1, \, \None)}_{gggg^*}(p_3^\ominus,p_2^\oplus, p_4^\oplus, q_2) = &c^{(1, \, \None)}_{g g g^* }(p_3,p_2,q_1)
\\+&
\frac{1}{2}\left(
\frac{
z(2-\zb)+X(z-\zb+1)
}{X|z-1|^2}
\right)L_0 \left(\frac{-t_1}{-t_2}\right)\,.
\end{split} 
\end{align}
\subsubsection{ \texorpdfstring{ $\{ g^\oplus, g^\ominus, g^\oplus, g^*\} $}{(g+,g-g+,g)}}\label{sec:N1_pmp}
%\subsubsection{$(p_3^\oplus, p_2^\ominus, p_4^\oplus, q_2)$}\label{sec:N1_3p2m4p}
We cannot obtain $c^{(1, \, \None)}_{gggg^*}(p_3^\oplus, p_2^\ominus,  p_4^\oplus, q_2)$ from a split helicity configuration, but must consider, for example,
$$a^{(1, \ \None)}_{5g}(p_1^\oplus, p_3^\oplus, p_2^\ominus, p_4^\oplus, p_5^\ominus)$$
in the forward NMRK limit. Again, we see that $V^{(1, \, \None)}$ is nothing other than the sum of $ggg^*$ vertices
\begin{align}
\mathrm{Re} \left[ 
V^{(1, \, \None)}(p_1^\oplus, p_3^\oplus, p_2^\ominus, p_4^\oplus, p_5^\ominus)
\right]
=
c^{(1, \ \None)}_{g g g^* }(p_2,p_3,q_1)+c^{(1, \ \None)}_{g g g^* }(p_5, p_1, -q_2)\,.
\end{align}
The forward NMRK limit of the remaining term is
\begin{align}
\begin{split}
&
G^{(1, \, \None)}(p_1^\oplus, p_3^\oplus, p_2^\ominus, p_4^\oplus, p_5^\ominus)
\toYinf
-
\frac{z |z|^2+z (z-1) X-(z-\zb+1)X^2}{2  |z-1|^2 (X+z)X}
L_0\left(\frac{-s_{23}}{-s_{51}}\right)
\\& 
\hspace{6 em}
-\frac{X z}{(X+z)^2}
Ls_{-1}\left(\frac{-s_{23}}{-s_{51}},\frac{-s_{24}}{-s_{51}} \right)
+\frac{z(X+\zb)}{|z-1|^2(X+z)}L_0\left(\frac{-s_{24}}{-s_{51}}\right)\,.
\end{split}
\end{align}
We note the presence of terms of transcendental weight 2 that appear in this vertex. This is unlike the factorised building blocks of the LO and NLO BFKL kernel and impact factors, where the terms of highest transcendental weight coincide with the $\Nfour$ building blocks. Whether such weight-2 terms survive phase-space integration, and therefore contribute to the NNLO impact factor, is a question that will be left to a future investigation. In any case these weight-2 terms are suppressed in the MRK limit, and indeed we find
\begin{align}
\begin{split}
 G^{(1, \, \None)}(p_1^\oplus, p_3^\oplus, p_2^\ominus, p_4^\oplus, p_5^\ominus)
&\toMRK
v^{(1, \, \None)}_{g^* g g^*}(-q_1, p_4, q_2)
\end{split}
\end{align}
as expected. 
Adding the contributions from $V^{(1, \, \None)}$ and $G^{(1, \, \None)}$ we obtain the vertex
\begin{align}
\begin{split}
c^{(1, \ \None)}(p_3^\oplus, p_2^\ominus, p_4^\oplus, q_2)
&=
-\frac{X z}{(X+z)^2}
Ls_{-1}\left(\frac{-t_1}{-t_2},\frac{-t_1^\prime}{-t_2} \right)
+\frac{z(X+\zb)}{|z-1|^2(X+z)}L_0\left(\frac{-t_1^\prime}{-t_2}\right)
\\&
-
\frac{z |z|^2+z (z-1) X-(z-\zb+1)X^2}{2  |z-1|^2 (X+z)X}
L_0\left(\frac{-t_1}{-t_2}\right)
\\&
+c^{(1, \ \None)}_{g g g^* }(p_2,p_3,q_1)
\,,
\label{eq:cN1_3p2m4p}
\end{split}
\end{align}
which has the MRK limit
\begin{align}
\begin{split}
c^{(1, \, \None)}_{gggg^*}(p_3^\oplus, p_2^\ominus, p_4^\oplus, q_2) \toXinf c^{(1, \ \None)}_{g g g^* }(p_2,p_3,q_1)+
v_{g^* g g^*}^{(1, \  \NoneX)}(q_1, p_4, q_2)\,.
\label{eq:cN1x_2m3p4p_X}
\end{split} 
\end{align}
As in \cref{sec:N1_mpp}, the other cyclic orderings of momenta can be obtained using the maps provided in \cref{sec:min}. For example, the map $P_{324}$ (\cref{eq:P243}) gives
\begin{align}
\begin{split}
c^{(1, \ \None)}_{gggg^*}(p_2^\oplus, p_3^\ominus, p_4^\oplus, q_2)
&=
\frac{z(X+z)}{X^2}
Ls_{-1}\left(\frac{-t_1}{-t_2},\frac{-s_{34}}{-t_2} \right)
+\frac{z(X+z)(1+X)}{X^2|z-1|^2}L_0\left(\frac{-s_{34}}{-t_2}\right)
\\&
+
\frac{2z^2+z(1+z)X+(z-\zb+1)(X+z)X}{2  |z-1|^2 X^2}
L_0\left(\frac{-t_1}{-t_2}\right)
\\&
+c^{(1, \ \None)}_{g g g^* }(p_2,p_3,q_1)
\,, \label{eq:cN1x_2p3m4p}
\end{split}
\end{align}
In the following, because the expressions for the vertices are lengthier, we will only display one representative ordering for each helicity configuration. Note that unlike previous expressions for vertices, we have not written the transcendental terms in \cref{eq:cN1x_2p3m4p} in a manifestly real way. Here and in the following, we choose to write each Mandelstam invariant with a minus sign, as in \cref{eq:cN1x_2p3m4p}, but it is implicit that the dispersive part of all factorised vertices should be taken, for the physical scattering region in question (here \cref{eq:outgoing}). Due to the simple analytic structure of the vertices, this convention is of little consequence for the present section, but in \cref{sec:c_qqbg}, where several distinct scattering processed are relevant, this convention will allow for simpler analytic continuation to those regions.

\subsection{Complex scalar contribution}
\label{sec:gs}
The analysis of amplitudes with a circulating complex scalar proceeds in close analogy to the $\NoneX$ case. Indeed, the divergent terms are given by
\begin{align}
   V^{(1, \, s)}=\frac{1}{3}V^{(1, \, \None)}+\frac{2}{9} \,.
\end{align}
By comparing to the vertex $c^{(1, \,s)}_{ggg^*}$ given in \cref{eq:c3_s}, we see that the identification of $V^{(1, \, s)}$ with impact factors proceeds in exact analogy with the $\NoneX$ case. The characteristic difference in the scalar case is the presence of non-cut-constructible rational terms. The split-helicity amplitude is still a relatively simple function,
\begin{align}
\begin{split}
    G^{(1, \, s)}(p_1^\ominus,p_2^\ominus,p_3^\oplus, p_4^\oplus, p_5^\oplus)&=
    \frac{1}{3}G^{(1, \, \None)}
    +\frac{I_{1234} I_{1245}\left(I_{1234}+I_{1245}\right)}{3 s_{12}^3 s_{51}^3} L_2\left(\frac{-s_{23}}{-s_{51}}\right) 
    \\&
    +\frac{I_{1235}^2}{3 s_{12}^2 s_{23} s_{51}}
    \left(1-\frac{s_{35}}{s_{12}}\right)
    +\frac{I_{1234} I_{1245}}{6 s_{12}^2 s_{23} s_{51}}\,,
\end{split}
\end{align}
which, as we will see in \cref{sec:s_mpp} leads to a relatively simple $g^\ominus g^\oplus g^\oplus g^* $ vertex.

The other helicity configuration is more involved,
\begin{align}
\begin{split}
    &G^{(1, \, s)}(p_1^\ominus,p_2^\oplus,p_3^\ominus, p_4^\oplus, p_5^\oplus)=
    \frac{2}{3} \frac{I_{1324}^3 I_{1342}}{s_{13}^4 s_{24} s_{51}^3} 
    L_2\left(\frac{-s_{23}}{-s_{51}}\right)
    +\frac{2}{3} \frac{I_{1352}^3 I_{1325}}{s_{13}^4 s_{25} s_{34}^3} 
    L_2\left(\frac{-s_{12}}{-s_{34}}\right)
    \\
    &
    \qquad
    -\frac{I_{1324}^2 I_{1342}^2}{s_{13}^4 s_{24}^2 s_{51}^2}
    \left[2 L s_1\left(\frac{-s_{23}}{-s_{51}}, \frac{-s_{34}}{-s_{51}}\right)+L_1\left(\frac{-s_{23}}{-s_{51}}\right)+L_1\left(\frac{-s_{34}}{-s_{51}}\right)\right] 
    \\
    &
    \qquad
    -\frac{I_{1325}^2 I_{1352}^2}{s_{13}^4 s_{25}^2 s_{34}^2}
    \left[2 L s_1\left(\frac{-s_{12}}{-s_{34}}, \frac{-s_{51}}{-s_{34}}\right)+L_1\left(\frac{-s_{12}}{-s_{34}}\right)+L_1\left(\frac{-s_{51}}{-s_{34}}\right)\right] 
    \\
    & 
    \qquad
    +\frac{1}{3 s_{51}^3} 
    L_2\left(\frac{-s_{34}}{-s_{51}}\right)
    \left[-\frac{I_{1325} I_{1342}\left(I_{1325}+I_{1342}\right)}{s_{13}^3}+2 \frac{I_{1342}^3 I_{1324}}{s_{13}^4 s_{24}}+2 \frac{I_{1325}^3 I_{1352}}{s_{13}^4 s_{25}}\right] 
    \\
    & 
    \qquad
    +\frac{I_{1325}+I_{1342}}{6 s_{13} s_{51}} 
    L_0\left(\frac{-s_{34}}{-s_{51}}\right)+\frac{I_{1325}^2 I_{1342}^2}{3 s_{13}^4 s_{23} s_{51} s_{34} s_{12}} 
    \\
    & 
    \qquad
    +\frac{I_{1324}^2 I_{1342}^2}{3 s_{13}^4 s_{23} s_{24} s_{34} s_{51}}
    +\frac{I_{1325}^2 I_{1352}^2}{3 s_{13}^4 s_{25} s_{12} s_{34} s_{51}}
    -\frac{I_{1342} I_{1325}}{6 s_{13}^2 s_{34} s_{51}}\,.
\end{split}
\end{align}
and, as we will see in \cref{sec:s_pmp}, this leads to a relatively complicated $g^\oplus g^\ominus g^\oplus g^* $ vertex.

Unlike the supersymmetric multiplets studied in \cref{sec:gN4} and \cref{sec:gN1X}, the all-plus
amplitude does not vanish for a circulating complex scalar~\cite{Bern:1993mq}:
\begin{align}
&A_{5}^{(1, \, s)}\left(1^{\oplus}, 2^{\oplus}, 3^{\oplus}, 4^{\oplus}, 5^{\oplus}\right)=  \frac{1}{6( 4 \pi)^2} \frac{\left(\sum_{i=1}^{5} s_{i-1, \, i} \, s_{i, \, i+1}\right)+\epsilon(1,2,3,4)}{\langle 12\rangle\langle 23\rangle\langle 34\rangle\langle 45\rangle\langle 51\rangle}+\cO(\epsilon)\,,
\label{eq:ppppp}
\end{align}
where the contracted Levi-Civita symbol can be written
\begin{equation}
    \epsilon (i,j,k,l)=
    [i\,j]\langle j\,k \rangle[k\,l]\langle l\,i \rangle
    -
    \langle i\,j \rangle[j\,k]\langle k\,l \rangle [l\,i]\,.
\end{equation}
Similarly, the single-minus amplitude is non-vanishing,
\begin{align}
\begin{split}
&A^{(1, \, s)}\left(1^{\ominus}, 2^{\oplus}, 3^{\oplus}, 4^{\oplus}, 5^{\oplus}\right)=  \frac{1}{3(4\pi)^2} \frac{1}{[12]\langle 23\rangle\langle 34\rangle\langle 45\rangle[51]}
\\
& \hspace{3 em} \times \bigg[
\left(s_{23}+s_{34}+s_{45}\right)[25]^2
-[24]\langle 43\rangle[35][25]
\\
& \hspace{6 em}
-\frac{[12][15]}{\langle 12\rangle\langle 15\rangle}\left(\langle 12\rangle^2\langle 13\rangle^2 \frac{[23]}{\langle 23\rangle}+\langle 13\rangle^2\langle 14\rangle^2 \frac{[34]}{\langle 34\rangle}+\langle 14\rangle^2\langle 15\rangle^2 \frac{[45]}{\langle 45\rangle}\right)\bigg]+\cO(\epsilon)
\end{split}
\end{align}
The NMRK limits of these amplitudes have already been investigated in ref.~\cite{Canay:2021yvm}, and the $ g^\oplus g^\oplus g^\oplus g^*$ vertex has been extracted. For completeness, in the following section we briefly repeat this analysis using the minimal set of variables of \cref{eq:min}.

\subsubsection{ \texorpdfstring{$\{ g^\oplus, g^\oplus, g^\oplus, g^*\} $}{(g+,g+,g+,g*)}}\label{sec:s_ppp}

We find the amplitude factorises according to \cref{eq:Regge_conjecture} as expected, e.g.
\begin{align}
A_5^{(1, \, s)}(2^{\oplus},3^{\oplus},4^{\oplus},5^{\oplus},1^{\ominus})&\toNMRK
2s \, C_{ggg^*}^{(1, \, s)}(p_2^\oplus,p_3^\oplus,p_4^\oplus, q_2) 
R^{(0)}_{g^*}(q_2)C^{(0)}_{g^*gg}(-q_2,p_5^{\oplus}, p_1^{\ominus})\,.
\end{align}
This allows us to identify the vertex
\begin{align}
\begin{split}
&C^{(1, \, s)}_{gggg^*}(p_2^\oplus,p_3^\oplus, p_4^\oplus, q_2) 
=
\\
&\frac{1}{q_{2\perp}}\left(\frac{q_{2\perp}^*}{q_{2\perp}}\right)
\frac{1}{X+1}\left(\frac{X^2 (z-1)^3 \zb}{z^2 (\zb-1) (X+z)}-\frac{X^2(z-1)^3 \left(X+ \zb\right)}{(X+1) z (\zb-1) (X+z)^2}+\frac{X (z-1)^2}{z (X+z)}\right)\,.
\label{eq:cs_2p3p4p}
\end{split} 
\end{align}
in agreement with \cite{Canay:2021yvm}. This vertex in turn factorises in the further MRK limit:
\begin{align}
\begin{split}
C^{(1, \, s)}_{gggg^*}(p_2^\oplus,p_3^\oplus, p_4^\oplus, q_2) 
\toMRK
&\frac{1}{q_{2\perp}}\left(\frac{q_{2\perp}^*}{q_{2\perp}}\right)
\frac{ \zb(z-1)^3}{ z^2 (\zb-1)}
\\&=
C^{(1, \, \None)}_{ggg^*}(p_2^\oplus,p_3^\oplus, q_1) 
R^{(0)}_{g^*}(q_1)
V^{(0)}_{g^*gg^*}(-q_1,p_4,q_2) \,.
\label{eq:cs_2p3p4p_MRK}
\end{split} 
\end{align}
Let us now consider the all-plus case, where \cref{eq:Regge_conjecture} reads
\begin{align}
\disp{A(2^{\oplus},3^{\oplus},4^{\oplus},5^{\oplus},1^{\oplus})}=
2s \, C_{ggg^*}(p_2^\oplus,p_3^\oplus,p_4^\oplus, q_2) 
R_{g^*}(q_2;s_{45})C_{g^*gg}(-q_2,p_5^\oplus, p_1^\oplus)\,.
\label{eq:ppppp_regge}
\end{align}
Given that both gluon-emission vertices in this expression begin at one loop, consistency with this factorised form would require this process to first contribute at leading power at order $\gs^7$,
\begin{align}
\disp{A^{(2)}(2^{\oplus},3^{\oplus},4^{\oplus},5^{\oplus},1^{\oplus})}=
2s \, C^{(1,\, s)}_{ggg^*}(p_2^\oplus,p_3^\oplus,p_4^\oplus, q_2) 
R^{(0)}_{g^*}(q_2)C^{(1,\, s)}_{g^*gg}(-q_2,p_5^\oplus, p_1^\oplus)\,.
\end{align}
Indeed, we find the one-loop amplitudes \cref{eq:ppppp} are power-supressed in the NMRK limit, consistent with \cref{eq:ppppp_regge}. 

\subsubsection{ \texorpdfstring{$\{ g^\ominus, g^\oplus, g^\oplus, g^*\} $}{(g-,g+,g+,g*)}}\label{sec:s_mpp}
Following the procedure of \cref{sec:N1_mpp}, we obtain
\begin{align}
\begin{split}
c^{(1, \, s)}_{gggg^*}(p_2^\ominus,p_3^\oplus, p_4^\oplus, q_2) 
&=
c^{(1, \ s)}_{g g g^* }(p_2,p_3,q_1)+
\frac{1}{6}\left(
\frac{
|z|^2+X(z-\zb+1)
}{X|z-1|^2}
\right)
L_0 \left(\frac{-t_1}{-t_2}\right)
\\&
-
\frac{z(\zb-1)(X+\zb) (|z|^2 +X (z-\zb+1))}{3 X^2 (|z-1|^2)^3}
L_2 \left(\frac{-t_1}{-t_2}\right)
\\&
-\frac{1}{3}\frac{z(\zb-1)X}{\zb(z-1)(1+X)^2}
+\frac{1}{6}\frac{X+\zb}{\zb(z-1)(1+X)}
\,.
\label{eq:cs_2m3p4p}
\end{split} 
\end{align}
In the MRK limit, we find 
\begin{align}
\begin{split}
c^{(1, \, s)}_{gggg^*}(p_2^\ominus,p_3^\oplus, p_4^\oplus, q_2) 
&\toXinf
c^{(1, \ s)}_{g g g^* }(p_2,p_3,q_1)+v^{(1, \, s)}_{g^*gg^*}(-q_{1},p_4,q_2)
\,,
\label{eq:cs_2m3p4p_X}
\end{split} 
\end{align}
as desired, where the Lipatov vertex with a complex scalar circulating in the loop is
  \begin{align}
  \begin{split}
		v^{(1, \, s)}_{g^*gg^*}(-q_{1},p_4,q_2)=&
   \frac{1}{3}v_{g^*gg^*}^{(1, \ \None)}(q_{1},p_4,q_2)
  -\frac{1}{6}\frac{|p_{4\perp}|^2}{q_{1\perp}^*q_{2\perp}}
  \\
   -&\frac{1}{3}|p_{4\perp}|^2 q_{1\perp}q_{2\perp}^*
		(|q_{1\perp}|^2+|q_{2\perp}|^2-2q_{1\perp}q_{2\perp}^*)
		\frac{L_2\left(\frac{|q_{1\perp}|^2}{|q_{2\perp}|^2}\right)}{(-|q_{2\perp}|^2)^3}
 \\
 =&
 \frac{1}{3}v_{g^*gg^*}^{(1, \ \None)}(q_{1},p_4,q_2)
  +\frac{1}{6}\frac{1}{\zb(z-1)}
  +\frac{1}{3}\frac{z(z- \zb +1)}{(|z-1|^2)^3}L_2\left(\frac{|q_{1\perp}|^2}{|q_{2\perp}|^2}\right)\,,
  \label{eq:cev1_n0}
	\end{split}
\end{align}
which can be inferred from e.g. ref.~\cite{DelDuca:1998cx}.

\subsubsection{ \texorpdfstring{$\{ g^\oplus, g^\ominus, g^\oplus, g^*\} $}{(g+,g-,g+,g*)}}\label{sec:s_pmp}
Following the procedure of \cref{sec:N1_pmp}, we obtain the vertex
\begin{align}
\begin{split}
c^{(1, \, s)}_{gggg^*}&(p_3^\oplus, p_2^\ominus, p_4^\oplus, q_2)
=
c^{(1, \ s)}_{g g g^* }(p_2,p_3,q_1)
-\frac{X+\zb}{6(1+X)(z-1)\zb}
\\&
+
\frac{X(1+z-\zb)+|z|^2}{6 X |z-1|^2}L_0\left(\frac{-t_1}{-t_2} \right)
+
\frac{X(X+\zb)(X(1+z-\zb)+|z|^2)}{3 X^2 (z-1)^3(\zb-1)^2}L_2\left(\frac{-t_1}{-t_2} \right)
\\&
+2\frac{ X z^2 (X+\zb)}{|z-1|^2 (X+z)^3}
\left[ 
L_0\left(\frac{-t_1}{-t_2}\right)
+L_0\left(\frac{-t_1^\prime}{-t_2}\right)
\right]
-\frac{2(X (z-\zb+1)+|z|^2)}{6 X |z-1|^2 }L_0\left(\frac{-t_1}{-t_2}\right)
\\&
-\left(
\frac{z (X+\zb)}{|z-1|^2(X+z)}
\right)^2
\left[ 
L_1\left(\frac{-t_1}{-t_2}\right)
+L_1\left(\frac{-t_1^\prime}{-t_2}\right)
\right]
+\frac{1}{(z-1)^2}
L_1\left(\frac{-t_1}{-t_2}\right)
\\&
-\frac{2}{3(z-1)^3}\left(
\frac{z (X+\zb) (X (1+z-\zb)+|z|^2)}{ X^2 (\zb-1)^2}
+\frac{ z^3 (X+\zb)^3}{ X^2  (\zb-1)^3 (X+z)}
-1
\right)L_2\left(\frac{-t_1}{-t_2}\right)
\\&
+\frac{2 z (X+\zb)^3}{3 (|z-1|^2)^3 (X+z)}
L_2\left(\frac{-t_1^\prime}{-t_2}\right)
-
\frac{2 z^2 X^2}{(X+z)^4}
Ls_{-1}\left(\frac{-t_1^\prime}{-t_2},\frac{-t_1}{-t_2} \right)
\\&
-\frac{z (X+\zb)^3}{3 (X+1)^2 |z-1|^2 \zb (X+z)}
-\frac{X (\zb -1)+|z|^2(1+X)}{3 (X+1) (z-1)|z|^2}
\,. \label{eq:c_ggg_s_pmp}
\end{split}
\end{align}
We have written this vertex in such a way to make the MRK limit transparent. In the $X\to \infty$ limit we note that only the first two lines of \cref{eq:c_ggg_s_pmp} contribute to leading power, and we can readily verify
\begin{align}
\begin{split}
c^{(1, \, s)}_{gggg^*}(p_3^\oplus,p_2^\ominus, p_4^\oplus, q_2) 
&\toMRK
c^{(1, \ s)}_{g g g^* }(p_2,p_3,q_1)+v^{(1, \ s)}_{g^*gg^*}(-q_{1},p_4,q_2)
\,.
\label{eq:cs_2m3p4p_MRK}
\end{split} 
\end{align}
As discussed in \cref{sec:gN4}, the tree-level amplitude for this colour ordering is leading in both MRK and $\MRKx$ limits. To obtain the $\MRKx$ limit of \cref{eq:c_ggg_s_pmp}, we simply note
\begin{equation}
  c^{(1, \, s)}_{gggg^*}(p_3^\oplus,p_2^\ominus, p_4^\oplus, q_2) = c^{(1, \, s)}_{gggg^*}(p_4^\oplus,p_2^\ominus, p_3^\oplus, q_2)   \,,
\end{equation}
which follows from \cref{eq:reflection} and the invariance of \cref{eq:c_ggg_s_pmp} under $p_5 \leftrightarrow p_1$. From this, it follows that
\begin{align}
\begin{split}
c^{(1, \, s)}_{gggg^*}(p_3^\oplus,p_2^\ominus, p_4^\oplus, q_2) 
&\toMRKx
c^{(1, \ s)}_{g g g^* }(p_2,p_4,q_1^\prime)+v^{(1, \ s)}_{g^*gg^*}(-q_{1}^\prime,p_3,q_2)
\,.
\label{eq:cs_2m3p4p_MRKx}
\end{split} 
\end{align}

%\begin{align}
%\begin{split}
%c^{(1, \ \scalar)}(p_3^\ominus, p_2^\oplus, p_4^\oplus, q_2)
%&=
%c^{(1, \ \scalar)}_{g g g^* }%(p_2,p_3,q_1)+%
%\frac{1}{6}\left(
%\frac{
%z(2-\zb)+X(z-\zb+1)
%}{X|z-1|^2}
%\right)L_0 \left(\frac{-s_{23}}{-s_{51}}\right)
%\\&
%+
%\frac{z(\zb-1)(1+X)(X+z) (z(2-\zb)+X(z-\zb+1))}{3 X^3 (|z-1|^2)^3}
%L_2 \left(\frac{-s_{23}}{-s_{51}}\right)
%\\&
%-\frac{1}{3}\frac{z(\zb-1)(1+X)}{\zb(z-1)X^2}
%-\frac{1}{6}\frac{X+z}{\zb(z-1)X}
%\,.
%\label{eq:cN1_2m3p4p}
%\end{split} 
%\end{align}
This concludes our discussion of primitive five-gluon amplitudes in the NMRK limit. The NMRK limit of the relevant partial amplitudes discussed in \cref{sec:simple} can be obtained from the results presented in this section, via the kinematic maps discussed in \cref{sec:min}, and the supersymmetric rearrangements in \cref{eq:Ag_decomp}.

\section{Colour-dressed one-loop vertex for the peripheral emission of two gluons}
\label{sec:cCgg}
In this section we combine the colour analysis of \cref{sec:colour-5} and the kinematic analysis of \cref{sec:2g}. 
Having analysed the NMRK limit of the necessary partial amplitudes, we insert them into \cref{eq:A1_g_simple} to find the NMRK limit of the colour-dressed amplitude.

Let us first consider \cref{eq:A1_g_simple} for MHV amplitudes. We find that the kinematic coefficients of the totally symmetric colour structures have the following form, for $m {\in} \{g, f\}$:
\begin{align}
\begin{split}
A^{(1, \, m)}_{d}&(2,3,4,5,1)\toNMRK 2s_{12} \, C^{(1, \, m)}_{d}(p_2,p_3,p_4,q_2) \, R^{(0)}_{g^*} (q_2) \, C^{(0)}_{g^* g g}(-q_2, p_5^{-\lambda_1}, p_1^{\lambda_1})
\label{eq:A1dm}
\end{split}
\end{align}
Here we have introduced the partial vertex
\begin{align}
\begin{split}
    C^{(1, \, m)}_{d}(p_2,p_3,p_4,q_2)= 
 &C^{(0)}_{g g g g^*}(p_2,p_3,p_4,q_2)
    \left( 
        c^{(1, \, m)}_{g g g g^*}(p_2,p_3,p_4,q_2)
         -c^{(1, \, m)}_{g g g g^*}(p_3,p_2,p_4,q_2)
    \right)
\\
+&C^{(0)}_{g g g g^*}(p_2,p_4,p_3,q_2)
    \left(
    c^{(1, \, m)}_{g g g g^*}(p_2,p_4,p_3,q_2)
    -c^{(1, \, m )}_{g g g g^*}(p_4,p_2,p_3,q_2)
    \right)  \,. \label{eq:Cd1}
\end{split}
\end{align}
We emphasise the fact that the one-loop corrections to $C_{g^* g g}(-q_2, p_5, p_1)$ have canceled in \cref{eq:A1dm}, and in the $m{=}g$ case, all large logarithmic corrections have also canceled. 

\begin{figure}[hbt]
\centering
     \includegraphics[scale=0.15]{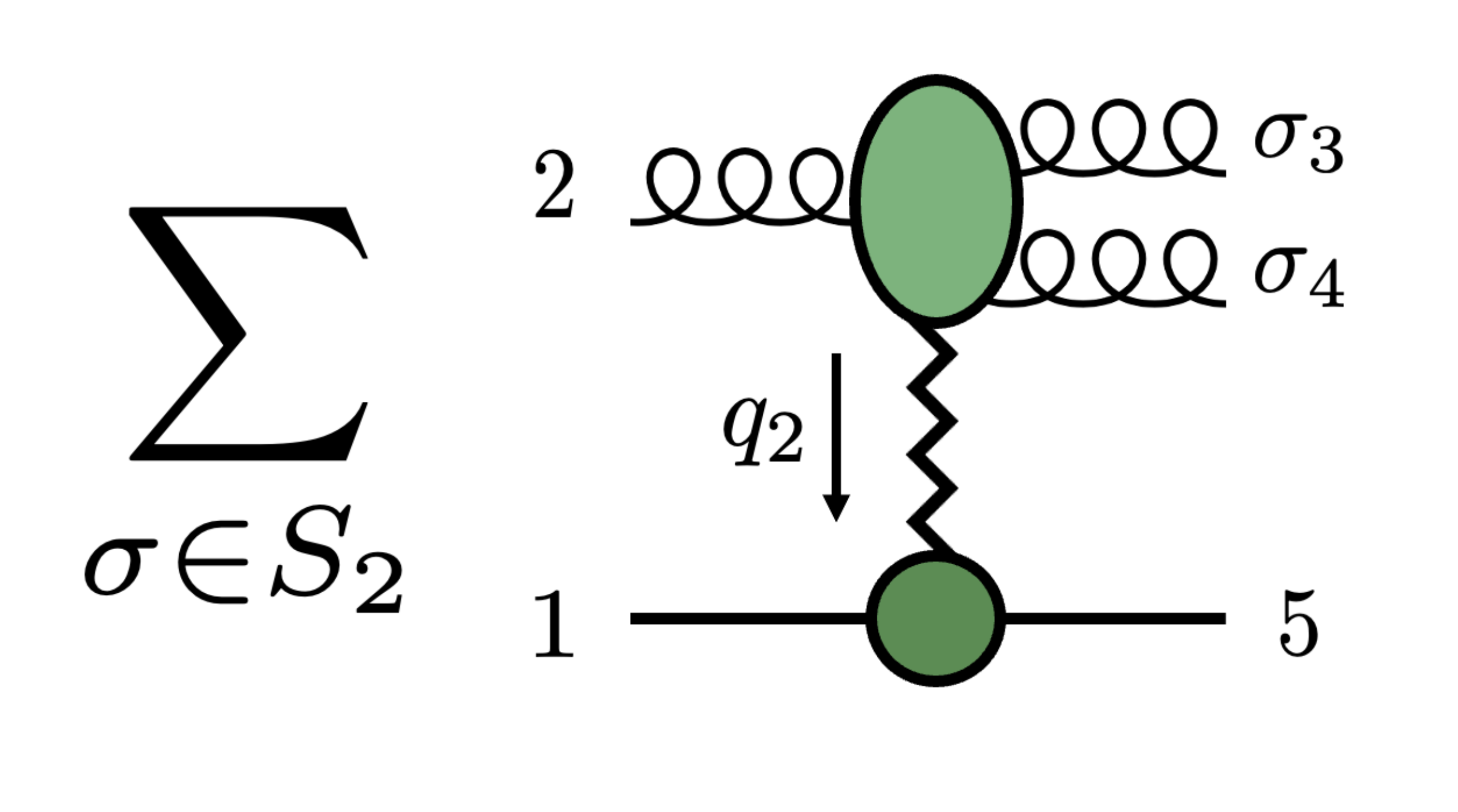}
    \caption{Illustration of \cref{eq:5g_permitted}, which is one possible form of all-orders factorisation of the dispersive part of five-parton amplitudes which (unlike the factorised form in \cref{subfig:Fac_NMRK}) is compatible with the one-loop result \cref{eq:A1_g_simple_3}. }
    \label{fig:Fac_NMRK_g}
\end{figure}

Performing a similar analysis of the kinematic coefficient of the tree-level colour structures we find, 
\begin{align}
\begin{split}
&A^{(1, \, m)}_{FF}(2,3,4,5,1)\toNMRK 2s_{12} \, R^{(0)}_{g^*}(q_2) \, C^{(0)}_{g^* g g}(-q_2, p_5^{-\lambda_1}, p_1^{\lambda_1}s)
\\
&\times
\Bigg\{
 C^{(1 \,, m)}_{FF}(p_2,p_3,p_4,q_2)
 +C^{(0)}_{g g g g^*}(p_2,p_3,p_4,q_2)
    \bigg( 
    r_{g^*}^{(1, \, m)}(q_2;p_4^+ p_5^-)+ c^{(1, \, m)}_{g^* g g}(-q_2,p_5,p_1)
    \bigg)\Bigg\} \,,\label{eq:A1FFm}
\end{split}
\end{align}
which again is valid for both $m {\in} \{g, f\}$. We have collected the dependence on the subset of momenta $\{p_2, p_3, p_4 \}$ with the definition
\begin{align}
\begin{split}
C^{(1 \,, m)}_{FF}(p_2,p_3,p_4,q_2)= 
 &C^{(0)}_{g g g g^*}(p_2,p_3,p_4,q_2)
    \bigg( 
        \frac{2}{3}c^{(1, \, m)}_{g g g g^*}(p_2,p_3,p_4,q_2)
         +\frac{1}{3}c^{(1, \, m)}_{g g g g^*}(p_3,p_2,p_4,q_2)
    \bigg)
\\
-&C^{(0)}_{g g g g^*}(p_2,p_4,p_3,q_2)
    \left(
    \frac{1}{3}c^{(1,\, m)}_{g g g g^*}(p_2,p_4,p_3,q_2)-\frac{1}{3}c^{(1\, m)}_{g g g g^*}(p_4,p_2,p_3,q_2)
    \right)  \,.
\label{eq:CFF1}
\end{split}
\end{align}
In contrast to \cref{eq:A1dm}, \cref{eq:A1FFm} has a large logarithmic term $r^{(1,m)}_{g^*}$, and importantly, the large scale of this logarithm depends on the colour ordering.  
The definitions of the partial vertices allow for easier discussion of the full one-loop amplitude, which becomes
\begin{align}
\begin{split}
  &\disp{\mathcal{A}^{(1)[-]}_{5}(2^{\lambda_2},3^{\lambda_3},4^{\lambda_4},5^{-\lambda_1},1^{\lambda_1})} \toNMRK g_s^5 \, 2s_{12} \, R^{(0)}_{g^*}(q_1) \, (-F^d_{a_5 a_1})C^{(0)}_{g^* g g}(-q_2, p_5^{-\lambda_1}, p_1^{\lambda_1})
  \\
  &\times 
  \Bigg\{
  \sum_{\sigma \in S_2} 
  (F^{a_{\sigma_3}} F^{a_{\sigma_4}})_{a_2 c_2} 
  \bigg[
      N_c \, C_{FF}^{(1, \, g)}(p_2, p_{\sigma_3}, p_{\sigma_4}, q_2) 
      +N_f \,  C_{FF}^{(1, \, f)}(p_2, p_{\sigma_3}, p_{\sigma_4}, q_2)
      \\
      & \hspace{2 em}
      +C^{(0)}_{gggg^*}(p_2,  p_{\sigma_3}, p_{\sigma_4}, q_2) 
          \bigg( N_c \, r_{g^*}^{(1, \, g)}(q_2; p_{\sigma_4}^+ p_5^-)
          \\&
          \hspace{12 em }
          +N_c \, c_{g^* g g}^{(1, \, g)}({-q_2}, p_5, p_1)
          +N_f \,  c_{g^* g g}^{(1, \, f)}({-q_2}, p_5, p_1)
      \bigg)
  \bigg]
 \\
  &\hspace{2 em}
  +d_A^{a_2 a_3 a_4 c_2} C_{d}^{(1, \, g)}(p_2, p_3, p_4, q_2) 
 +2 \, d_F^{a_2 a_3 a_4 c_2} C_{d}^{(1, \, f)}(p_2, p_3, p_4, q_2)
  \Bigg\}+\cO(\epsilon)\,.
  \label{eq:A1_g_simple_2}
\end{split}
\end{align}
The form of \cref{eq:A1_g_simple_2} is not compatible with the straightforward factorisation given in \cref{eq:Regge_conjecture}. Due to fact that different large logarithms multiply the two different colour orderings of $C^{(0)}_{gggg^*}$, it is not possible to simply define a colour-dressed one-loop vertex of the form
$$\cC^{(1)}_{gggg^*}(p_2, p_3, p_4, q_2)\,.$$
Rather, we define colour-dressed building blocks for each colour structure in \cref{eq:A1_g_simple_2}. For an individual tree-level colour structure we define
\begin{equation}
    \cC_{FF}(p_2, p_3, p_4, q_2)= \cC^{(0)}_{FF}(p_2, p_3, p_4, q_2)+\cC_{FF}^{(1)}(p_2, p_3, p_4, q_2)+ \mathcal{O}(\gs^6)\,,
\end{equation}
with tree-level and one-loop coefficients
\begin{align}
  \cC^{(0)}_{FF}(p_2, p_3, p_4, q_2) &= \gs^2 \, (F^{a_{3}} F^{a_{4}})_{a_2 c_2}  \cC^{(0)}_{gggg^*}(p_2, p_3, p_4, q_2)\,,
  \\
  \begin{split}
  \cC_{FF}^{(1)}(p_2, p_3, p_4, q_2) &=  c_\Gamma \gs^4
   (F^{a_{3}} F^{a_{4}})_{a_2 c_2} 
  \Big(
      N_c \, C_{FF}^{(1, \, g)}(p_2, p_3, p_4, q_2) \\
      & \qquad \qquad \qquad \qquad \,
      +N_f \, C_{FF}^{(1, \, f)}(p_2, p_3, p_4, q_2)
  \Big)\,, \label{eq:cCFF}
  \end{split}
\end{align}
respectively. In terms of this notation, we note the colour-dressed tree-level vertex, \cref{eq:cC0gg}, can be written
\begin{equation}
     \cC^{(0)}_{gggg^*}(p_2, p_3, p_4, q_2) = \sum_{\sigma \in S_2} \cC^{(0)}_{FF}(p_2, p_{\sigma_3}, p_{\sigma_4}, q_2)\,.
\end{equation}
Similarly, for the symmetric colour structures which first appear at one loop, we define
\begin{equation}
    \cC_{d}(p_2, p_3, p_4, q_2)= \cC_{d}^{(1)}(p_2, p_3, p_4, q_2) + \mathcal{O}(\gs^6) \,,
\end{equation}
which starts at order $\gs^4$,
\begin{align}
  \cC_{d}^{(1)}(p_2, p_3, p_4, q_2) &=  c_\Gamma \, \gs^4 \left(
   d_A^{a_2 a_3 a_4 c_2} C_{d}^{(1, \, g)}(p_2, p_3, p_4, q_2) 
 +2 \, d_F^{a_2 a_3 a_4 c_2} C_{d}^{(1, \, f)}(p_2, p_3, p_4, q_2)
 \right)\,. \label{eq:cCd}
\end{align}
With these colour-dressed definitions \cref{eq:A1_g_simple_2} can be written in the more compact from
\begin{align}
\begin{split}
  &\frac{1}{2s}\disp{\mathcal{A}^{(1)[-]}_{5}(2^{\lambda_2},3^{\lambda_3},4^{\lambda_4},5^{-\lambda_1},1^{\lambda_1})} \toNMRK\\
  &\left( 
  \mathcal{C}_{FF}^{(1)}(p_2, p_3, p_4, q_2)
  +\mathcal{C}_{FF}^{(1)}(p_2, p_4, p_3, q_2)
  +\mathcal{C}_{d}^{(1)}(p_2, p_3, p_4, q_2)
  \right)
  \mathcal{R}_{g^*}^{(0)}(q_2) \mathcal{C}_{g^* g g}^{(0)}(-q_2,p_5, p_1)
  \\
  &+\left(
   \mathcal{C}^{(0)}_{FF}(p_2, p_3, p_{4}, q_2) \mathcal{R}_{g^*}^{(1)}(q_2;s_{45}) 
   +
   \mathcal{C}^{(0)}_{FF}(p_2, p_4, p_{3}, q_2) \mathcal{R}_{g^*}^{(1)}(q_2;s_{35}) 
   \right)
   \mathcal{C}_{g^* g g}^{(0)}(-q_2,p_5, p_1)
\\
  &+\mathcal{C}^{(0)}_{g g g g^*}(p_2, p_3, p_4, q_2) \mathcal{R}_{g^*}^{(0)}(q_2) \mathcal{C}_{g^* g g}^{(1)}(-q_2,p_5, p_1)+\cO(\epsilon)
\,.\label{eq:A1_g_simple_3}
\end{split}
\end{align}
This is as far as the analysis of the MHV one-loop pure-gluon amplitudes can carry us. We conclude this section with a brief discussion on how the one-loop result \cref{eq:A1_g_simple_3} may fit into an all-orders expansion describing $2 \to 3$ amplitudes in the NMRK to NLL accuracy.
While, as we have discussed, a single-term factorisation formula such as \cref{eq:Regge_conjecture} is not compatible with the requirement of a single vertex which itself factorises into the known $\MRK $ and $\MRKx$ limits, one possibility which \emph{is} permitted by the one-loop result \cref{eq:A1_g_simple_3} is 
\begin{align}
\begin{split}
 \disp{ \mathcal{A}^{[-]}_{5}(2^{\lambda_2},3^{\lambda_3},4^{\lambda_4},5^{-\lambda_1},1^{\lambda_1})} \toNMRK & 2s \, \sum_{\sigma \in S_2} 
 \left(
  \cC_{FF}(p_2, p_{\sigma_3}, p_{\sigma_4}, q_2)+\frac{1}{2}\cC_{d}(p_2, p_{\sigma_3}, p_{\sigma_4}, q_2) 
  \right)
  \\
  &\times
  \mathcal{R}_{g^*}(q_2;p_{\sigma_4}^+p_5^-)
  \,
  \mathcal{C}_{g^* g g}(-q_2,p_5,p_1)\,,
  \label{eq:5g_permitted}
\end{split}
\end{align}
which is depicted in \cref{fig:Fac_NMRK_g}.

Of course, whether the term $\cC_d$ recieves logarithmic corrections at higher orders, let alone the form they may take, cannot be informed by this one-loop analysis. Much can be learned, therefore, from investigating the NMRK limit two-loop five-gluon amplitudes. 

Finally, we turn to the simpler case of a single negative helicity gluon. We obtain the compact results
\begin{align}
\begin{split}
    C_{FF}^{(1, \, s)}(p_2^\oplus, p_3^\oplus, p_4^\oplus, q_2)&=C_{FF}^{(1, \, g)}(p_2, p_3, p_4, q_2)
    =
    -C_{FF}^{(1, \, f)}(p_2, p_3, p_4, q_2)
    \\
    &=\frac{1}{q_{2 \perp}}\left(\frac{q_{2 \perp}^*}{q_{2 \perp}}  \right)
\frac{X(z-1)^3 \left(X^3 \zb+X^2 (z+1) \zb+X z (z+1)+z^2\right)}{ (X+1)^2 z^2 (\zb-1) (X+z)^2}\,,
\end{split}
\\
\begin{split}
    C_{d}^{(1, \, s)}(p_2^\oplus, p_3^\oplus, p_4^\oplus, q_2)&=C_{d}^{(1, \, g)}(p_2, p_3, p_4, q_2)
    =-C_{d}^{(1, \, f)}(p_2, p_3, p_4, q_2)
    \\
    &=-\frac{1}{q_{2 \perp}}\left(\frac{q_{2 \perp}^*}{q_{2 \perp}}\right)
  \frac{3  X (z-1)^2 (z-\zb)}{ (X+1) z (\zb-1) (X+z)}\,.
\end{split}
\end{align}

The one-loop amplitude in the NMRK is then
\begin{align}
\begin{split}
  &\mathcal{A}^{(1)[-]}_{5}(2^\oplus,3^\oplus,4^\oplus,5^{-\lambda_1},1^{\lambda_1}) \toNMRK\gs^5 \,2s_{12} \,R^{(0)}_{g^*}(q_1)(-F^d_{a_5 a_1}) C^{(0)}_{g^* g g}(-q_2, p_5^{-\lambda_1},p_1^{\lambda_1})
  \\
  &\times 
  \Bigg\{
 \sum_{\sigma \in S_2}(F^{a_{\sigma_3}} F^{a_{\sigma_4}})_{a_2 c_2} 
  \  (N_c-N_f)C_{FF}^{(1, \, s)}(p_2^\oplus, p_{\sigma_3}^\oplus, p_{\sigma_4}^\oplus, q_2) \,
  \\
  &\quad +
  \left(d_A-2 \, d_F\right)^{a_2 a_3 a_4 c_2} C_{d}^{(1, \, s)}(p_2^\oplus, p_3^\oplus, p_4^\oplus, q_2)
  \Bigg\}+\cO(\epsilon)\,.
  \label{eq:A1_g_plus}
\end{split}
\end{align}
This amplitude is compatible with \cref{eq:Regge_conjecture}, and we can define the one-loop colour-dressed vertex
\begin{align}
\begin{split}
    \mathcal{C}^{(1)}_{ggg^*}(p_2^{\oplus}, p_3^{\oplus}, p_4^{\oplus}, q_2)&=
 \sum_{\sigma \in S_2}(F^{a_{\sigma_3}} F^{a_{\sigma_4}})_{a_2 c_2} 
  \  (N_c-N_f)C_{FF}^{(1, \, s)}(p_2^\oplus, p_{\sigma_3}^\oplus, p_{\sigma_4}^\oplus, q_2) \,
  \\
  &\quad +
  \left(d_A-2 \, d_F\right)^{a_2 a_3 a_4 c_2} C_{d}^{(1, \, s)}(p_2^\oplus, p_3^\oplus, p_4^\oplus, q_2)\,.
  \end{split}
\end{align}
This colour-dressed vertex has the MRK limit
\begin{equation}
   \mathcal{C}^{(1)}_{ggg^*}(p_2^{\oplus}, p_3^{\oplus}, p_4^{\oplus}, q_2)
   \toMRK
   \mathcal{C}^{(1)}_{ggg^*}(p_2^{\oplus}, p_3^{\oplus}, q_1)
   \mathcal{R}_{g^*}^{(0)}(q_1)
   \mathcal{V}_{g^* g g^*}^{(0)}(-q_1, p_4^\oplus, q_2),
\end{equation}
as required by the known compatibility of the amplitude with \cref{eq:Regge_5}.

In the remainder of this paper, we perform an analogous investigation of the one-loop amplitudes with an external quark-antiquark pair.

\section{One-loop primitive vertices for the peripheral emission of a quark and gluon}
\label{sec:c_qqbg}
In this section we take the NMRK limit of one-loop primitive amplitudes of three gluons and a quark-antiquark pair. We use the amplitudes provided by ref.~\cite{Bern:1994fz}. Expressions are given in that paper for the following N$^{-1}$MHV helicity configurations
$$
  \{ \qb^\ominus, q^\oplus, g^\oplus,g^\oplus, g^\oplus\}, \quad
    \{ \qb^\ominus, g^\oplus, q^\oplus,   g^\oplus, g^\oplus\},
$$
which in the NMRK limit only give rise to the one-loop 
$\{g^\oplus, g^\oplus, g^\oplus, g^* \}$ vertex discussed in \cref{sec:s_ppp}, and the MHV configurations
\begin{align*}
    \{ \qb^\ominus, q^\oplus, g^\ominus,g^\oplus, g^\oplus\}, \quad
    \{ \qb^\ominus, q^\oplus,  g^\oplus,g^\ominus, g^\oplus\}, \quad
    &\{ \qb^\ominus, q^\oplus,  g^\oplus, g^\oplus,g^\ominus\}, \quad
    \{ \qb^\ominus, g^\ominus, q^\oplus, g^\oplus,g^\oplus\}\\
    \{ \qb^\ominus, g^\oplus, q^\oplus, g^\ominus, g^\oplus\}, \quad
    &\{ \qb^\ominus, g^\oplus, q^\oplus, g^\oplus,g^\ominus\}.
\end{align*}
Let us briefly review how any amplitude can be related to one of the above by discrete transformations. We note that parity conjugation, via complex conjugation, can be used to map to a single-minus or two-minus amplitude, and charge conjugation,
\begin{equation}
    A_n^{(1, \, m_L)}(\cdots, i_{q}^{\lambda}, \cdots,  j_{\qb}^{(-\lambda)}, \cdots) 
    =
    A_n^{(1, \, m_R)}(\cdots, i_{\qb}^{\lambda}, \cdots,  j_{q}^{(-\lambda)}, \cdots) \,,
\end{equation}
can be used to map to an amplitude where the negative-helicity fermion is the antiquark. Cyclicity of the amplitude means the first argument can always be set to be the antiquark. Finally, the reflection identity, \cref{eq:reflection_q}, can be used to bring the quark into either the second or third entry.

A supersymmetric decomposition analogous to \cref{eq:susy} can be used to organise one-loop amplitudes with an external quark-antiquark pair. We follow ref.~\cite{Bern:1994fz} in relating $g$ and $f$ amplitudes to a $\cN{=}1$ vector multiplet,
\begin{equation}
A_n^{(1, \ \NoneV)}=A_n^{(1, \, g_L)}+A_n^{(1, \, g_R)}+A_n^{(1, \, f_L)}+A_n^{(1, \, f_R)}\,.
\label{eq:susy_q}
\end{equation}
The left-hand side of this equation can be immediately obtained from the corresponding pure-gluon amplitude via a supersymmetric Ward identity, reducing the number of independent amplitudes which we need to analyse. In \cref{sec:fin_4} we explain how a supersymmetric Ward identity leads to the equivalence of the one-loop corrections to $ggg^*$ and $\qb q g^*$ vertices in $\NoneV$, \cref{eq:c3q_N1V}. The same reasoning applies here for five-parton amplitudes in the NMRK limit. For example, from the supersymmetric Ward identity
\begin{equation}
    a_5^{(1, \, \NoneV)}(2_\qb, 3_q, 4,5,1)= a_5^{(1,\, \NoneV )}(2, 3, 4,5,1) \,,\label{eq:susy_1loop_5}
\end{equation}
it follows that
\begin{equation}
 c_{\qb q g g^*}^{(1, \, \NoneV)}(p_2^{\lambda_2}, p_3^{\lambda_3}, p_4^{\lambda_4}, q_2)=  c_{g g g g^*}^{(1, \, \NoneV)}(p_2^{\lambda_2}, p_3^{\lambda_3}, p_4^{\lambda_4}, q_2)\,,\label{eq:cq_N1V}
\end{equation}
with analogous relations for different orderings of the external particle species.

For each configuration listed above, explicit amplitudes are provided in ref.~\cite{Bern:1994fz} for particle content $\NoneV$, $g_L$, ${\None}_L$ and $s_L$.
We remind the reader that the ${\None}_R$ and $s_R$ amplitudes vanish for these configurations according to \cref{eq:tadpole} and \cref{eq:bubble}. The fermion contribution can be obtained via the decomposition
\begin{equation}
A_n^{(1, \, f_L)}=A_n^{(1, \, {\NoneX}_L)}
-A_n^{(1, \, {s}_L)}
\,,
\label{eq:susy_fL}
\end{equation}
and the $g_R$ contribution can be subsequently obtained via \cref{eq:susy_q}, i.e.
\begin{equation}
A_n^{(1, \, g_R)}=
A_n^{(1, \ \NoneV)}-A_n^{(1, \, g_L)}-A_n^{(1, f_L)}\,.
\label{eq:susy_q_dupe}
\end{equation}
While such organisational principles have been long used in the study of one-loop amplitudes, this work is the first time extensive use has been made these principles to organise factorised building blocks in a high-energy limit. As such, we needed to re-analyse the Regge limit of one-loop $\qb q g g$ amplitudes in this framework, which is included in \cref{sec:rederiving}.
We find that a separate treatment of $m_L$ and $m_R$ primitive amplitudes, already useful for one-loop amplitudes, becomes even more useful when studying one-loop corrections. This is because requiring the external fermion line to pass to the left or right of the loop essentially limits the loop corrections to affect certain building blocks. We see this already in \cref{eq:aR1g_4_v2}, where $g_R$ amplitudes in the Regge limit only contribute to the $\qb q g^*$ vertex, in line with our intuition.
%for $\sum_{i=2}^4 \lambda_i=\oplus$.

The organisation of this section is similar to \cref{sec:2g}, where we first categorise the peripheral-emission vertices by internal particle content, then by helicity configuration, where a minimal set of primitive vertices is given by the three cyclic orderings 
$$
\{\qb^\ominus, q^\oplus, g^\oplus , g^*\}\,,
\qquad 
\{\qb^\ominus,  g^\oplus , q^\oplus,g^*\}\,,
\qquad
\{g^\oplus , \qb^\ominus, q^\oplus, g^*\}\,.
$$
%and finally by the cyclic ordering of momenta.
\subsection{\texorpdfstring{$\Nfour$}{N=4} contribution}
\label{sec:N4_qq}
Although in practice we use a $\NoneV$ multiplet decomposition, \cref{eq:susy_q_dupe}, to organise the amplitudes and vertices, it is worthwhile to analyse the amplitudes with a circulating $\Nfour$ multiplet for two reasons. Firstly, ref.~\cite{Bern:1994fz} writes the $g_L$ and $\NoneV$ amplitudes in terms of the function \cref{eq:VN4_5}, and this proves convenient also for the factorised expressions. Secondly, it provides us with the simplest context to study the key difference between the pure-gluon case and the case with an external quark-antiquark pair. 

Recall the discussion of \cref{eq:VN4_5} in \cref{sec:N4_mhv}. To be concrete, let us study the physical scattering 
$$q(p_2) \, g(p_1) \to q(p_3)\, g(p_4) \, g(p_5). $$
The treatment of the ordering $\{p_2, p_3, p_4, q_2 \}$ proceeds similarly to the pure gluon case, and by a SUSY Ward identity we obtain
\begin{align}
\begin{split}
  \mathrm{Re}\left[  a_5^{(1, \, \Nfour) }(2_\qb,3_g,4,5,1) \right] \toNMRK 
  &c_{\qb q g g^*}^{(1, \, \Nfour)}(p_2,p_3,p_4,q_2)
  \\
  +
  &r_{g^*}^{(1, \, \Nfour)}(q_2;p_4^+ p_5^-)
  +
  c_{g^* g g}^{(1, \, \Nfour)}(-q_2,p_5,p_1) 
  \end{split}
\end{align}
with primitive vertex
\begin{align}
\begin{split}
 c_{\qb q g g^*}^{(1, \, \Nfour)}(p_2,p_3,p_4,q_2)=&c_{g g g^*}^{(1, \, \Nfour)}(p_2,p_3,q_1)  
 +r_{g^*}^{(1, \, \Nfour)}(q_1;s_{34})\\
 +&v^{(1, \, \Nfour)}(t_1,\frac{s_{34} \, p_4^+}{(p_3^++p_4^+)},t_2)\,.
 \label{eq:234_N4_q}
 \end{split}
\end{align}
Likewise, for the second colour-ordering in \cref{eq:A1_q_simple}
we write
\begin{align}
\begin{split}
  \mathrm{Re}\left[  a_5^{(1, \, \Nfour) }(4,2_\qb,3_q,5,1) \right] \toNMRK 
  &c_{g \qb  q g^*}^{(1, \, \Nfour)}(p_4,p_2,p_3,q_2)
  \\
  +
  &r_{g^*}^{(1, \, \Nfour)}(q_2;p_4^+ p_5^-)
  +
   c_{g^* g g}^{(1, \, \Nfour)}(-q_2,p_5,p_1)\,, 
   \end{split}
\end{align}
with primitive vertex
\begin{align}
\begin{split}
 c_{ g \qb q g^*}^{(1, \, \Nfour)}(p_4,p_2,p_3,q_2)=&c_{gg g^*}^{(1, \, \Nfour)}(p_3,p_2,q_1)  
 +r_{g^*}^{(1, \, \Nfour)}(q_1;s_{24})\\
 +&v^{(1, \, \Nfour)}(t_1,\frac{s_{24} \, p_4^+}{(-p_2^++p_4^+)},t_2)\,.
 \label{eq:423_N4_q}
 \end{split}
\end{align}
In this form, it is transparent that the correct large logarithms are generated by the $\qb q g g^*$ and $g \qb  q g^*$ vertices in the MRK limit. This follows from the $U(1)$-decoupling identities, e.g.
\begin{equation}
C^{(0)}_{\qb q g g^*}(p_2^\oplus, p_3^\ominus, p_4^\oplus, q_2)+C^{(0)}_{\qb g q g^*}(p_2^\oplus, p_4^\oplus,  p_3^\ominus, q_2)
    +C^{(0)}_{g \qb qg^*}(p_4^\oplus, p_2^\oplus, p_3^\ominus, q_2)=0\,,\label{eq:U1_dupe_q-qg}
\end{equation}
which in the MRK limit becomes
\begin{equation}
C^{(0)}_{\qb q g g^*}(p_2^\oplus, p_3^\ominus, p_4^\oplus, q_2)=-
    C^{(0)}_{g \qb qg^*}(p_4^\oplus, p_2^\oplus, p_3^\ominus, q_2)\,,\label{eq:U1_dupe_q-qg_MRK}
\end{equation}
due to the power-suppression of the second term in \cref{eq:U1_dupe_q-qg}. The one-loop corrections, \cref{eq:234_N4_q,eq:423_N4_q} on the other hand become equal in the MRK limit, due to the limiting behaviour
\begin{equation}
\frac{s_{24} \, p_4^+}{(-p_2^++p_4^+)}\underset{\MRK}{=} \frac{s_{34} \, p_4^+}{(p_3^++p_4^+)}\,. \label{eq:eta_MRK}
\end{equation}
Thus the kinematic coefficients of the first two colour-structures in \cref{eq:A1_q_simple} are equal and opposite, as required by compatibility with \cref{eq:Regge_5}. 

We also need to consider the $\delta \Delta$ coefficient, \cref{eq:A_delta}. This must be free from large logarithms in the NMRK and must vanish in the MRK limit. To verify the former condition, we can write, (merely for convenience to separate out a large logarithmic term and a backward $g^*gg$ vertex)
\begin{align}
\begin{split}
  \mathrm{Re}\left[  a_5^{(1, \, \Nfour) }(2_\qb,4,3_q,5,1) \right] \toNMRK 
  &c_{ \qb g q g^*}^{(1, \, \Nfour)}(p_2,p_4,p_3,q_2)
  \\
  +
  &r_{g^*}^{(1, \, \Nfour)}(q_2;p_3^+ p_5^-)
  +
   c_{g^* g g}^{(1, \, \Nfour)}(-q_2,p_5,p_1)\,, 
   \end{split}
\end{align}
with vertex
\begin{align}
\begin{split}
 c_{\qb  g q g^*}^{(1, \, \Nfour)}(p_2,p_4,p_3,q_2)=&c_{gg g^*}^{(1, \, \Nfour)}(p_2,p_4,q_1^\prime)  
 +r_{g^*}^{(1, \, \Nfour)}(q_1^\prime;s_{43})\\
 +&v^{(1, \, \Nfour)}(t_1^\prime,\frac{s_{43} \, p_3^+}{(p_4^++p_3^+)},t_2)\,.
 %=
 %&c_{g g g^*}^{(1, \, \Nfour)}(p_3,p_4,-(p_3+p_4)) 
% +r_{g^*}^{(1, \, \Nfour)}(-(p_3+p_4);s_{42})\\
% +&v^{(1, \, \Nfour)}(t_1^\prime,\frac{s_{42} \, p_2^+}{(p_4^++p_2^+)},t_2)
 \label{eq:243_N4_q}
 \end{split}
\end{align}
All large logarithmic behaviour in the $m{=}g$ amplitudes can be conveniently written in terms of the $\Nfour$ function \cref{eq:VN4_5}, and only the $m{=}g_L$ amplitudes have large logarithms in the NMRK limit. Therefore the LL behaviour of \cref{eq:A_delta} in the NMRK limit coincides with that of
\begin{align}
\begin{split}
&\disp{A^{(1, \, \Nfour)}_{\delta \Delta}(p_2,p_3,p_4,p_5,p_1)}
\toNMRK
C^{(1, \, \Nfour)}_{\delta \Delta}
(p_2,p_3,p_4,q_2)R_{g^*}^{(0)}(q_2)C^{(0)}_{g^*gg}(-q_2,p_5,p_1)
\,. \label{eq:N4_dD}
\end{split}
\end{align}
Here we have defined the partial vertex
\begin{align}
\begin{split}
&C^{(1, \, \Nfour)}_{\delta \Delta}
(p_2,p_3,p_4,q_2)=
\\& \quad
		C^{(0)}_{ \qb q g g^*}(p_2, p_3, p_4, q_2)
\\&
		\quad\times\bigg(
			 c^{(1, \, \Nfour)}_{\qb q g g^*}(p_2, p_3, p_4, q_2)
    %\\& \hspace{-5 em}
			-c^{(1, \, \Nfour)}_{\qb g q g^*}(p_2, p_4, p_3, q_2)
            +r^{(1,\, g)}_{g^*}(q_2, p_4^+ p_5^-)
			-r^{(1,\, g)}_{g^*}(q_2, p_3^+ p_5^-)
		\bigg)
\\
&\quad+C^{(0)}_{g \qb q g^*}(p_4, p_2, p_3, q_2)
\\
		&\quad\times\bigg(
			 c^{(1, \, \Nfour)}_{g \qb q  g^*}(p_4, p_2, p_3, q_2)
      % \\& \hspace{-5 em}
			-c^{(1, \, \Nfour)}_{\qb g q g^*}(p_2, p_4, p_3, q_2)
			+r^{(1,\, g)}_{g^*}(q_2, p_4^+ p_5^-)
			-r^{(1,\, g)}_{g^*}(q_2, p_3^+ p_5^-)
   \bigg)\,. \label{eq:C_N4_dD}
\end{split}
\end{align}
We note that in \cref{eq:N4_dD}, all dependence on the one-loop corrections to the backward $g^*ggg$ vertex cancels, so that this partial amplitude only contributes to the forward physics. We note that this partial vertex is free from large logarithms, because the difference of Reggeisation factors
$$r^{(1,\, g)}_{g^*}(q_2, p_4^+ p_5^-)
			-r^{(1,\, g)}_{g^*}(q_2, p_3^+ p_5^-)\,,$$
depends only on the ratio $X=p_3^+/p_4^+$ which is finite in the NMRK limit. Finally, we note that due to \cref{eq:U1_dupe_q-qg_MRK} and \cref{eq:eta_MRK}, the entire partial vertex vanishes in the MRK limit as expected.

\subsection{\texorpdfstring{$\NoneX$}{N=1X} contribution}
\label{sec:N1_qqb}
In this section, we repeat the analysis of \cref{sec:gN1X} but for amplitudes with an external quark-antiquark pair. 
\subsubsection{\texorpdfstring{$\{ \qb^\ominus, q^\oplus,g^\oplus,g^*\}$}{(qb-,q+,g+,g*)}}
\label{sec:N1_qbm_qp_gp}
The amplitudes we can use to derive these vertices have a very simple form, for example,
\begin{align}
    \begin{split}
    a_{5}^{(1, \, {\None}_L)}(2_{\bar{q}}^\ominus,3_q^\oplus,4^\oplus,5^\oplus,1^\ominus)
=&
    -\frac{(X+z)(Y(\zb-1)-1)}{X+XY|z-1|^2+|z|^2}
    L_0\left(\frac{-s_{23}}{-s_{51}}\right)\,. \label{eq:a5L_2m3p4p5m1p_gen}
    \end{split}
\end{align}
This amplitude is finite in $\epsilon$, and has the simple NMRK limit
\begin{align}
    \begin{split}
    a_{5}^{(1, \, {\None}_L)}(2_{\bar{q}}^\ominus,3_q^\oplus,4^\oplus,5^\oplus,1^\ominus)
\toNMRK&
    -\frac{(X+z)}{X(z-1)}
    L_0\left(\frac{-s_{23}}{-s_{51}}\right)\,. \label{eq:a5L_2m3p4p5m1p}
    \end{split}
\end{align}
Unlike the amplitudes in \cref{sec:gN1X}, the form of this amplitude is not suggestive of a separation into central and peripheral parts. To define a $\qb q g g^*$ vertex, we  must therefore massage the amplitude into such a form. We use \cref{eq:c3fL} to add and subtract
\begin{equation}
  c^{(1, \, {\None}_L)}_{\qb q g^*}( p_2^\ominus,p_3^\oplus, q_1)
    +  c_{g^*gg}^{(1, \, \None)}( -q_2, p_5^\ominus, p_1^\oplus)=\frac{1}{2}\log \left( \frac{-s_{23}}{-s_{51}} \right)\,.
\end{equation}
This brings the amplitude into a more familiar form
\begin{align}
    \begin{split}
    a_{5}^{(1, \, {\None}_L)}(2_{\bar{q}}^\ominus,3_q^\oplus,4^\oplus,5^\ominus,1^\oplus)
\toNMRK&
 c^{(1, \, {\None}_L)}_{\qb q g^*}( p_2^\ominus,p_3^\oplus, q_1)
 \\
    +&\frac{X(1+z-\zb)-z(\zb-2)}{2 X|z-1|^2}
    L_0\left(\frac{-s_{23}}{-s_{51}}\right)
    \\
    +&c_{g^*gg}^{(1, \, {\None}_L)}( -q_2, p_5^\ominus, p_1^\oplus)
    \,.
    \end{split}
\end{align}
Now subtracting the $g^*gg$ vertex, we are left with the two-parton peripheral-emission vertex 
\begin{align}
    \begin{split}
   c^{(1, \, {\None}_L)}_{\qb q g g^*}( p_2^\ominus,p_3^\oplus,p_4^\oplus, q_2)
=&
 c^{(1, \, {\None}_L)}_{\qb q g^*}( p_2^\ominus,p_3^\oplus, q_1)
    +\frac{X(1+z-\zb)-z(\zb-2)}{2 X|z-1|^2}
    L_0\left(\frac{-t_1}{-t_2}\right)
    \,,
    \end{split}
\end{align}
which has the trivial MRK limit
\begin{align}
    \begin{split}
   c^{(1, \, {\None}_L)}_{\qb q g g^*}( p_2^\ominus,p_3^\oplus,p_4^\oplus, q_2)
\toXinf&
c^{(1, \, {\None}_L)}_{\qb q g^*}(p_2^\ominus,p_3^\oplus, q_1)
    +v_{g^* g g^*}^{(1, \, \None)}(-q_1, p_4, q_2 )
    \,.
    \end{split}
\end{align}
As mentioned previously, we cannot use the $\MRKx$ limit to check this function, as the tree-level amplitude which multiplies it is power-suppressed in this limit.

\subsubsection{\texorpdfstring{$\{\qb^\ominus, g^\oplus, q^\oplus, g^* \}$}{(qb-,g+,q+,g*)}}
The derivation of the $\qb^\ominus g^\oplus q^\oplus g^* $ vertex is subtle, because the amplitudes from which we start are zero:
\begin{equation}
    A_5^{(1,\, {\NoneX}_L)}(\qb^\ominus, g^\oplus, q^\oplus, g^{\oplus},g^{\ominus})=A_5^{(1,\, {\NoneX}_L)}(\qb^\ominus, g^\oplus, q^\oplus, g^{\ominus},g^{\oplus})=0\,.
\end{equation}
We emphasise that this identity is not a consequence of \cref{eq:bubble}, which only applies to their ${\NoneX}_R$ counterparts. However, we should proceed as in all previous examples, and define a $c^{(1,\, {\NoneX}_L)}_{\qb g q g^*}$ vertex by subtracting the known $c^{^{(1,\, {\NoneX})}}_{g^* g g}$ vertex. This is the same logic we followed in \cref{sec:fin_4}, which led us to re-derive the known one-loop $\qb q g^*$ vertex in QCD, \cref{eq:IF_q_2}. In the present case we obtain
\begin{align}
    c_{\qb g q g^*}^{(1,\, {\NoneX}_L)}(p_2^\ominus, p_3^\oplus, p_4^\oplus, q_2) &= -c_{g^* g g}^{(1,\, {\NoneX})}(-q_2, p_5^{\lambda_5}, p_1^{\lambda_1})\\
    &=-\frac{1}{2\epsilon}
	\left(\frac{\mu^2}{-s_{234}}\right)^\epsilon
    -1
    \,,
\end{align}
where in the last line we have made use of the equality $s_{51}=s_{234}$. We note that the tree-level pre-factor of this one-loop function vanishes in both the MRK and $\MRKx$ limits.

\subsubsection{\texorpdfstring{$\{g^\oplus, \qb^\ominus, q^\oplus, g^* \}$}{(g+,qb-,q+,g*)}}
As in \cref{sec:N1_qbm_qp_gp}, the NMRK limits of the relevant amplitudes are again very simple, for example
\begin{align}
    \begin{split}
    a_{5}^{(1,\, {\NoneX}_L)}(2_{\bar{q}}^\ominus,3_q^\oplus,5^\ominus,1^\oplus,4^\oplus)
\toNMRK&
    \frac{1}{(z-1)}
    L_0\left(\frac{-s_{23}}{-s_{51}}\right)\,.
    \label{eq:a5N1L_b}
    \end{split}
\end{align}
We remark that this is simply $X/(X+z)$ times the amplitude in \cref{eq:a5L_2m3p4p5m1p}.
Recalling the NMRK behaviour of the tree level amplitudes,
\begin{align}
\frac{A^{(0)}(4^\oplus,2_{\bar{q}}^\ominus,3_q^\oplus,5^\ominus,1^\oplus)}{ A^{(0)}(2_{\bar{q}}^\ominus,3_q^\oplus,4^\oplus,5^\ominus,1^\oplus)}
    \toNMRK
   \frac{X}{X+z}
\end{align}
we see that
\begin{align}
A^{(1, \ {\NoneX}_L)}(4^\oplus,2_{\bar{q}}^\ominus,3_q^\oplus,5^\ominus,1^\oplus)+A^{(1, \ {\NoneX}_L)}(2_{\bar{q}}^\ominus,3_q^\oplus,4^\oplus,5^\ominus,1^\oplus)
    \toNMRK
  0\,,
\end{align}
which is reminiscent of the identity \cref{eq:furry}, which followed from Furry's theorem, but for an off-shell gluon with momentum $p_5+p_1$.

The vertex resulting from \cref{eq:a5N1L_b} is
\begin{align}
    \begin{split}
   c^{(1, \, {\NoneX}_L)}_{g \qb q  g^*}(p_4^\oplus, p_2^\ominus,p_3^\oplus, q_2)
=&
 c^{(1, \, {\NoneX}_L)}_{\qb q g^*}( p_2^\ominus,p_3^\oplus, q_1)
    +\frac{X(1+z-\zb)+|z|^2}{2 X|z-1|^2}
    L_0\left(\frac{-t_1}{-t_2}\right)
    \,,
    \end{split}\label{eq:c1_N1x_4p2m3p}
\end{align}
which has the transparent MRK limit
\begin{align}
    \begin{split}
    c^{(1, \, {\NoneX}_L)}_{g \qb q  g^*}(p_4^\oplus, p_2^\ominus,p_3^\oplus, q_2)
\toXinf&
c^{(1, \, {\NoneX}_L)}_{\qb q g^*}( p_2^\ominus,p_3^\oplus, q_1)
    +v_{g^* g g^*}^{(1, \, \None)}(-q_1, p_4, q_2 )
    \,.
    \end{split}
\end{align}
Once again, the tree-level coefficient of \cref{eq:c1_N1x_4p2m3p} is power supressed in the $\MRKx$ limit. This concludes our extraction of the necessary two-parton emission vertices with a circulating ${\NoneX}_L$ multiplet.

\subsection{Complex scalar contribution}
\label{sec:s_qqb}
The procedure to obtain these vertices proceeds similarly to \cref{sec:N1_qqb} and we simply collect the primitive vertices here.
%\subsubsection{\texorpdfstring{$\{\qb^\ominus, q^\oplus, g^\oplus, g^* \}$}{(qb-,q+,g+,g*)}}
The $\qb^\ominus q^\oplus g^\oplus  g^*$ vertex is
\begin{align}
\begin{split}
 c_{\qb q g g^*}^{(1, \ s_L)}(p_2^\ominus, p_3^\oplus, p_4^\oplus,q_1)  
&=
 c_{\qb q g^*}^{(1, \ s_L)}(p_2, p_3,q_1)
+\frac{1}{6}\frac{X(1+z-\zb)-z(\zb-2)}{ X |z-1|^2}L_0\left(\frac{-t_1}{-t_2}\right)
\\
&+\frac{1}{3}\frac{z|X+z|^2 (X(1+z-\zb)+|z|^2)}{ X^3 (z-1)^3(\zb-1)^2}L_2\left(\frac{-t_1}{-t_2}\right)
-\frac{|X+z|^2}{6X(1+X)(z-1)\zb}
\,. \label{eq:c1_sL_2m3p4p}
\end{split}
\end{align}
The $\qb^\ominus g^\oplus q^\oplus  g^*$ vertex is
%\subsubsection{\texorpdfstring{$\{\qb^\ominus, g^\oplus, q^\oplus, g^* \}$}{(qb-,g+,q-,g*)}}
\begin{equation}
    c_{\qb g q  g^*}^{(1,\, s_L)}(p_2^\ominus, p_4^\oplus, p_3^\oplus, q_2) = -c_{g^* g g}^{(1,\, {s})}(-q_2, p_5^{\lambda_5}, p_1^{\lambda_1})\,,
\end{equation}
which we recall is a transcendental function that depends only on the kinematic scale $s_{51}=s_{234}$.
%\subsubsection{\texorpdfstring{$\{g^\oplus, \qb^\ominus, q^\oplus, g^* \}$}{(g+,qb-,q+,g*)}}
Finally, the $  g^\oplus \qb^\ominus q^\oplus  g^*$ vertex is
\begin{align}
\begin{split}
 c_{g \qb q g^*}^{(1, \ s_L)}(p_4^\oplus,p_2^\ominus, p_3^\oplus, q_2)  
&=
 c_{\qb q g^*}^{(1, \ s_L)}(p_2, p_3,q_1)
+\frac{1}{6}\frac{X(1+z-\zb)+|z|^2}{ X |z-1|^2}L_0\left(\frac{-t_1}{-t_2}\right)
\\
&+\frac{1}{3}\frac{z(X+\zb)^2 (X(1+z-\zb)+|z|^2)}{ X^3 (z-1)^3(\zb-1)^2}L_2\left(\frac{-t_1}{-t_2}\right)
-\frac{(X+\zb)^2}{6X(1+X)(z-1)\zb}
\,.\label{eq:c1_sL_4p2m3p}
\end{split}
\end{align}
%\begin{align}
%    \begin{split}
%    a_{5L}^{(1, \ s)}(2_{\bar{q}}^\ominus,3_q^\oplus,5^\ominus,1^\oplus,4^\oplus)
%\toNMRK&
%    \frac{2}{3}\frac{1}{(z-1)^3}L_2\left(\frac{s_{23}}{s_{51}}\right)
%    +\frac{1}{(z-1)^2}L_1\left(\frac{s_{23}}{s_{51}}\right)
%    \\+&\frac{1}{3}\frac{X(1+z)(\zb-1)}{(1+X)(z-1)|z|^2}
%    \end{split}
%\end{align}
We note that, similar to the correspondence between the $\NoneX$ and $s$ vertices in the pure-gluon case, \cref{sec:gN1X}, the vertices presented in this section are similar to the vertices in \cref{sec:N1_qqb}, but the vertices have in addition an $L_2$ function and a purely rational term. We note that, in contrast to the $L_{s_{-1}}$ functions in, e.g. \cref{eq:c_ggg_s_pmp}, there are no instances of transcendental functions of weight two.

We note that, as expected, the MRK limits of \cref{eq:c1_sL_2m3p4p} and \cref{eq:c1_sL_4p2m3p} are both equal to
\begin{equation}
    c_{\qb q  g^*}^{(1,\, s_L)}(p_2^\ominus, p_3^\oplus, q_1)  
    +v_{g^* g g^*}^{(1, \, s)}(-q_1, p_4, q_2 )
    \,. \label{eq:sL_MRK}
\end{equation}
The equivalence of these two vertices in the MRK limit is an observation that we will make use of in \cref{sec:cC_qqbg}.

\subsection{Gluon contribution}
Again, the analysis proceeds similarly to \cref{sec:N1_qqb} and we simply collect the $g_L$ primitive vertices here. The $g_R$ vertices can be obtained via the supersymmetric decomposition \cref{eq:susy_q}. 
The $\qb^\ominus q^\oplus g^\oplus g^* $ vertex is
%\subsubsection{\texorpdfstring{$\{\qb^\ominus, q^\oplus, g^\oplus, g^* \}$}{(qb-,q+,g+,g*)}}
\begin{align}
\begin{split}
 c_{\qb q g g^*}^{(1, \ g_L)}(p_2^\ominus,  p_3^\oplus, p_4^\oplus, q_2) 
 &=
 c_{\qb q g^*}^{(1, \ g_L)}(p_2, p_3, q_1)+r_{g^*}^{(1, \ g)}(t_1;s_{34})
+v^{(1, \, \Nfour)}\left(t_1,\frac{s_{34} \, p_4^+}{(p_3^++p_4^+)},t_2\right)
\\
&+\left(c_{\qb q g g^*}^{(1, \, s_L)}(p_2, p_3, p_4, q_2) - c_{\qb q  g^*}^{(1, \, s_L)}(p_2, p_3, q_1)
\right)
\\
&
-4 \frac{X(1+z-\zb)+z}{2X|z-1|^2}L_0\left(\frac{-t_1}{-t_2}\right)
\\
&-\frac{z}{X}L_{s_{-1}}\left(\frac{-t_1}{-t_2},\frac{-t_1^\prime}{-t_2}\right)
-\frac{z}{2 X (z-1)^2}L_1\left(\frac{-t_1}{-t_2}\right)\,.
\label{eq:c1_gL_2m3p4p}
\end{split}
\end{align}
It is instructive to take the MRK limit of this vertex. Recalling the result in \cref{eq:sL_MRK}, we see that in the MRK limit, the second line simply reduces to the Lipatov vertex with a circulating scalar, \cref{eq:cev1_n0}. The third line becomes $-4$ times the $\NoneX$ Lipatov vertex, \cref{eq:N1X_Lipatov}. The fourth line is power suppressed in the MRK limit. Adding these contributions together, and applying the supersymmetric organisation \cref{eq:Ag_decomp} to the Lipatov vertices, we obtain
\begin{equation}
   c_{\qb q g g^*}^{(1, \ g_L)}(p_2^\ominus,  p_3^\oplus, p_4^\oplus, q_2) \toMRK  c_{\qb q  g^*}^{(1,\, g_L)}(p_2^\ominus, p_3^\oplus, q_1)  
    +r_{g^*}^{(1, \ g)}(t_1;s_{34})+v_{g^* g g^*}^{(1, \, g)}(-q_1, p_4, q_2 )
    \,.
\end{equation}
In contrast, the $g_R$ vertex only contributes to the peripheral-emission vertex in the MRK,
\begin{equation}
   c_{\qb q g g^*}^{(1, \ g_R)}(p_2^\ominus,  p_3^\oplus, p_4^\oplus, q_2) \toMRK  c_{\qb q  g^*}^{(1,\, g_R)}(p_2^\ominus, p_3^\oplus, q_1)  
    \,.
\end{equation}
This is an intuitive result. By requiring the external fermion line to pass to the right of the loop, we essentially restrict the loop correction to only dress the peripheral emission in the MRK.

%\subsubsection{\texorpdfstring{$\{\qb^\ominus, g^\oplus, q^\oplus, g^* \}$}{(qb-,g+,q+,g*)}}
The $\qb^\ominus  g^\oplus q^\oplus g^* $ vertex is
\begin{align}
\begin{split}
 c_{\qb g q  g^*}^{(1, \ g_L)}(p_2^\ominus, p_4^\oplus, p_3^\oplus,  q_2) 
 =&
 c_{\qb g q g^*}^{(1, \ \Nfour)}(p_2, p_4, p_3, q_1)
 +\frac{1}{\epsilon^2}
\left(
        \frac{\mu^2}{-s_{24}}
\right)^{\epsilon}
     +\frac{1}{\epsilon^2}
\left(
        \frac{\mu^2}{-s_{34}}
    \right)^{\epsilon}
-\frac{3}{2}
\left(
        \frac{\mu^2}{-s_{24}}
\right)
\\
&
\hspace{-8 em}
+L_{s_{-1}}\left(\frac{-s_{24}}{-t_2},\frac{-s_{34}}{-t_2}\right)
-\frac{2z}{z-1}L_0\left(\frac{-s_{24}}{-t_2}\right)
+\frac{z^2}{2 (z-1)^2}L_1\left(\frac{-s_{24}}{-t_2}\right)
-3\,,
\label{eq:c_gL_2m4p3p}
\end{split}
\end{align}
with $\Nfour$ vertex as in \cref{eq:243_N4_q}. Recalling the comment beneath \cref{eq:cN1x_2p3m4p}, we emphasise that only the dispersive part of \cref{eq:c_gL_2m4p3p} for the physical scattering 
$q(p_2) \to q(p_3) g(p_4)$
should be taken. The tree-level coefficient of this vertex vanishes in both MRK and $\MRKx$ limits.

%Note that, for example, the dispersive part for the physical scattering of $g(-p_2) \to q(p_3) \qb(p_4)$ differs to this by $\pi^2/2$.
%\subsubsection{\texorpdfstring{$\{g^\oplus,\qb^\ominus,  q^\oplus, g^* \}$}{(g+,qb-,q+,g*)}}
Finally, the $g^\oplus  \qb^\ominus  q^\oplus g^* $ vertex is
\begin{align}
\begin{split}
 &c_{g \qb q g^*}^{(1, \ g_L)}(p_4^\oplus,p_2^\ominus,  p_3^\oplus, q_2) 
 =
 c_{\qb q g^*}^{(1, \ g_L)}(p_2, p_3, q_1)+r_{g^*}^{(1, \ g)}(t_1;s_{24})
+v^{(1, \, \Nfour)}\left(t_1,\frac{-s_{24} \, p_4^+}{(p_3^++p_4^+)},t_2\right)
\\
&+(c_{g \qb q g^*}^{(1, \ s_L)}(p_4, p_2, p_3, q_2) - c_{\qb q  g^*}^{(1, \ s_L)}(p_2, p_3, q_1))
\\
&
+\left(
\frac{X (z+\zb-1)-2 z (X+\zb)+4 X (\zb-1)+|z|^2}{2 X |z-1|^2}
-\frac{z (2 X+z) (X+\zb)}{|z-1|^2 (X+z)^2}
\right)L_0\left(\frac{-t_1}{-t_2}\right)
\\&
+\left(
\frac{X^2 \left((\zb-1)^2-z^2\right)+X z \left(-2 (z+1) \zb+\zb^2+1\right)-z^2 \zb^2}{2 X (z-1)^2 (\zb-1)^2 (X+z)}
\right)L_1\left(\frac{-t_1}{-t_2}\right)
\\
&+\left( -\frac{X z^2}{(X+z)^3}+\frac{2 X z}{(X+z)^2}+\frac{z}{X+z}\right)L_{s_{-1}}\left(\frac{-t_1}{-t_2},\frac{-t_1^\prime}{-t_2}\right)
\\&
-
\left(
\frac{z (X+\zb) \left(-5 X^2+X z (4 \zb-7)-5 X \zb+2 z^2 (\zb-1)-3 z \zb\right)}{2 (z-1)^2 (\zb-1)^2 (X+z)^2}
\right)
L_1\left(\frac{-t_1^\prime}{-t_2}\right)
\\&
+
\frac{X^2 (4 z+\zb-1)+2 X z (z+3 \zb-1)+z^2 (3 \zb-1)}{2 (z-1) (\zb-1) (X+z)^2}\,.
\label{eq:c1_gL_4p2m3p}
\end{split}
\end{align}
Following the same procedure as discussed above \cref{eq:c1_gL_2m3p4p}, we find this vertex has the MRK limit
\begin{equation}
   c_{g \qb q  g^*}^{(1, \ g_L)}(p_4^\oplus,p_2^\ominus,  p_3^\oplus,  q_2) \toMRK  c_{\qb q  g^*}^{(1,\, g_L)}(p_2^\ominus, p_3^\oplus, q_1)  
    +r_{g^*}^{(1, \ g)}(t_1;p_3^+ p_4^-)+v_{g^* g g^*}^{(1, \, g)}(-q_1, p_4, q_2 )
    \,.
\end{equation}
Again, the $g_R$ vertex only contributes to the peripheral-emission vertex in the MRK, that is, 
\begin{equation}
   c_{g \qb q  g^*}^{(1, \ g_R)}(p_4^\oplus,p_2^\ominus,  p_3^\oplus,  q_2)  \toMRK  c_{\qb q  g^*}^{(1,\, g_R)}(p_2^\ominus, p_3^\oplus, q_1)  
    \,.
\end{equation}
%\subsection{$g_R$ contribution}
%\subsubsection{$\{\qb^\ominus, q^\oplus, g^\oplus, g^* \}$}
%\subsubsection{$\{\qb^\ominus, g^\oplus, q^\oplus,  g^* \}$}
%\subsubsection{$\{g^\oplus, \qb^\ominus, q^\oplus, g^* \}$}
We have now obtained all the colour-ordered two-parton emission vertices required in \cref{eq:A1_q_simple}, and we can progress to studying the NMRK limit of the colour-dressed amplitudes.
\section{Colour-dressed one-loop vertex for the peripheral emission of a quark and gluon}
\label{sec:cC_qqbg}
In \cref{sec:colour-5} we analysed the colour structure of the one-loop $\qb q ggg$ amplitudes at leading power in the NMRK limit. In the previous section we analysed the factorisation properties of the primitive amplitudes in the NMRK limit. 
Following the procedure of \cref{sec:cCgg}, we now assemble the NMRK limit of the colour-dressed amplitude. 

As in \cref{sec:cCgg}, it is useful to consider colour-dressed vertices for each of the colour structures in \cref{eq:A1_q_simple}.
For the tree-level colour structures we define
\begin{equation}
    \cC_{TT}(p_2, p_3, p_4, q_2)= \cC^{(0)}_{TT}(p_2, p_3, p_4, q_2)+\cC_{TT}^{(1)}(p_2, p_3, p_4, q_2)+ \mathcal{O}(\gs^6)\,,
\end{equation}
where the tree-level terms  are taken from \cref{eq:cC0qqb},
\begin{align}
\cC^{(0)}_{TT}(p_2, p_3, p_4, q_2)&=\gs^2
  (T^{a_{4}} T^{c_2})_{\imath_3 \bar{\imath}_2} 
  \ C^{(0)}_{\qb q g g^*}(p_2^{\lambda_2}, p_{3}^{\lambda_3}, p_{4}^{\lambda_{4}}, q_2)\\
  \cC^{(0)}_{TT}(p_4, p_2, p_3, q_2)&=\gs^2(T^{c_2} T^{a_{4}} )_{\imath_3 \bar{\imath}_2} 
  \ C^{(0)}_{\qb q g g^*}(p_{4}^{\lambda_{4}}, p_2^{\lambda_2}, p_{3}^{\lambda_{3}}, q_2) 
\end{align}
such that
\begin{align}
\begin{split}
  \mathcal{C}^{(0)}_{\qb q g g^*}(p_2^{\lambda_2}, p_3^{\lambda_3}, p_4^{\lambda_4}, q_2) =
  \cC^{(0)}_{TT}(p_2, p_3, p_4, q_2)+\cC^{(0)}_{TT}(p_4, p_2, p_3, q_2)
  \,.
  \label{eq:cC0qqb_2}
\end{split}
\end{align}
The NMRK limit of the one-loop amplitudes lead us to define the one-loop correction to these colour-dressed vertices,
\begin{align}
\begin{split}
&\mathcal{C}^{(1)}_{TT}(p_2,p_3,p_4,q_2)=c_\Gamma g_s^4
        \big( T^{a_4} T^{d}\big)_\ibi 
		C^{(0)}_{ \qb q g g^*}(p_2, p_3, p_4, q_2)
  \\
  & \ \times
		\left(
			N_c \, c^{(1, \, g_L)}_{ \qb q g g^*}(p_2, p_3, p_4, q_2)
-\frac{1}{N_c} \, c^{(1, \, g_R)}_{ \qb q g g^*}(p_2, p_3, p_4, q_2)
+N_f \, c^{(1, \, f_L )}_{ \qb q g g^*}(p_2, p_3, p_4, q_2)
		\right)\,,
\end{split}
\\
\begin{split}
&\mathcal{C}^{(1)}_{TT}(p_4,p_2,p_3,q_2)=c_\Gamma g_s^4
        \big(  T^{d} T^{a_4}\big)_\ibi 
		C^{(0)}_{ \qb q g g^*}(p_4,p_2,p_3,q_2)
  \\
  & \ \times
		\left(
			N_c \, c^{(1, \, g_L)}_{ \qb q g g^*}(p_4,p_2,p_3,q_2)
-\frac{1}{N_c} \, c^{(1, \, g_R)}_{ \qb q g g^*}(p_4,p_2,p_3,q_2)
+N_f \, c^{(1, \, f_L )}_{ \qb q g g^*}(p_4,p_2,p_3,q_2)
		\right)\,.
\end{split}
\end{align}
Similarly we define a colour-dressed vertex for the new colour structure which first appears at one loop,
\begin{equation}
    \mathcal{C}^{(1)}_{\delta \Delta}(p_2,p_3,p_4,q_2)=\mathcal{C}^{(1)}_{\delta \Delta}(p_2,p_3,p_4,q_2)+\cO(\gs^4)\,.
\end{equation}
As in \cref{sec:N4_qq}, the partial amplitude \cref{eq:A_delta} for $m{\in}\{g_L, g_R\}$ only contributes to the forward vertex and we find
\begin{align}
\begin{split}
\mathcal{C}^{(1)}_{\delta \Delta}(p_2,p_3,p_4,q_2)=
     c_\Gamma g_s^4 \,  \delta_\ibi \Delta_{d a_4} \,
\Big(
		C^{(1, \,  g_L)}_{\delta \Delta}(p_2,p_3,p_4,q_2)
  +
  C^{(1, \,  g_R)}_{\delta \Delta}(p_2,p_3,p_4,q_2)
\Big)\,, \label{eq:cCddD}
\end{split}
\end{align}
with
\begin{align}
\begin{split}
C^{(1, \, g_L)}_{\delta \Delta}(p_2,p_3,p_4,q_2)=
&C^{(0)}_{ \qb q g g^*}(p_2, p_3, p_4, q_2)
		\bigg(
			 c^{L(1, \, g)}_{\qb q g g^*}(p_2, p_3, p_4, q_2)
			-c^{L(1, \, g)}_{\qb g q g^*}(p_2, p_4, p_3, q_2)
   \\
   &  \hspace{12 em} 
			+r^{(1,\, g)}_{g^*}(q_2, p_4^+ p_5^-)
			-r^{(1,\, g)}_{g^*}(q_2, p_3^+ p_5^-)
   \bigg)
\\
+&C^{(0)}_{g \qb q g^*}(p_4, p_2, p_3, q_2)
		\bigg(
			 c^{L(1, \, g)}_{g \qb q  g^*}(p_4, p_2, p_3, q_2)
			-c^{L(1, \, g)}_{\qb g q g^*}(p_2, p_4, p_3, q_2)
   \\
   & \hspace{12 em} 
			+r^{(1,\, g)}_{g^*}(q_2, p_4^+ p_5^-)
			-r^{(1,\, g)}_{g^*}(q_2, p_3^+ p_5^-)
		\bigg)\,, \label{eq:g_dD}
\end{split}
\end{align}
and
\begin{align}
\begin{split}
C^{(1, \, g_R)}_{\delta \Delta}(p_2,p_3,p_4,q_2)=
&C^{(0)}_{ \qb q g g^*}(p_2, p_3, p_4, q_2)
		\bigg(
			 c^{(1, \, g_R)}_{\qb q g g^*}(p_2, p_3, p_4, q_2)
			-c^{(1, \, g_R)}_{\qb g q g^*}(p_2, p_4, p_3, q_2)
   \bigg)
\\
+&C^{(0)}_{g \qb q g^*}(p_4, p_2, p_3, q_2)
		\bigg(
			 c^{(1, \, g_R)}_{g \qb q  g^*}(p_4, p_2, p_3, q_2)
			-c^{(1, \, g_R)}_{\qb g q g^*}(p_2, p_4, p_3, q_2)
		\bigg)\,.
\end{split}
\end{align}
As in \cref{eq:N4_dD}, \cref{eq:g_dD} is free from large logarithms. With these definitions we can write the NMRK limit of the one-loop amplitudes in the compact form
\begin{align}
\begin{split}
&\frac{1}{2s}\disp{\mathcal{A}_5^{(1 )[-]}
\left(\bar{q}_2, q_3, g_4, g_5, g_1\right)}
\\
&\toNMRK 
\bigg(
	\mathcal{C}^{(1)}_{TT}(p_2,p_3,p_4,q_2)+\mathcal{C}^{(1)}_{TT}(p_4,p_2,p_3,q_2)+\mathcal{C}^{(1)}_{\delta \Delta}(p_2,p_3,p_4,q_2)
\bigg)
\mathcal{R}^{(0)}_{g^*}(q_2)\mathcal{C}^{(0)}_{g^* g g }(-q_2, p_5, p_1)
\\&
\quad \ 
+
\bigg(
	\mathcal{C}^{(0)}_{TT}(p_2,p_3,p_4,q_2)+\mathcal{C}^{(0)}_{TT}(p_4,p_2,p_3,q_2)
\bigg)
\mathcal{R}^{(1)}_{g^*}(q_2;p_4^+p_5^-)\mathcal{C}^{(0)}_{g^* g g }(-q_2, p_5, p_1)
\\&
\quad \ 
+
\bigg(
	\mathcal{C}^{(0)}_{TT}(p_2,p_3,p_4,q_2)+\mathcal{C}^{(0)}_{TT}(p_4,p_2,p_3,q_2)
\bigg)
\mathcal{R}^{(0)}_{g^*}(q_2)\mathcal{C}^{(1)}_{g^* g g }(-q_2, p_5, p_1)+\cO(\epsilon)\,.
\label{eq:final_qq}
\end{split}
\end{align}
This is as far as the analysis of one-loop amplitudes can take us. However, \cref{eq:final_qq} is still an interesting result.
Unlike the pure-gluon case, \cref{eq:A1_g_simple_2}, \cref{eq:final_qq} \emph{is} compatible with \cref{eq:Regge_conjecture} at NLL, that is, with an all-orders factorisation of the form
\begin{align}
\begin{split}
\disp{\mathcal{A}_5^{[-]}
\left(\bar{q}_2, q_3, g_4, f_5, f_1\right)}
\toNMRK 
&
2s \,\mathcal{C}_{\bar{q}qgg^*}(p_2,p_3,p_4,q_2)
\mathcal{R}_{g^*}(q_2; p_4^+p_5^-)\mathcal{C}_{g^* f_5 f_1 }(-q_2, p_5, p_1)\,,
\label{eq:cA_NMRK_qqb}
\end{split}
\end{align}
where the one-loop correction to \cref{eq:cC0qqb} is simply given by the sum of the three colour-dressed vertices
\begin{align}
\begin{split}
C^{(1)}_{\bar{q}qgg^*}(p_2,p_3,p_4,q_2)=
	\mathcal{C}^{(1)}_{TT}(p_2,p_3,p_4,q_2)+\mathcal{C}^{(1)}_{TT}(p_4,p_2,p_3,q_2)+\mathcal{C}^{(1)}_{\delta \Delta}(p_2,p_3,p_4,q_2)\,.
 \label{eq:cC1q}
\end{split}
\end{align}
\begin{figure}[hbt]
\centering
    \subfigure[]{\includegraphics[width=0.25\textwidth]{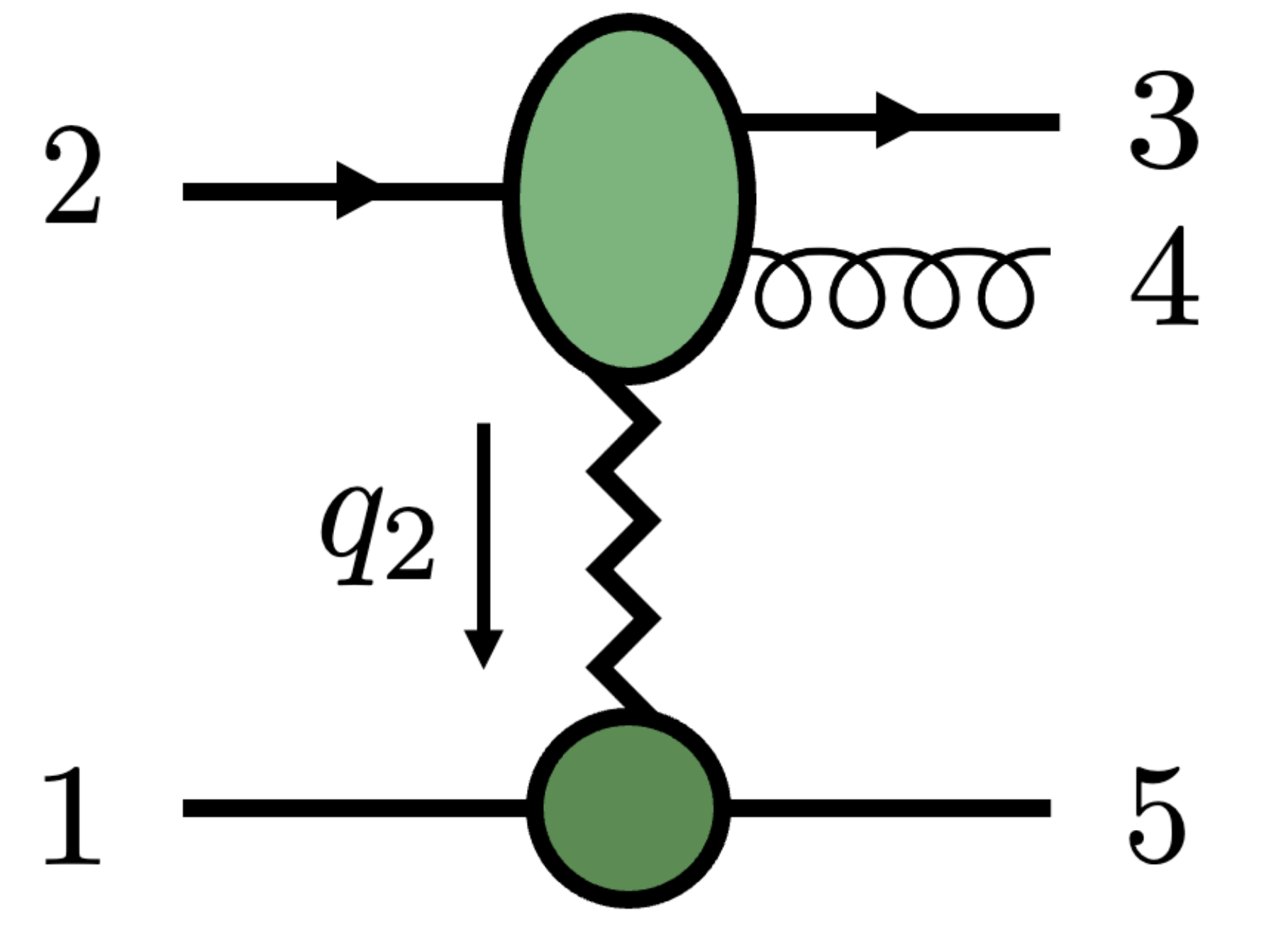}\label{subfig:Fac_NMRK_qg}}
    \qquad \qquad
    \subfigure[]{\includegraphics[width=0.25\textwidth]{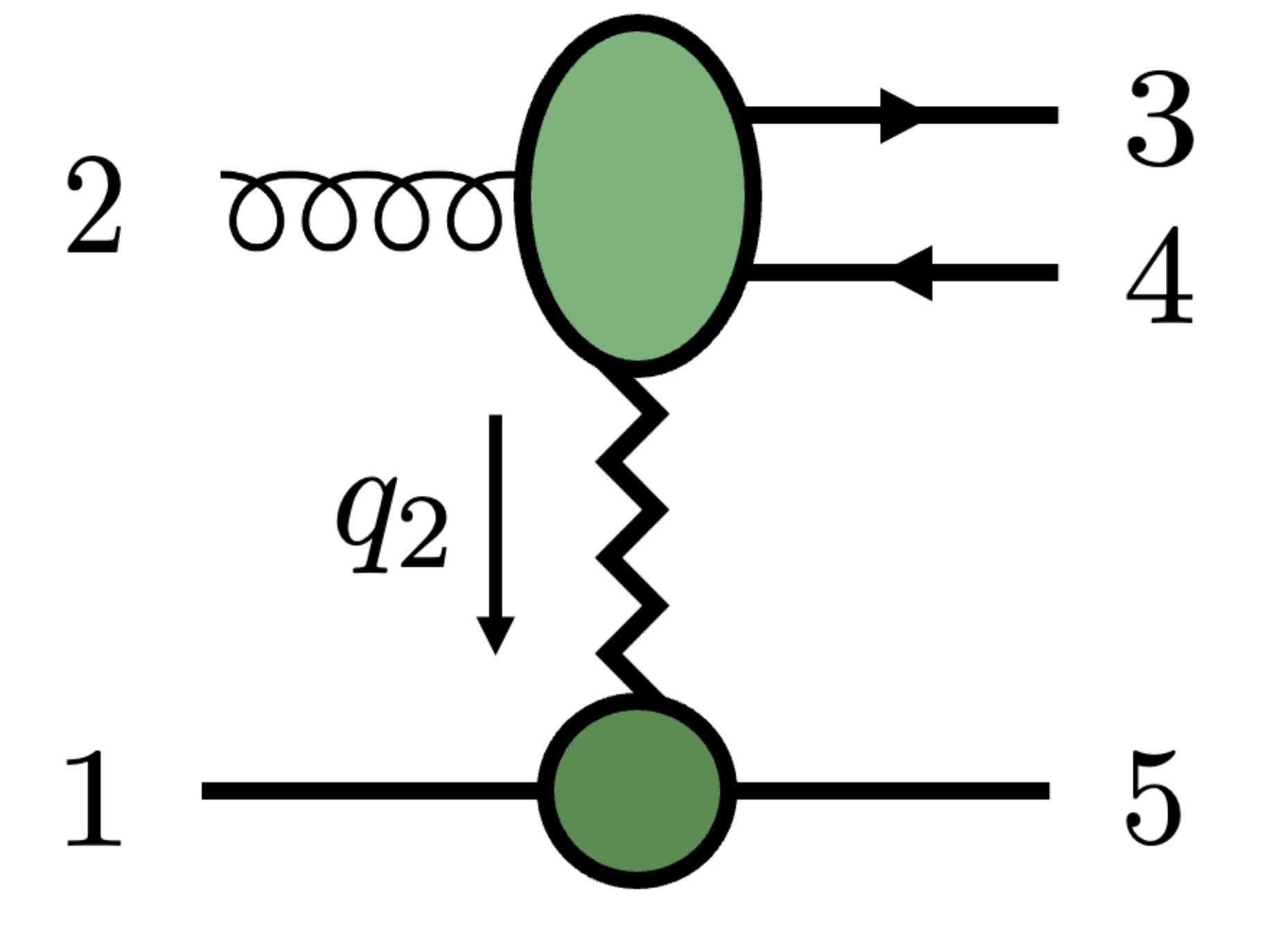}\label{subfig:Fac_NMRK_qq}}
    \caption{Illustration of (a) the all-orders factorisation in \cref{eq:cA_NMRK_qqb}, and (b) the same, but where the gluon is taken to be the physically incoming parton. 
    This is the simplest form of all-orders factorisation of the dispersive part of five-parton amplitudes, and is compatible with the one-loop result \cref{eq:final_qq}.}
    \label{fig:Fac_NMRK_q}
\end{figure}

This simplest possibility is illustrated in \cref{fig:Fac_NMRK_q}. As per our comments in \cref{sec:cCgg}, it is an interesting question whether the new colour structure \cref{eq:cCddD} receives large logarithmic corrections at the next loop order, and, if so, whether they are simply in the form of the one-loop Regge trajectory, as in \cref{eq:cA_NMRK_qqb}. As for the pure-gluon case, this can be tested with the recently computed two-loop five-parton amplitudes~\cite{Agarwal:2023suw,DeLaurentis:2023izi}.

\section{Conclusions}
\label{sec:conc}
In this paper we have analysed the dispersive parts of the one-loop $ggggg$ and $\qb q ggg$ ampltiudes in the NMRK limit \cref{eq:NMRK_5}. The final results are presented in a compact notation in \cref{eq:A1_g_simple_3} and \cref{eq:final_qq} respectively. In both cases, to leading power, the amplitudes comprise solely of the exchange of a colour octet in the $t_2$ channel. \Cref{eq:final_qq} is compatible with an all-orders factorisation of the form \cref{eq:Regge_conjecture}, which defines the colour dressed one-loop $\qb q g g^*$ vertex \cref{eq:cC1q}. The pure-gluon case is not quite as simple \cref{eq:A1_g_simple_3}, but the one-loop amplitude is compatible with an analogous factorisation of each colour structure, one such possibility being given in  \cref{eq:5g_permitted}. We have defined a colour-dressed one-loop $gggg^*$ vertex for each colour structure in \cref{eq:cCFF} and \cref{eq:cCd}. 

As reviewed in the introduction, there has been much progress towards extending the BFKL approach to NNLL accuracy. In order to make predictions for jet cross sections at this accuracy, it is necessary to compute the jet impact factors to NNLO. The vertices extracted in this paper are necessary ingredients for the real-virtual contribution to the NNLO impact factors. The new totally symmetric colour structures appearing in the one-loop $g g g g^*$ vertex do not survive interference with the tree-level amplitude and therefore are not relevant for the impact factors. The same is not true for the $\qb q g g^*$ vertex, where the new one-loop colour structure in \cref{eq:cCddD} has non-zero interference with the tree-level colour structures.

The analysis in this paper was performed to $\cO(\epsilon^0)$. To obtain the NNLL impact factor for the inclusive production of a jet it will be necessary to obtain the one-loop two-parton emission vertices to higher powers of $\epsilon$ in IR singular limits, up to $\cO(\epsilon^2)$ in the soft-collinear limit. The vertices presented here have a more immediate phenomenological application if the two peripheral partons are required to form separate jets.

The present work also sets some groundwork for the extraction of the one-loop two-parton central emission vertices in QCD, which are necessary ingredients for extending the BFKL approach to NNLL accuracy. The colour structure of the $g^* ggg^*$ and $g^* \qb q g^*$ central-emission vertices is expected to be the same as the $gggg^*$ and $\qb q g g^*$ peripheral-emission vertices presented here.

While this analysis of one-loop amplitudes allows for the extraction of one-loop vertices which themselves factorise into known quantities in the further MRK limit, a one-loop analysis cannot provide information on the logarithmic corrections these vertices may receive at higher orders. A particularly interesting question is whether the new colour structures encountered in this one-loop analysis recieve large logarithmic corrections at the next loop order, and if so, whether these large logarithms are simply governed by the one-loop Regge trajectory. The results of this paper therefore provide strong motivation to perform a similar investigation of the NMRK limit of the recently computed two-loop five parton amplitudes. 

\section{Acknowledgements}
The author acknowledges useful discussions with Vittorio del Duca, Einan Gardi and Jenni Smillie and thanks them  for a careful reading of the manuscript. The author thanks Maxim Nefedov and Andreas van Hameren for helpful comments and discussions of this topic at the Low-x 2023 workshop. This work was funded by the ERC Starting Grant 715049 ‘QCDforfuture’. EB is supported by the Royal Society through Grant URF\textbackslash R1\textbackslash201500. For the purpose of open access, the author has applied a Creative Commons Attribution (CC BY) licence to any Author Accepted Manuscript version arising from this submission.
\appendix

\section{Kinematics}
\label{sec:kin}
\subsection{General kinematics}
\label{sec:kin_gen}

In this paper we consider the scattering processes of $2 \to 2$ and $2 \to 3$ massless partons. 
%We always take the incoming partons to have momenta $-p_1$ and $-p_2$. 
The amplitudes in refs.~\cite{Bern:1993mq,Bern:1994fz} are given in the unphysical region where all momenta are taken as out-going. We analytically continue to the
physical region where $p_1^0<0$ and $p_2^0<0$. In summary, the two scattering regions we consider are the $n{=}4$ and $n{=}5$ case of
\begin{equation}
    \label{eq:outgoing}
p_i^2=0 \ \ \forall \ i \in \{1,\cdots, n\},
\qquad
\sum_{i=1}^n p_i=0\,, 
\qquad 
p_1^{0}, p_2^{0} <0\,, 
\qquad 
p_i^{0} >0  \ \ \forall i \in \{3,\cdots, n\}, 
\end{equation}
We parameterise the momenta with light-cone coordinates 
$p^{\pm}= p^0\pm p^z $, and
complex transverse coordinates $p_{\perp} = p^x + i p^y$, with scalar
product,
\begin{equation}
2 \, (p\cdot q) = p^+q^- + p^-q^+ - p_{\perp} q^*_{\perp} - p^*_{\perp} q_{\perp}\,.
\label{eq:scalprod}
\end{equation} 
We take momenta $p_1$ and $p_2$ to be along the $-$ and $+$ lightcone directions respectively. Explicitly, the four-momenta are
\begin{align}
\begin{split}
p_1 &= \left(\tfrac{p_1^-}{2}, 0, 0,\tfrac{-p_1^-}{2} \right) 
     \equiv  \left(0, p_1^-; 0\right)\,,  \\
p_2 &= \left(\tfrac{p_2^+}{2}, 0, 0,\tfrac{p_2^+}{2} \right) 
     \ \ \equiv  \left(p_2^+ , 0;  0 \right)\,  ,\label{eq:in}\\
p_i &= \left( \tfrac{p_i^+ + p_i^- }{2}, 
                {\rm Re}[p_{i\perp}],
                {\rm Im}[p_{i\perp}], 
                \tfrac{p_i^+ - p_i^- }{2} \right)
                %\nonumber\\&\equiv& 
                \equiv
   \left( p_i^+,  p_i^-; 
p_{i\perp} \right) \quad \forall i \in \{3,\cdots, n\},
\end{split}
\end{align}
in terms of cartesian and lightcone coordinates respectively. We also make use of rapidity, which in terms of lightcone coordinates is
\begin{equation}
    y_i = \frac{1}{2}\log\left( \frac{p_i^+}{p_i^-} \right).
\end{equation}
In terms of the lightcone components the Mandelstam invariants may be written
\begin{align}
s &= 2 \, p_1\cdot p_2 = p_2^+ p_1^-\nonumber\\ 
s_{2i} &= 2 \, p_2\cdot p_i = \sum_{i=3}^5 p_2^+ p_i^- \label{eq:inv}\\ 
s_{1i} &= 2 \, p_1\cdot p_i = \sum_{i=3}^5 p_i^+ p_1^- \nonumber\\
s_{ij} &= 2 \, p_i\cdot p_j = p_i^+p_j^- + p_i^-p_j^+ - p_{i\perp} p^\ast_{j\perp} - p^\ast_{i\perp} p_{j\perp}\,, \nonumber
\end{align}
with $3\le i, j \le 6$.

For the momenta in \cref{eq:in}, we find the following right-handed spinor products,
$\langle k \, p \rangle = \overline{u}^{\ominus}(k) u^{\oplus}(p)$,
in the notation of ref.~\cite{DelDuca:1999iql},
\begin{align}
\begin{split}
\langle p_1 \, p_2\rangle 
&= \sqrt{p_2^+p_1^-}\, ,\\
\langle p_2 \, p_i\rangle &= - i \sqrt{\frac{-p_2^+}{p_i^+}}\, p_{i\perp}\, ,\label{spro}\\ 
\langle p_i \, p_1\rangle &=
i \sqrt{-p_1^- p_i^+}\, ,\\ 
\langle p_i \, p_j\rangle &= p_{i\perp}\sqrt{\frac{p_j^+}{p_i^+}} - p_{j\perp}
\sqrt{\frac{p_i^+}{p_j^+}}\, , \\ 
\end{split}
\end{align}
with $3\le i, j \le 6$, where we have used the mass-shell condition,
\begin{equation}
|p_{i\perp}|^2 = p_i^+ p_i^-\,.
\label{eq:masssh}
\end{equation}
Left-handed spinor products, $[k \, p] = \overline{u}^{\oplus}(k) u^{\ominus}(p)$, are given by complex conjugation,
\begin{equation}
[k \, p] = {\rm sign}(k^0 p^0) \langle p \, k\rangle^\ast\,.
\end{equation}
Spinor products are antisymmetric,
\begin{equation}
\langle k\, p\rangle = - \langle p \, k\rangle\,,\qquad [k \, p] = -[p \, k]\,.
\end{equation}
We also use the currents, $[p| \gamma^{\mu} |k\rangle$ and $\langle p|\gamma^{\mu}|k]$,
which are related by charge conjugation,
\begin{equation}
[p| \gamma^{\mu} |k\rangle=\langle k|\gamma^{\mu}|p]\,,
\end{equation}
and complex conjugation,
\begin{equation}
[p| \gamma^{\mu} |k\rangle^{*}  = {\rm sign}(k^0 p^0) [k| \gamma^{\mu} |p\rangle \,.
\end{equation}
Through the Fierz rearrangement,
\begin{equation}
 \langle k|\gamma^{\mu}|p]\langle v|\gamma^{\mu}|q] = 2 \langle k \, v\rangle [q \, p]\,,
\end{equation}
and the Gordon identity,
\begin{equation}
 [p| \gamma^{\mu} |p\rangle= \langle p|\gamma^{\mu}|p]= 2 p^{\mu}\,,
\end{equation}
we obtain that
\begin{eqnarray}
&& \langle k|\slashed{q}|p] = \langle k \, q\rangle [q \, p]\,, \nonumber\\ 
&&  [p| \slashed{q} |k\rangle = [p \, q] \langle q \, k\rangle \,.
\end{eqnarray}

The phase conventions of the amplitudes studied in this paper follow from the choice of polarisation vectors,
\begin{equation}
    \epsilon_\mu^\oplus(p,r)=-\frac{[p|\gamma_\mu|r\rangle }{\sqrt{2}\langle p \, r \rangle}\,,
    \qquad \qquad
    \epsilon_\mu^\ominus(p,r)=\frac{\langle p|\gamma_\mu|r] }{\sqrt{2}[ p \, r ]}\,,\label{eq:polarisation}
\end{equation}
where $r$ is an arbitrary lightlike reference momentum.

\subsection{Kinematic regions}
\label{sec:kin_Regge}
\begin{figure}[hbt]
\centering
     \includegraphics[scale=0.2]{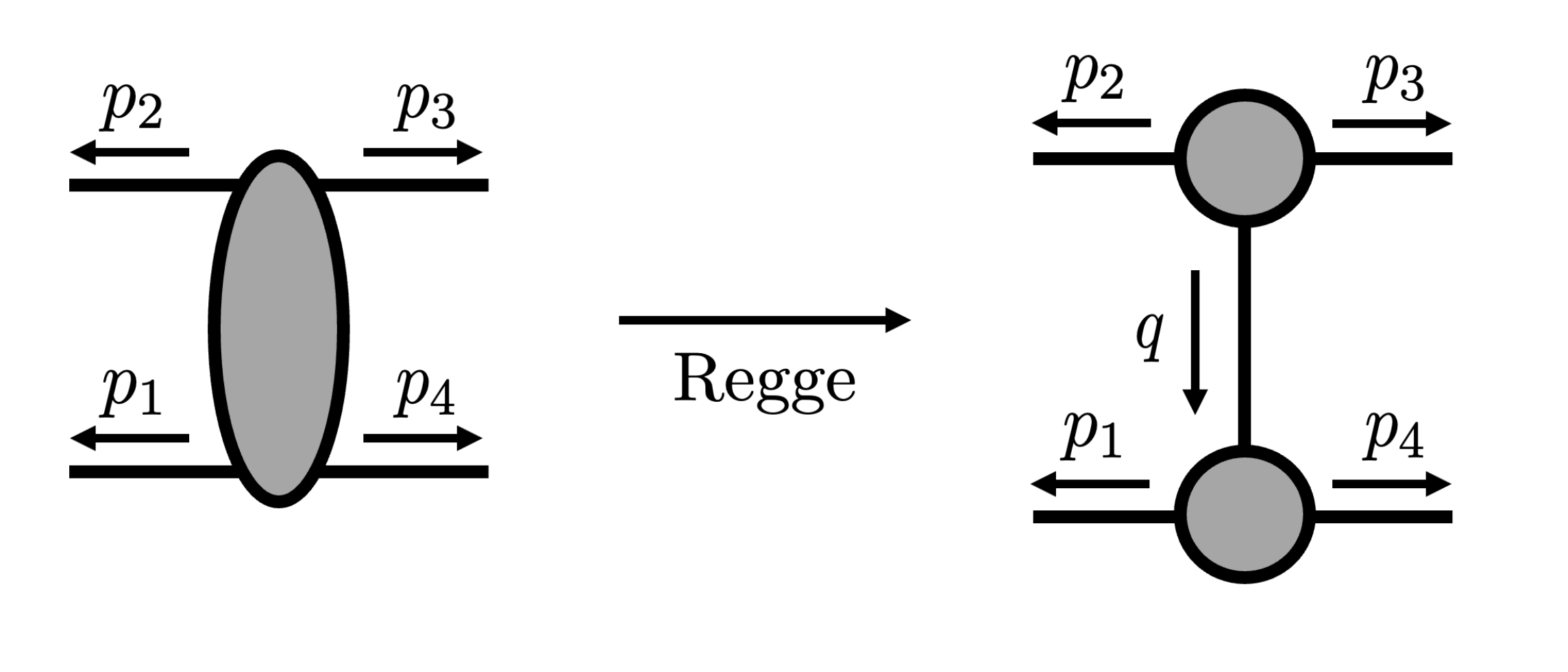}
    \caption{Illustration of the Regge-kinematic region (right), obtained from general kinematics (left) by taking the limit \cref{eq:Regge}. The grey blobs group final-state momenta where no strict rapidity ordering is assumed.}
    \label{fig:Regge}
\end{figure}

In this appendix we introduce the kinematic regions that are relevant for this paper, namely the Regge limit of $2\to 2$ scattering and the MRK and NMRK limits of $2\to 3$ scattering. The high-energy limit, or Regge limit of $2\to 2$ scattering is given by the limit $s\gg -t$, where $s = s_{12}$ and $t = s_{23}$.
%\begin{equation}
%    s = s_{12}\,, \qquad \qquad t = s_{23}\,.
%\end{equation}
In terms of the rapidity and transverse momenta of the final-state partons, this limit can be expressed
\begin{equation}
    y_3 \gg y_4\,, \qquad \qquad |p_{3\perp}| \sim |p_{4\perp}|\,, \label{eq:Regge}
\end{equation}
where the transverse momenta are finite. The momentum flowing in the $t$-channel is denoted $q=-p_2-p_3=p_4+p_1$, as depicted in \cref{fig:Regge}.

The MRK limit is the straightforward generalisation of this limit to the multi-parton case. In this paper we only consider the MRK limit of $2\to 3$ scattering,
\begin{equation}
    y_3 \gg y_4 \gg y_5\,, \qquad \qquad |p_{3\perp}| \sim |p_{4\perp}| \sim |p_{5\perp}|\,. \label{eq:MRK_5}
\end{equation}
The momenta flowing in the $t$-channels are $q_1=-p_2-p_3$ and $q_2=p_5+p_1$, that is, $t_1=q_1^2$ and $t_2=q_2^2$.

\begin{figure}[hbt]
\centering
     \includegraphics[scale=0.4]{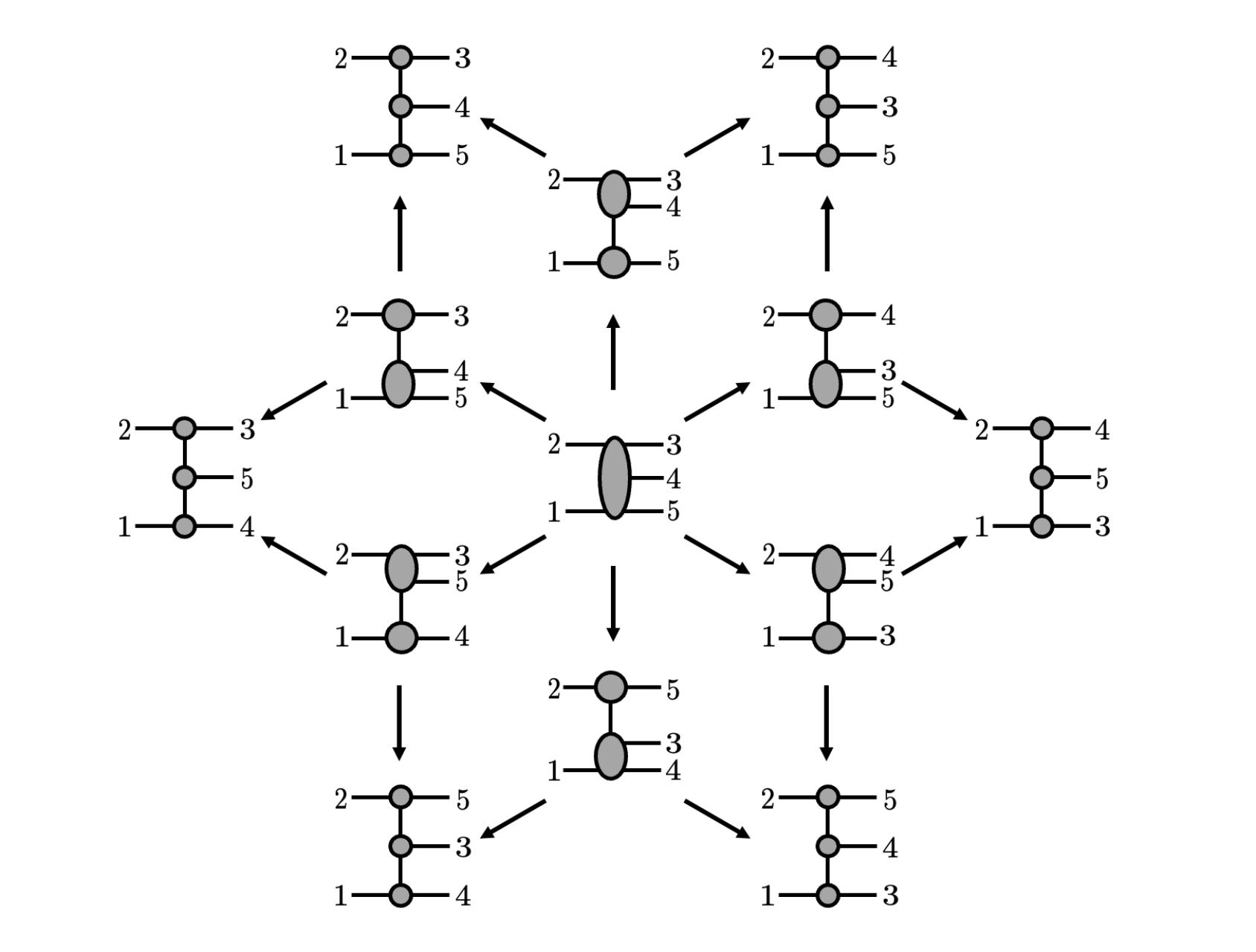}
    \caption{Illustration of the 3 forward NMRK, 3 backward NMRK and 3! MRK limits which can be obtained from a $2\to 3$ scattering process in general kinematics. }
    \label{fig:NMRK_gen}
\end{figure}

\begin{figure}[hbt]
\centering
     \includegraphics[scale=0.3]{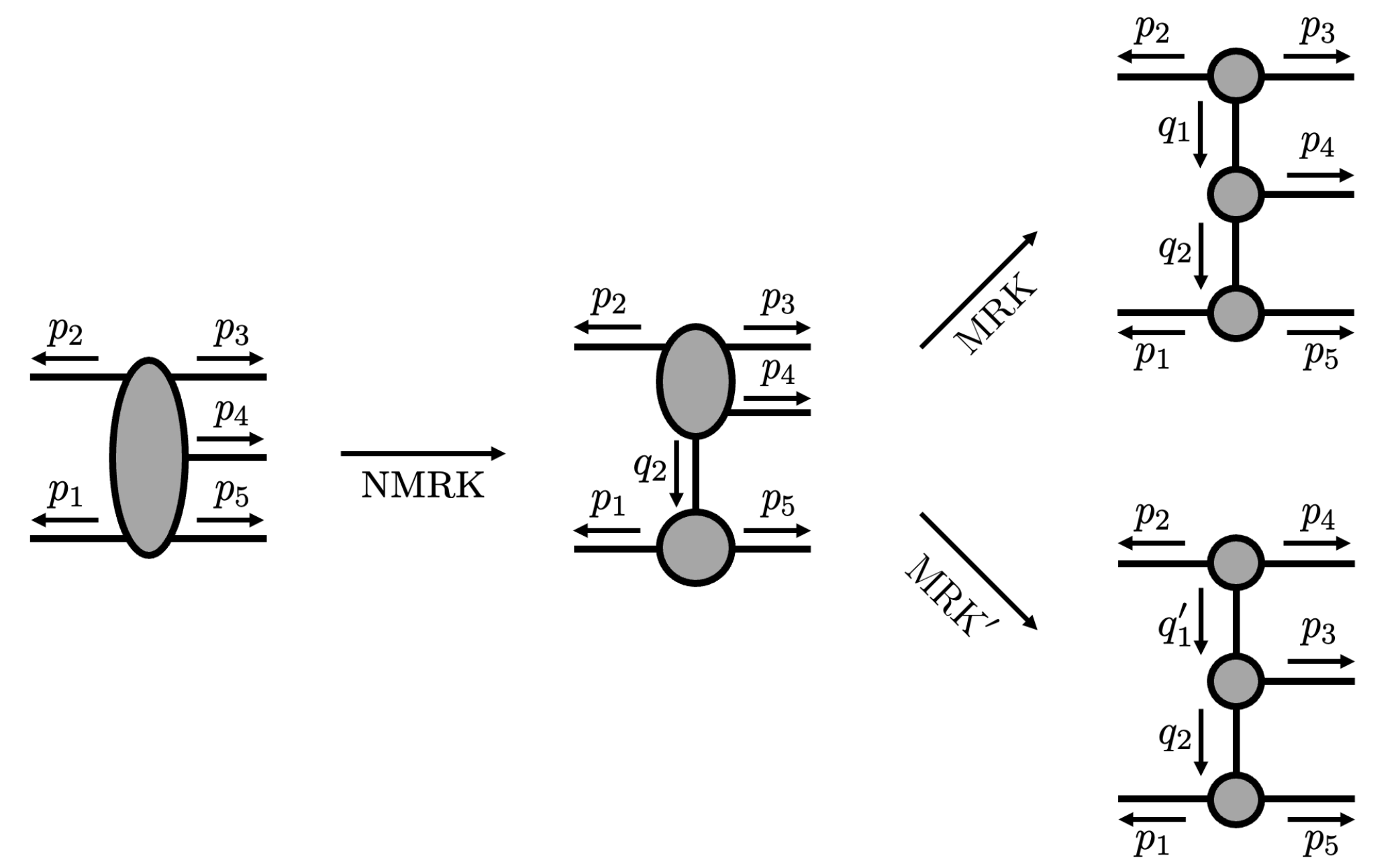}
    \caption{Illustration of the heirarchy of limits relevant for this paper. Starting from general kinematics (left) the forward NMRK region (centre) is reached by taking the limit \cref{eq:NMRK_5}. The further MRK and $\MRK'$ regions can be reached by taking the limits \cref{eq:MRK_5} or \cref{eq:MRKx} respectively, which are the rapidity boundaries of the NMRK region.}
    \label{fig:NMRK}
\end{figure}

As stated in the introduction, the NMRK limit (also known as a quasi-multi-Regge kinematic or QMRK limit) is given by relaxing one of these strict rapidity orderings. 
The focus of this paper is on $2\to 3$ scattering, which has a rich structure of nested NMRK and MRK limits, illustrated in \cref{fig:NMRK_gen}. We restrict to the canonical ordering of particles, and, without loss of generality, only consider the forward NMRK limit, that is, 
\begin{equation}
    y_3 \sim y_4 \gg y_5\,, \qquad \qquad |p_{3\perp}| \sim |p_{4\perp}| \sim |p_{5\perp}|\,. \label{eq:NMRK_5}
\end{equation}
We note that two different MRK limits can be reached from \cref{eq:NMRK_5}. In addition to the canonical MRK limit \cref{eq:MRK_5}, the limit 
\begin{equation}
    y_4 \gg y_3 \gg y_5\,, \qquad \qquad |p_{3\perp}| \sim |p_{4\perp}| \sim |p_{5\perp}|\,, \label{eq:MRKx}
\end{equation}
is also a boundary of the forward NMRK limit. We will refer to the region \cref{eq:MRKx} as the $\MRKx$ limit, and likewise denote the relevant $t$-channel invariant by a prime, that is, $t_1^\prime=(q_1^\prime)^2$ with $q_1^\prime=-p_2-p_4$.
%It is fun to notice that, more generally, the N$^m$MRK limits of form vertices of a hypercube.
The relationship between the MRK, $\MRK'$ and forward NMRK limits is depicted in \cref{fig:NMRK}. 

In practice, to obtain the (N)MRK limit of rational functions, we first implement the explicit representation of spinor products given in \cref{sec:kin_gen}, and then map to the minimal set of variables described in \cref{sec:min}.

\FloatBarrier

\subsection{A minimal set of variables for five-parton scattering}
\label{sec:min}

In this section, following ref.~\cite{Byrne:2022wzk}, we introduce a minimal set of dimensionless variables for $2 \to 3$ scattering, which encode transverse momentum conservation and which make it straightforward to take Regge limits of rational functions. The longitudinal degrees of freedom are given by the ratios of lightcone coordinates,
\begin{subequations}
    \label{eq:min}
\begin{equation}
   X=\frac{p_3^+}{p_4^+}\,, \qquad \qquad Y=\frac{p_4^+}{p_5^+}\,, \qquad \qquad X,Y \in \mathbb{R}\,,
\end{equation}
and the remaining transverse degree of freedom is given by \footnote{After completion of this work, the author became aware of a very similar treatment of transverse momenta in ref.~\cite{Schwartz:2018obd}, including a Lorentz invariant representation of the transverse variables given in appendix D of that reference.}
\begin{equation}
z=\frac{q_{1\perp}}{p_{4 \perp}}=-\frac{p_{3\perp}}{p_{4 \perp}}\,, 
\qquad \qquad z \in \mathbb{C}
\,. 
\label{eq:min_z}
\end{equation}
\end{subequations}
In practise, we can move to the minimal parameterisation by eliminating all but one transverse coordinate, e.g.
\begin{equation}
    p_{4 \perp} \to -\frac{1}{1-z} \, q_{2 \perp}, \qquad q_{1 \perp} \to -\frac{z}{1-z} \,q_{2 \perp}\,. \label{eq:minz_q2}
\end{equation}
We see that the variable $z$ automatically, imposes transverse momentum conservation, i.e. the relation
\begin{equation}
    \frac{-q_{1 \perp}+ p_{4 \perp}+q_{2 \perp}}{q_{2 \perp}}=0\,
\end{equation}
is manifest.
For Lorentz invariant quantities with non-zero mass dimension, $q_{2\perp}$ will provide the dimensionfull scale. 

The parameterisation \cref{eq:min} allows for simple permutation of $p_3 \leftrightarrow p_4$,
\begin{equation}
    X \underset{P_{12435}}{\to } \frac{1}{X}\,, \qquad Y \underset{P_{12435}}{\to } X Y \,, \qquad z \underset{P_{12435}}{\to } \frac{1}{z}\,.
    \label{eq:P12435}
\end{equation}
Our choice of frame with $p_{1 \perp}=p_{2 \perp}=0$ might suggest it not possible to perform the analogous permutation of $p_2 \leftrightarrow p_3$. However, we can write the minimal variables in a Lorentz invariant form,
\begin{align}
   X =  \frac{s_{13}}{s_{14}} \,, \qquad  
   Y =  \frac{s_{14}}{s_{15}} \,, \qquad  
    z= -\frac{\langle 2| 3|1]}{\langle 2| 4|1]}\,.
    %\frac{s_{13}s_{23}}
    %{  [1 \ 2] \langle 2 \ 4\rangle [ 4 \ 3]  \langle 3 \ 1\rangle - s_{13}s_{42}}\,.
\end{align}
These expressions permit us to obtain maps for arbitrary permutations of momenta, for example, $p_2 \leftrightarrow p_3$,
\begin{equation}
    X \underset{P_{13245}}{\to } -\left(\frac{1+Y + XY}{Y}\right)\,, 
    \qquad 
    Y \underset{P_{13245}}{\to } Y
    \,, 
    \qquad 
    z \underset{P_{13245}}{\to } 
    \frac{z (1+Y+ X Y)}{Y (z+X)}
    \,.
    \label{eq:P13245}
\end{equation}
or  $p_2 \leftrightarrow p_4$,
\begin{equation}
    X \underset{P_{14325}}{\to } -\left(\frac{X Y}{1+Y + XY}\right)\,, 
    \qquad 
    Y \underset{P_{14325}}{\to } -(1+ Y+ X Y)  \,, 
    \qquad 
    z \underset{P_{14325}}{\to } 
    \frac{Y (z+X)}{(1+ Y+ X Y)}
    \,.
    \label{eq:P14325}
\end{equation}
It can be checked that such maps satisfy the required group structure, e.g.
\begin{align}
    P_{13245}^2 =  P_{12435}^2=  P_{14325}^2 &= \mathrm{id} \,, \\
    P_{12435} \circ  P_{13245}  \circ  P_{14325} \circ  P_{13245} &=\mathrm{id} \,.
\end{align}
A key advantage of the minimal set \cref{eq:min} is that they make it simple to take generalised Regge limits. For example, the MRK limit is obtained by taking $X,Y \to \infty$ 
at the same rate, keeping $z$ finite, while the forward NMRK limit is given by 
$Y \to \infty$ with $X$ and $z$ finite.

In this paper, we are particularly interested in the leading power of amplitudes in the forward NMRK limit, in which we can obtain factorised vertices that depend only on $X$ and $z$. For dimensionless rational functions, the maps \cref{eq:P13245,eq:P14325,eq:P12435} commute with this limit, that is, within the limit, the permutation of momenta $p_2$, $p_3$ and $p_4$ is given by the $X\to \infty$ limit of these above maps, which are respectively
\begin{align}
    \{X, z \} &\underset{P_{243}}{\to } \bigg\{ \frac{1}{X}, \frac{1}{z} \bigg\} \,, \label{eq:P243}\\
    \{X, z \} &\underset{P_{324}}{\to } \bigg\{
    -(1+X), \frac{z(1+X)}{z+X}  \bigg\}
    \label{eq:P324}
    \,, \\
    \{X, z \} &\underset{P_{432}}{\to } 
    \bigg\{
    -\frac{X}{1+X}, \frac{z+X}{1+X}\bigg\}
    \label{eq:P432}\,.
\end{align}

\section{Special Functions}
\label{sec:special}
In this section we list the special functions used in this paper. 
%The only divergent special function we us is the massless bubble integral,
%\begin{equation}
%    K_0(x)=\frac{1}{\epsilon}+2 - \log(-x)+\mathcal{O}(\epsilon)\,.
%    \label{eq:K0}
%\end{equation}
We use the following weight-1 functions, which are non-singular as $x\to1$ \cite{Bern:1993mq}
\begin{align}
 L_0(x)&=\frac{\log (x)}{1-x}\,, \\
 L_1(x)&=\frac{\log (x)+1-x}{(1-x)^2} \,,\\
 L_2(x)&=\frac{1}{(1-x)^3}\left[\log (x)-\frac{x}{2}+\frac{1}{2 x}\right] \,,
\end{align}
and we use the following weight-2 functions, which are non-singular as $x+y\to1$
\begin{align}
 L s_{-1}(x, y)&= \Li(1-x)+\Li(1-y)+\log (x) \log (y)-\frac{\pi^2}{6}\,,
 \\
L s_1(x, y)&=  \frac{1}{(1-x-y)^2}\bigg[
L s_{-1}(x, y)
+(1-x-y)\left[L_0(x)+L_0(y)\right]\bigg]\,.
\end{align}
For $x>0$, the argument of the $L_i$ functions can be inverted using
\begin{align}
    L_0\left(\frac{1}{x}\right)&=x \, L_0(x)\,,\\
    L_1\left(\frac{1}{x}\right)&=-x^2 \, L_1(x)-x\,,\\
    L_2\left(\frac{1}{x}\right)&=x^3 \, L_2(x)\,.
\end{align}
It is often necessary to write the $L_i$ functions in terms of functions with a higher or lower index so that the resulting coefficients are free from unphysical singularities. To this end we list explicitly the identities
\begin{align}
    L_0\left(x\right)&= (1-x) \, L_1(x)-1
     =(1-x)^2 \, L_2\left(x\right)
    -\left(\frac{1+x}{2x} \right)
    \,,\\
    L_1\left(x\right)&=(1-x) \, L_2(x)-\frac{1}{2x}\,.
\end{align}
We analytically continue these functions by giving all Mandelstam invariants a small positive imaginary part, $s \to s +i0$. For logarithms of ratios of invariants, this prescription yields
\begin{equation}
    \log\left(\frac{-s_1}{-s_2} \right)=\log\left(\left|\frac{s_1}{s_2}\right|\right)-i\pi \left( \Theta(s_1)-\Theta(s_2) \right)
\end{equation}
The analytic continuation of the dilogarithm is given in ref.~\cite{vanHameren:2005ed}.
%and
%\begin{align}
%    L_1\left(x\right)&= \frac{1}{1-x}\,  \big(L_0(x)+1 \big)
%    \,,\\
%    L_2\left(x\right)&=\frac{1}{1-x} \, \left(L_1(x)+\frac{1}{2x} \right)
%    =
%    \frac{1}{(1-x)^2} \, \left(L_0(x)+\frac{1}{2x} \right)
%    \,.
%\end{align}

%\begin{align}
%L s_1(x, y)&=  \frac{1}{(1-x-y)^2}\bigg[
%\Li(1-x)+\Li(1-y)+\log (x) \log (y)-\frac{\pi^2}{6}
%\\
%& 
%+(1-x-y)\left[L_0(x)+L_0(y)\right]\bigg]
%\end{align}

\section{Review of tree-level two-parton peripheral-emission vertices}
\label{sec:C0234}
In this appendix we provide the kinematic coefficients that comprise the two-parton peripheral-emission vertices \cref{eq:cC0gg} and \cref{eq:cC0qqb} and we review some of their properties.

The pure-gluon vertices~\cite{Fadin:1989kf,DelDuca:1995ki}
\begin{align}
    C^{(0)}_{g g g g^*}(p_2^\oplus, p_3^\oplus, p_4^\oplus,q_2)&=0\,,\\
    C^{(0)}_{g g g g^*}(p_2^\ominus, p_3^\oplus, p_4^\oplus,q_2)&=
    \frac{1}{q_{2 \perp}}\frac{(z-1)^2X}{z(z+X)}\,,\\
    C^{(0)}_{g g g g^*}(p_2^\oplus, p_3^\ominus, p_4^\oplus,q_2)&=
    \frac{1}{q_{2 \perp}}\frac{(z-1)^2X^3}{z(z+X)(1+X)^2}\,.
  %  C^{(0)}_{g g g g^*}(p_2^\oplus, p_3^\oplus, p_4^\ominus,q_2)&=
% \frac{1}{q_{2 \perp}}\frac{(z-1)^2X}{z(z+X)(1+X)^2}
\end{align}
constitute a minimal set from which all others can be obtained through parity conjugation or the crossing of momenta as in \cref{sec:min}.
The peripheral-emission vertices inherit identities from the parent partial amplitudes in general kinematics, for example, in the NMRK limit, the reflection identity for amplitudes
\begin{equation}
    A^{(0)}_5(2^{\lambda_2},3^{\lambda_3},4^{\lambda_4},5^{\lambda_5},1^{\lambda_1})=(-1)^5
    A^{(0)}_5(4^{\lambda_4},3^{\lambda_3},2^{\lambda_2},1^{\lambda_1},5^{\lambda_5})\,,
\end{equation}
together with the lower-point identity
\begin{equation}
   C^{(0)}_{g g g^*}(-q_2, p_5^{\lambda_5}, p_1^{\lambda_1})=-C^{(0)}_{g g g^*}(-q_2, p_1^{\lambda_1}, p_5^{\lambda_5})\,,
\end{equation}
gives rise to an analogous reflection identity for the $gggg^*$ vertices,
\begin{align}
    C^{(0)}_{g g g g^*}(p_2^{\lambda_2}, p_3^{\lambda_3}, p_4^{\lambda_4},q_2)=
    C^{(0)}_{g g g g^*}(p_4^{\lambda_4}, p_3^{\lambda_3}, p_2^{\lambda_2},q_2)\,.
\end{align}
Similarly, the NMRK limit of the $U(1)$ decoupling identity
\begin{equation}
    A^{(0)}_5( 2, 3, 4, 5, 1)
    +
    A^{(0)}_5( 2, 4, 3, 5, 1)
    +
    A^{(0)}_5( 4, 2, 3, 5, 1)
    +
    A^{(0)}_5( 2, 3, 5, 4, 1)
    =0\,,
    \, \label{eq:U1}
\end{equation}
leads to the vertex identity
\begin{align}
    C^{(0)}_{g g g g^*}(p_2^{\lambda_2}, p_3^{\lambda_3}, p_4^{\lambda_4},q_2)+
    C^{(0)}_{g g g g^*}(p_3^{\lambda_3}, p_2^{\lambda_2}, p_4^{\lambda_4} ,q_2)+
     C^{(0)}_{g g g g^*}(p_3^{\lambda_3}, p_4^{\lambda_4}, p_2^{\lambda_2},q_2)=0\,,
     \label{eq:U1_v}
\end{align}
due to the power supression of the final term in \cref{eq:U1}.
Supersymmetric Ward identities \cite{Grisaru:1976vm} are also inherited by the vertices, allowing us to straightforwardly obtain peripheral-emission vertices with an external quark-antiquark pair~\cite{DelDuca:1996km}. 
%$0 \to \qb q g g^*$. 
For example, consider the supersymmetric Ward identity
\begin{equation}
    A_5^{(0)}(2_\qb^\ominus,3_q^\oplus,4^\oplus,5^\oplus,1^\ominus)=
    \frac{\langle 1 \ 3 \rangle}{\langle 1 \ 2 \rangle}  A_5^{(0)}(2^\ominus,3^\oplus,4^\oplus,5^\oplus,1^\ominus)\,.
        \label{eq:susy1}
\end{equation}
In the forward NMRK limit this yields the vertex relation
\begin{equation}
  C^{(0)}_{\qb q g g^*}(p_2^{\ominus}, p_3^{\oplus}, p_4^{\oplus},q_2)= -i\sqrt{\frac{X}{(1+X)}}  C^{(0)}_{g g g g^*}(p_2^{\ominus}, p_3^{\oplus}, p_4^{\oplus},q_2)\,.
   \label{eq:susy1_C}
\end{equation}
Likewise, the supersymmetric Ward identity
\begin{equation}
    A_5^{(0)}(2^\oplus, 3_\qb^\ominus,4_q^\oplus,5^\oplus,1^\ominus)=
    \frac{\langle 1 \ 4 \rangle}{\langle 1 \ 3 \rangle}  A_5^{(0)}(2^\oplus,3^\ominus,4^\oplus,5^\oplus,1^\ominus)\,,
    \label{eq:susy2}
\end{equation}
yields the vertex relation
\begin{equation}
  C^{(0)}_{g \qb q g^*}(p_2^{\oplus}, p_3^{\ominus}, p_4^{\oplus},q_2)= \sqrt{\frac{1}{X}}  C^{(0)}_{g g g g^*}(p_2^{\oplus}, p_3^{\ominus}, p_4^{\oplus},q_2)\,.
  \label{eq:susy2_C}
\end{equation}
We note that the simple ratios of spinor products in \cref{eq:susy1,eq:susy2} encode both the difference in little-group weight of the amplitudes and the correct power-dependence predicted by Regge theory (that is, in the MRK limit the vertex \cref{eq:susy2_C} is suppressed by a factor $X^\frac{1}{2}$ compared to the pure-gluon vertex because it is governed by an effective $t$-channel exchange of a spin-$\frac{1}{2}$ particle).

\section{Re-deriving the one-loop vertex for the peripheral emission of a quark or gluon}
\label{sec:rederiving}
In this appendix, we re-derive the one-loop vertices for the peripheral emission of a quark, $\cC_{\qb q g^*}^{(1)}$ (\cref{eq:IF_q}), or gluon, $\cC_{ggg^*}^{(1)}$ (\cref{eq:IF_g,eq:IF_pp}), using the same techniques and notation that we use in the main text for the analysis of five-parton one-loop amplitudes. The vertices $\cC_{ggg^*}^{(1)}$ and $\cC_{\qb q g^*}^{(1)}$ have previously been extracted from one-loop four-parton amplitudes~\cite{DelDuca:1998kx}. However, this paper preceded the development of the DDM bases for one-loop amplitudes (\cref{eq:DDM_ng,eq:DDM_qqbn}), so we re-derive the vertices here utilising this more recent development. While ref.~\cite{DelDuca:1998kx} did show that a supersymmetric decomposition of the pure-gluon amplitudes could be used to derive \cref{eq:IF_g}, that paper only explicitly gave results for factorised building blocks in QCD and did not explicitly state the intermediate supersymmetric one-loop building blocks. For this paper, explicit expressions for these supersymmetric building blocks are essential. In the main text, we make heavy use of colour-stripped supersymmetric two-parton emission vertices. To define these objects, we will need to subtract from one-loop amplitudes the lower point building blocks in order to define the two-parton peripheral-emission vertices, so we collect the necessary expressions here. We also demonstrate their compatibility with \cref{eq:IF_g,eq:IF_q,eq:IF_pp}. Furthermore, to derive the $\cC_{\qb q g^*}^{(1)}$ vertex, ref.~\cite{DelDuca:1998kx} started from the four-quark amplitudes presented in ref.~\cite{Kunszt:1993sd}. In contrast, we start from the one-loop supersymmetric $q \qb g g$ amplitudes presented in ref.~\cite{Bern:1994fz}, in close analogy to the analysis of one-loop $q \qb g g g$ amplitudes in the main text of this paper. The derivation of $\cC_{\qb q g^*}^{(1)}$ in this appendix therefore lays a foundation for the interpretation of the one-loop, colour-stripped $(1,m_L)$ and $(1,m_R)$ building blocks which we build upon in \cref{sec:c_qqbg}.

\subsection{Colour structure of one-loop four-parton amplitudes in the Regge limit}
\label{sec:colour-4}
In \cref{sec:colour_n}, we review the DDM colour decomposition for $n$-parton amplitudes in general kinematics. In this section we specify to the four-parton case and show how the colour structure simplifies in the Regge limit.

Let us first consider the physical scattering of gluons $g_1(p_1) \  g_2(p_2) \to g_3(p_3) \  g_4(p_4)$, with all outgoing momenta \cref{eq:outgoing}, in the Regge limit \cref{eq:Regge}. We specify to the $n{=}4$ case of \cref{eq:DDM_ng},
\begin{align}
\begin{split}
\mathcal{A}_{4}^{(1)}(g_1,g_2,g_3,g_4)&= \gs^4  \sum_{\sigma \in S_{3} / R} \tr 
\left(F^{a_{\sigma_1}} F^{a_{\sigma_2}} F^{a_{\sigma_3}} F^{a_{\sigma_4}}\right) 
A^{(1, \, g)}_{4}
\left({\sigma_1}, {\sigma_2},{\sigma_3},{\sigma_4}\right) \\
 &+ \Nf \ \gs^4  \sum_{\sigma \in S_4 } \tr \left(T^{a_{\sigma_1}} T^{a_{\sigma_2}} T^{a_{\sigma_3}} T^{a_{\sigma_4}}\right) 
 A^{(1, \, f)}_{4}
\left({\sigma_1}, {\sigma_2},{\sigma_3},{\sigma_4}\right)\,.
 \label{eq:DDM_4g}
 \end{split}
\end{align}
We are interested in the leading behaviour of this amplitude in the Regge limit. We note that all primitive amplitudes with colour-orderings in which $a_1$ and $a_4$ are not adjacent are power supressed in this limit. The structure of the amplitude \cref{eq:DDM_4g} in the Regge limit therefore simplifies to 
%\begin{align}
%\begin{split}
%\mathcal{A}_{4}^{(1,\ \QCD)}(g_1,g_2,g_3,g_4)\to
%\gs^4  
%\Big\{
%\tr 
%\left(F^{a_{2}} F^{a_{3}} F^{a_{4}} F^{a_{1}}\right) 
%&A^{(1, \ g)}_{4}
%\left(2,3,4,1\right) 
%\\
%+\tr 
%\left(F^{a_{2}} F^{a_{3}} F^{a_{1}} F^{a_{4}}\right) 
%&A^{(1, \ g)}_{4}
%\left(2,3,1,4\right) 
%\Big\}
%\\ 
% + N_f \ \gs^4  
%\Big\{ 
%\left[
% \tr \left(T^{a_2} T^{a_3} T^{a_4} T^{a_1}\right) + 
% \tr \left(T^{a_3} T^{a_2} T^{a_1} T^{a_4}\right)
% \right]
% &A^{(1, \ f)}_{4}
%\left(2,3,4,1\right)  
%\\
%\left[
% \tr \left(T^{a_2} T^{a_3} T^{a_1} T^{a_4}\right) + 
% \tr \left(T^{a_3} T^{a_2} T^{a_4} T^{a_1}\right)
% \right]
% &A^{(1, \ f)}_{4}
%\left(2,3,1,4\right) 
%\Big\}
% \,,
% \label{eq:DDM_4g}
% \end{split}
%\end{align}
\begin{align}
\begin{split}
\mathcal{A}_{4}^{(1)}(g_1,g_2,g_3,g_4)\toRegge
\gs^4  
\sum_{\sigma \in S_2}
\tr 
\left(F^{a_{2}} F^{a_{3}} F^{a_{\sigma_4}} F^{a_{\sigma_1}}\right) 
&A^{(1, \, g)}_{4}
\left(2,3,\sigma_4,\sigma_1\right) 
\\ 
 + N_f \ \gs^4  \sum_{\sigma \in S_2}
\big[
 \tr \left(T^{a_2} T^{a_3} T^{a_{\sigma_4}} T^{a_{\sigma_1}}\right) + 
 \tr \left(T^{a_3} T^{a_2} T^{a_{\sigma_1}} T^{a_{\sigma_4}}\right)
 \big]
 &A^{(1, \, f)}_{4}
\left(2,3,\sigma_4,\sigma_1\right)  
 \,,
 \label{eq:DDM_4g_Regge}
 \end{split}
\end{align}
where we have made use of the reflection identity \cref{eq:reflection}.  The observation that the terms in \cref{eq:DDM_4g_Regge} are pairwise related by $\{p_4, \lambda_4, a_4\} \leftrightarrow \{p_1, \lambda_1, a_1\}$ naturally suggests we consider amplitudes of definite signature in the $t$-channel, 
\begin{align}
\begin{split}
    A_4^{(1, \, m)[\pm]}(2^{\lambda_2},3^{\lambda_3},4^{\lambda_4},1^{\lambda_1})
    =\frac{1}{2}\Big(
      &A_4^{(1, \, m)}(2^{\lambda_2},3^{\lambda_3},4^{\lambda_4},1^{\lambda_1})\\
\pm &A_4^{(1, \, m)}(2^{\lambda_2},3^{\lambda_3},1^{\lambda_1},4^{\lambda_4})
    \Big)\,.
    \end{split}\label{eq:sig4}
\end{align}
The signature-odd amplitude has a particularly simple structure. Using identities \cref{eq:f}, \cref{eq:Jacobi} and the trace identities
\begin{subequations}
\begin{align}
    \tr \left(T^a T^b T^c\right) &=  \frac{1}{2}\left(d^{abc}+ i  f^{abc}\right) \,, \\
    \tr \left(F^a F^b F^c\right) &=  i N_c \ f^{abc}\,,
\end{align}
\end{subequations}
we see that the signature-odd amplitude consists of a single colour structure, namely that of a colour-octet exchange in the $t$-channel:
\begin{align}
\begin{split}
\mathcal{A}_{4}^{(1)[-]}(g_1,g_2,g_3,g_4)\to
\gs^4  F^d_{a_2 a_3}F^d_{a_4 a_1}
\Big\{ N_c \
& A^{(1, \, g)[-]}_{4}
\left(2,3,4,1\right) \\
 + \Nf \
 &A^{(1, \, f)[-]}_{4}
\left(2,3,4,1\right) 
\Big\}
 \,.
 \label{eq:DDM_4g_Regge_odd}
 \end{split}
\end{align}

Let us now perform a similar study of the physical scattering of a gluon and quark, $g_1(p_1) \ q_2(p_2) \to q_3(p_3) \ g_4(p_4)$. The $n{=}4$ case of \cref{eq:DDM_qqbn} reads 
\begin{align}
\begin{split}
\mathcal{A}_4^{(1)}
\left(\bar{q}_2, q_3, g_4, g_1\right)
=  \gs^4
\Bigg\{
    \sum_{\sigma \in S_2}  
    \bigg[
        &\left(T^{c_2} T^{c_1}\right)_\ibi
        (F^{a_{\sigma_4}} F^{a_{\sigma_1}})_{c_1 c_2} 
        A_{4}^{(1, \, g_R)}
        \left(2_{\bar{q}}, \sigma_4, \sigma_1, 3_q \right) 
        \\
         +&(T^{c_2} T^{a_{\sigma_4}} T^{c_1})_\ibi
         (F^{a_{\sigma_1}})_{c_1 c_2} 
         A_4^{(1, \, g_R)}
         \left(2_{\bar{q}}, \sigma_1, 3_q, \sigma_4\right) 
         \\
         +&(T^{c_2} T^{a_{\sigma_4}} T^{a_{\sigma_1}} T^{c_1})_\ibi 
         \Delta_{c_1 c_2} 
         A_4^{(1, \, g_R)}
         \left(2_{\bar{q}}, 3_q, \sigma_4, \sigma_1\right)
     \bigg] 
     \\
    + 
    \frac{\Nf}{N_c}
    \bigg[
        \sum_{\sigma \in S_{2}} 
        &N_c
        (T^{a_{\sigma_4}}T^{a_{\sigma_1}})_\ibi
        A_{4 ; 1}^{(1, \, f)}
        \left(2_{\bar{q}}, 3_q ; \sigma_4, \sigma_1 \right)
        \\
        +\sum_{\sigma \in S_{2}/Z_2} &\tr(T^{a_{\sigma_4}}T^{a_{\sigma_1}})
        \delta_\ibi
        A_{4 ; 3}^{(1, \, f)}
        \left(2_{\bar{q}}, 3_q ; \sigma_4, \sigma_1 \right)
    \bigg]
\Bigg\}\,.
\label{eq:DDM_4_q}
\end{split}
%Raise the equation label: (https://tex.stackexchange.com/questions/301150/equation-number-in-long-equation)
\raisetag{3\normalbaselineskip}
\end{align}
As was shown in ref.~\cite{Bern:1994fz}, the partial amplitude $A_{4;3}^{(1, \, f)}$ is in fact zero. It is instructive to recall why this is the case. Using \cref{eq:partial_j}, we expand the partial amplitude in terms of primitive amplitudes,
\begin{align}
\begin{split}
A_{4 ; 3}^{(1, \, f)}\left(2_{\bar{q}}, 3_q ; 4, 1 \right) &= 
A_{4 }^{(1, \, f_R)}\left(2_{\bar{q}}, 3_q , 4, 1 \right)
+
A_{4 }^{(1, \, f_R)}\left(2_{\bar{q}}, 3_q , 1, 4 \right)
\\
&+
A_{4 }^{(1, \, f_R)}\left(2_{\bar{q}}, 4,  3_q,1\right)
+
A_{4 }^{(1, \, f_R)}\left(2_{\bar{q}}, 1, 3_q, 4\right)
\\
&+
A_{4 }^{(1, \, f_R)}\left(2_{\bar{q}}, 4, 1, 3_q\right)
+
A_{4 }^{(1, \, f_R)}\left(2_{\bar{q}}, 1, 4, 3_q\right)\,.
\label{eq:A43q}
\end{split}
\end{align}
The first two lines vanish due to \cref{eq:tadpole} and \cref{eq:bubble} respectively. The final two primitive amplitudes consist solely of Feynman diagrams with closed fermion triangles. Therefore, Furry's theorem implies
\begin{equation}
    A_{4 }^{(1, \, f_R)}\left(2_{\bar{q}}, 4, 1, 3_q\right)
=-
A_{4 }^{(1, \ f_R)}\left(2_{\bar{q}}, 1, 4, 3_q\right)\,. \label{eq:furry}
\end{equation}
Thus, we see that \cref{eq:A43q} is zero, and \cref{eq:DDM_4_q} is simply expressed in terms of primitive amplitudes.

As for the pure-gluon case above, let us now utilise the fact that primitive amplitudes that do not have $p_4$ and $p_1$ cyclically adjacent are power suppressed in the Regge limit. This observation leads to the simplified colour structure
\begin{align}
\begin{split}
\mathcal{A}_4^{(1)}
\left(\bar{q}_2, q_3, g_4, g_1\right)
\toRegge  \gs^4
    \sum_{\sigma \in S_2}  
    \bigg[
        &\left(T^{c_2} T^{c_1}\right)_\ibi
        (F^{d_{\sigma_4}} F^{d_{\sigma_1}})_{c_1 c_2} 
        A_{4}^{(1, \, g_R)}
        \left(2_{\bar{q}}, \sigma_4, \sigma_1, 3_q \right) 
         \\
         +&(T^{c_2} T^{d_{\sigma_4}} T^{d_{\sigma_1}} T^{c_1})_\ibi 
         \delta_{c_1 c_2} 
         A_4^{(1, \, g_R)}
         \left(2_{\bar{q}}, 3_q, \sigma_4, \sigma_1\right)
         \\
        &+N_f \ 
        (T^{\sigma_4}T^{\sigma_1})_\ibi
        A_{4 ; 1}^{(1, \, f_L)}
        \left(2_{\bar{q}}, 3_q , \sigma_4, \sigma_1 \right)
    \bigg]\,.
\end{split}
\raisetag{1\normalbaselineskip}
\end{align}
Finally, by using the identities 
\begin{align}
    \left(
    T^aT^b
    \right)_{ij}F^d_{ba}&=N_c \ T^d_{ij} \,,\\
    \left(
    T^aT^dT^a
    \right)_{ij}&=-\frac{1}{N_c} \ T^d_{ij} \,,
\end{align}
we find that the signature-odd amplitude consists of only $t$-channel octet exchange,
\begin{align}
\begin{split}
\mathcal{A}_4^{(1)[-]}
\left(\bar{q}_2, q_3, g_4, g_1\right)
\toRegge  \gs^4 T^d_\ibi F^d_{a_4 a_1}
    \Big\{
        N_c \
        &A_{4}^{(1, \, g_L)[-]}
        \left(2_{\bar{q}}, 3_q, 4, 1\right) 
         \\
        -\frac{1}{N_c} \ 
         &A_4^{(1, \, g_R)[-]}
         \left(2_{\bar{q}}, 3_q, 4,1\right)
         \\
        +\Nf \ 
        &A_{4 }^{(1, \, f_L)[-]}
        \left(2_{\bar{q}}, 3_q , 4,1 \right)
    \Big\}\,,
    \label{eq:A4q-}
\end{split}
\end{align}
as in \cref{eq:DDM_4g_Regge_odd}. We have used the reflection identity \cref{eq:reflection_q} to write the primitive amplitudes in the form provided by \cite{Bern:1994fz}. In the next section, we analyze these leading-power amplitudes in the Regge limit.
\subsection{One-loop primitive vertices for the peripheral emission of a gluon or quark}
\label{sec:A4loop_Regge}
Having studied the colour structure of amplitudes to leading power in the Regge limit, we now turn our attention to the structure of the leading-power primitive amplitudes.
As usual, we start with the pure-gluon case, where \cref{eq:DDM_4g} requires us to obtain the leading behaviour of
$$A_{4}^{(1, \, g)[-]}(2,3,4,1) \qquad \qquad \textrm{and} \qquad  \qquad A_{4}^{(1, \, f)[-]}(2,3,4,1)\,.$$
We will obtain these via supersymmetric one-loop amplitudes, according to \cref{eq:gf-susy}.

At one-loop, it is useful to analyse the qualitatively-different MHV and non-MHV amplitudes separately. The non-MHV amplitudes are zero at tree-level, and at one-loop they are finite, purely rational functions. We study the Regge limit of four-parton non-MHV amplitudes in \cref{sec:div_4}, where we find they give rise to helicity-violating peripheral-emission vertices. The one-loop MHV amplitudes contain divergences in $\epsilon$ and contain products of rational and transcendental functions. We study the Regge limit of four-parton MHV amplitudes in \cref{sec:div_4}, where we find they give rise to helicity-conserving peripheral-emission vertices.
\subsubsection{Helicity-violating vertices}
\label{sec:div_4}
The finite one-loop four-gluon amplitudes are the \emph{all-plus} amplitudes \cite{Bern:1991aq,Kunszt:1993sd},
\begin{subequations}
\begin{align}
A_4^{(1,\,\Nfour)}\left(1^\oplus, 2^\oplus, 3^\oplus, 4^\oplus\right) & =0\,, \\
A_4^{(1,\,\NoneX)}\left(1^\oplus, 2^\oplus, 3^\oplus, 4^\oplus\right) & =0\,, \\
A_4^{(1,\,s)}\left(1^\oplus, 2^\oplus, 3^\oplus, 4^\oplus\right) & =\frac{1}{3 (4\pi)^2} \frac{s_{12} s_{23}}{\langle 12\rangle\langle 23\rangle\langle 34\rangle\langle 41\rangle}\,,
\label{eq:all-plus}
\end{align}
\end{subequations}
and the \emph{single-minus} amplitudes,
\begin{subequations}
\begin{align}
A_4^{(1,\,\Nfour)}\left(1^\ominus, 2^\oplus, 3^\oplus, 4^\oplus\right) & =0\,, \\
A_4^{(1,\,\NoneX)}\left(1^\ominus, 2^\oplus, 3^\oplus, 4^\oplus\right) & =0\,, \\
A_4^{(1,\,s)}\left(1^\ominus, 2^\oplus, 3^\oplus, 4^\oplus\right) & =\frac{1}{3 (4\pi)^2} \frac{\langle 24\rangle[24]^3}{[12]\langle 23\rangle\langle 34\rangle[41]}\,.
\label{eq:one-minus}
\end{align}
\end{subequations}
The all-plus amplitudes are power suppressed in the Regge limit, while the single-minus amplitudes have leading-power contributions, e.g.,
\begin{align}
    A_4^{(1,\,s)}(2^\ominus,3^\oplus,4^\oplus, 1^\oplus) &\toRegge \frac{1}{3 (4\pi)^2} \frac{s}{t}\,,\\
    A_4^{(1,\,s)}(2^\oplus,3^\oplus,4^\oplus, 1^\ominus) &\toRegge \frac{1}{3 (4\pi)^2} \frac{s}{t}\left( \frac{q_{\perp}^*}{q_\perp}\right)^2\,.
\end{align}
Both these sets of observations are compatible with Regge factorisation of one-loop colour-ordered amplitudes, discussed in \cref{sec:Regge}. The four-parton analogue of \cref{eq:Regge_CO} is
\begin{align}
\begin{split}
\mathrm{Disp} \left[ A_4
\left(f_2^{\lambda_2}, f_3^{\lambda_3}, f_4^{\lambda_3}, f_1^{\lambda_1}\right) \right]
\toNMRK 2 s \ & C_{f_2 f_3 g^*}(p_2^{\lambda_2}, p_3^{\lambda_3}, q) \\
	 \times & R_{g^*}(q; s_{34}) \\
  \times &
C_{g^* f_4 f_1 }(-q, p_4^{\lambda_4}, p_1^{\lambda_1})
\label{eq:Regge_CO_4}
\end{split}
\end{align}
with expansion of the colour-ordered vertices,
\begin{align}
    C_{ggg^*}(p_2^{\lambda_2}, p_3^{\lambda_3}, q) =  \gs \ C^{(0)}_{ggg^*}(p_2^{\lambda_2}, p_3^{\lambda_3}, q) +  \gs^3 \ C^{(1)}_{ggg^*}(p_2^{\lambda_2}, p_3^{\lambda_3}, q)  + \cO (\gs^5)\,.
\end{align}
Expanding \cref{eq:Regge_CO_4} to order $\gs^6$ we obtain 
\begin{align}
\begin{split}
   \frac{1}{2s}\disp{A_4^{(1)}(2,3,4,1)} \toRegge 
   &C^{(1)}_{ggg^*}(p_2^{\lambda_2}, p_3^{\lambda_3}, q)
   R_{g^*}^{(0)}(q; s_{34})
   C_{g^* f_5 f_1 }^{(0)}(-q, p_4^{\lambda_4}, p_1^{\lambda_1})
   \\ 
   + 
   &
   C^{(0)}_{ggg^*}(p_2^{\lambda_2}, p_3^{\lambda_3}, q)
   R_{g^*}^{(1)}(q; s_{34})
   C_{g^* f_5 f_1 }^{(0)}(-q, p_4^{\lambda_4}, p_1^{\lambda_1})
   \\ 
   + 
   &
   C^{(0)}_{ggg^*}(p_2^{\lambda_2}, p_3^{\lambda_3}, q)
   R_{g^*}^{(0)}(q; s_{34})
   C_{g^* f_5 f_1 }^{(1)}(-q, p_4^{\lambda_4}, p_1^{\lambda_1})\,.
   \label{eq:Regge_CO_4_loop}
\end{split}
\end{align}
The fact the tree-level all-plus and single-minus amplitudes vanish implies
\begin{equation}
    C^{(0)}_{ggg^*}(p_2^\oplus, p_3^\oplus, q) =  C^{(0)}_{g^* gg}(-q, p_5^\oplus, p_1^\oplus) =  0\,.
    \label{eq:Cpp0}
\end{equation}
Subsequently, the ansatz \cref{eq:Regge_CO_4} correctly predicts the amplitudes \cref{eq:all-plus} should be power-suppressed in the Regge limit. 
Likewise, for the single-minus case, \cref{eq:Cpp0} implies only the first or last line of \cref{eq:Regge_CO_4_loop} will contribute at leading power, and we can extract \cite{DelDuca:1998cx}
\begin{align}
    C_{ggg^*}^{(1,\,s)}(p_2^\oplus, p_3^\oplus, q)&=\frac{1}{3 (4\pi)^2} \left(\frac{-q_{\perp}^*}{q_\perp}\right)\,, 
\\
C_{ggg^*}^{(1,\,s)}(p_4^\oplus, p_1^\oplus, q)&=\frac{1}{3 (4\pi)^2}\,.
\end{align}
Just as in \cref{eq:C0}, the vertices are antisymmetric under $p_2\leftrightarrow p_3$ or $p_4\leftrightarrow p_1$, and the parity conjugate is given by complex conjugation. The fact that only the scalar term in \cref{eq:one-minus} is non-zero leads to the neat property
\begin{equation}
   C_{ggg^*}^{(1,\,s)}(p_2^\oplus, p_3^\oplus, q)=C_{ggg^*}^{(1,\,g)}(p_2^\oplus, p_3^\oplus, q)=-C_{ggg^*}^{(1,\,f)}(p_2^\oplus, p_3^\oplus, q) \,.
\end{equation}
\subsubsection{Helicity-conserving vertices}
\label{sec:fin_4}
Let us now proceed with the analysis of MHV amplitudes, using the normalisation \cref{eq:mhv_amplitude}.
For these amplitudes, \cref{eq:Regge_CO_4_loop} can then be written
\begin{align}
\begin{split}
 &\frac{1}{2s}\disp{ A_4^{(1, \, m)}(2^{\lambda_2},3^{-\lambda_2},4^{-\lambda_1},1^{\lambda_1}) }\toRegge 
    c_\Gamma \, C^{(0)}_{ f_2 f_3 g^*}(p_2^{\lambda_2}, p_3^{-\lambda_2}, q)
   \\ 
   &
   \hspace{3 em}
   \times 
   R_{g^*}^{(0)}(q; s_{34})
   \times \Big[
   c^{(1, \, m )}_{f_2 f_3 g^*}(p_2^{\lambda_2}, p_3^{-\lambda_2}, q)
   + r_{g^*}^{(1, \,  m)}(q; s_{34})
   +c_{g^* f_4 f_1 }^{(1, \, m)}(-q, p_4^{-\lambda_1}, p_1^{\lambda_1})
   \Big]
   \\
   &
   \hspace{3 em}
   \times 
   C_{g^* f_4 f_1 }^{(0)}(-q, p_4^{-\lambda_1}, p_1^{\lambda_1})\,.
   \label{eq:Regge_CO_A4_loop_MHV}
\end{split}
\end{align}
We see that the factorisation ansatz \cref{eq:Regge_CO} implies the Regge limit of one-loop coefficients will simply be given by a sum of universal, process independent functions which depend on a subset of the momenta of the process:
\begin{align}
\begin{split}
 \mathrm{Re}\left[ a_4^{(1 \, m)}(2,3,4,1) 
 \right]
 \toRegge 
   c^{(1,\, m )}_{f_2 f_3 g^*}(p_2^{\lambda_2}, p_3^{-\lambda_2}, q)
   + r_{g^*}^{(1,\, m)}(q; s_{34})
   +c_{g^* f_4 f_1 }^{(1,\, m)}(-q, p_4^{-\lambda_1}, p_1^{\lambda_1})\,.
\label{eq:Regge_CO_a4_loop_MHV}
\end{split}
\end{align}
The simplest one-loop amplitudes are pure-gluon amplitudes in $\Nfour$ sYM, where the one-loop corrections, $a^{(1, \, \Nfour)}$, are helicity-independent and purely transcendental~\cite{Green:1982sw}
\begin{align}
a_4^{(1, \ \Nfour)}(2^{\lambda_2},3^{-\lambda_2},4^{-\lambda_1},1^{\lambda_1}) = -\frac{2}{\epsilon^2}\left[\left(\frac{\mu^2}{-s_{12}}\right)^\epsilon+\left(\frac{\mu^2}{-s_{23}}\right)^\epsilon\right]+\ln ^2\left(\frac{-s_{12}}{-s_{23}}\right)+\pi+\cO(\epsilon)\,.
\label{eq:VN4_4}
\end{align}
In the following it should be understood that we only work up to $\cO(\epsilon^0)$.
As observed in ref.~\cite{Bartels:2008ce}, this function admits an \emph{exact} decomposition into one-loop building blocks. In particular, in the physical region \cref{eq:outgoing}, the real part can be written:
\begin{equation}
 \real{a_4^{(1, \, \Nfour)}(2,3,4,1)}=c_{ggg^*}^{(1, \, \Nfour)}(p_2,p_3,q)  +r_{g^*}^{(1, \, \Nfour)}(q;s_{12})+c_{g^*gg}^{(1, \, \Nfour)}(-q, p_4,p_1)\,.
\end{equation}
Here, the real (thus parity invariant), one-loop correction to the peripheral-emission vertex is 
\begin{equation}
 c_{ggg^*}^{(1, \, \Nfour)}(2^{\lambda_2},3^{-\lambda_2},q_1)
 =
\left(\frac{\mu^2}{-s_{23}}\right)^\epsilon	\left( -\frac{2}{\epsilon^2}  + \frac1{\epsilon} \log\left(\frac{\tau}{-s_{23}}\right)
	+ \frac{\pi^2}{2} \right)\,,\label{eq:c3_N4}
\end{equation}
and the large logarithmic term
\begin{equation}
 r_{g^*}^{(1, \, \Nfour)}(q;s)=\frac{\alpha^{(1)}(t)}{N_c  \ c_\Gamma}\log \left( \frac{s}{\tau}\right) = \frac{2 }{\epsilon}\left(\frac{\mu^2}{-t}\right)^\epsilon \log\left( \frac{s}{\tau}\right)\,,
 \label{eq:r}
\end{equation}
is the colour-stripped one-loop truncation of the Reggeised gluon in the $t$ channel.

The primitive amplitudes with a $\NoneX$ multiplet or complex scalar circulating in 
the loop are not helicity independent. 

For the `split' helicity configuration, the one-loop amplitudes are given by
\begin{align}
    V^{(1, \, \NoneX)}(1^\oplus,2^\oplus,3^\ominus, 4^\ominus)&=\frac{1}{\epsilon}\left(\frac{\mu^2}{-s_{23}}\right)^\epsilon+2\,,
    \\
    G^{(1, \, \NoneX)}(1^\oplus,2^\oplus,3^\ominus, 4^\ominus)&= 0 \,,
    \\
    V^{(1, \, s)}(1^\oplus,2^\oplus,3^\ominus, 4^\ominus)&=
    \frac{1}{3}V^{(1, \ \NoneX)}+\frac{2}{9}\,,
    \\
    G^{(1, \, s)}(1^\oplus,2^\oplus,3^\ominus, 4^\ominus)&=0
    \,,
    \label{eq:VN1}
\end{align}
where we use the notation defined in \cref{eq:PG}. 
These amplitudes have no logarithmic dependence on $s_{12}$, and trivially admit an exact decomposition into peripheral-emission vertices,
 \begin{align}
 \real{a_4^{(1, \ \None)}
 (2^{\oplus},3^{\ominus},4^{\ominus},1^{\oplus})}
 &=
 c_{g g g^*}^{(1, \ \None)}(p_2,p_3,q) 
 +c_{g g g^*}^{(1, \ \None)}(p_4,p_1,-q) \,,
 \\
\real{ a_4^{(1, \ s)}
  (2^{\oplus},3^{\ominus},4^{\ominus},1^{\oplus})}
 &=
 c_{g g g^*}^{(1, \ s)}(p_2,p_3,q)+c_{g g g^*}^{(1, \ s)}(p_4,p_1,-q)\,,
 \end{align}
where we simply take the vertices to be half the one-loop functions, i.e.,
 \begin{align}
 c_{g g g^*}^{(1, \ \None)}(p_2,p_3,q)&=\frac{1}{2\epsilon}
	\left(\frac{\mu^2}{-s_{23}}\right)^\epsilon
    +1,\label{eq:c3_N1}\\
 c_{g g g^*}^{(1, \ s)}(p_2,p_3,q)&=\frac{1}{3}c^{(1, \ \None)}_{g g g^*}(p_2,p_3,q)+\frac{1}{9}\,.
 \label{eq:c3_s}\end{align}
Note the lack of any logarithmic dependence on $s$ implies
\begin{equation}
    r_{g^*}^{(1, \, \NoneX)}=r^{(1, \, s)}_{g^*}=r^{(1, \, f)}_{g^*}=0\label{eq:no_trajectory}\,,
\end{equation}
and correspondingly, the scale $\tau$ does not enter into \cref{eq:c3_N1,eq:c3_s}. This fact also implies the well-known equivalence of gluon Regge trajectories in QCD and supersymmetric theories with spin-1 particles,
\begin{equation}
    r_{g^*}^{(1, \, \Nfour)}=r_{g^*}^{(1, \, \NoneV)}=r_{g^*}^{(1, \, g)} \label{eq:equiv_trajectory}\,.
\end{equation}
The \emph{alternating-helicity} amplitudes are considerably more complicated (see ref.~\cite{Bern:1994cg}). Of course, in the Regge limit, the split-helicity amplitudes coincide with the alternating-helicity amplitudes, e.g.,
\begin{align}
 \real{a_4^{(1, \, \None)}
 (2^{\oplus},3^{\ominus},4^{\oplus},1^{\ominus})}
 &\toRegge
 c_{g g g^*}^{(1, \, \None)}(p_2^{\oplus},p_3^{\ominus},q) 
 +c_{g^* g g}^{(1, \, \None)}(-q,p_4^{\oplus},p_1^{\ominus}) \,,
 \\
 \real{a_4^{(1, \, s)}
  (2^{\oplus},3^{\ominus},4^{\oplus},1^{\ominus})}
 &\toRegge
 c_{g g g^*}^{(1, \, s)}(p_2^{\oplus},p_3^{\ominus},q)
 +c_{g^* g g}^{(1, \, s)}(-q,p_4^{\oplus},p_1^{\ominus})\,.
 \end{align}
Via the decomposition \cref{eq:gf-susy}, the peripheral-emission vertices \cref{eq:c3_N4,eq:c3_N1,eq:c3_s} are sufficient to construct the vertices relevant for QCD, 
\begin{align}
 c_{ggg^*}^{(1, \ g)}(p_2^{\lambda_2},p_3^{-\lambda_2},q)&=
 \left(\frac{\mu^2}{-s_{23}}\right)^\epsilon
 \left( 
    -\frac{2}{\epsilon^2}  
    -\frac{11}{6\epsilon}
    + \frac1{\epsilon} \log\left(\frac{\tau}{-s_{23}}\right)
	+ \frac{\pi^2}{2}  
\right)\,,\label{eq:c3_g}\\
 c_{ggg^*}^{(1, \ f)}(p_2^{\lambda_2},p_3^{-\lambda_2},q)&=
 \left(\frac{\mu^2}{-s_{23}}\right)^\epsilon
  \left( 
    \frac{1}{3\epsilon}
   +\frac{5}{9}
\right)\,.
 \label{eq:c3_f}
\end{align}
For completeness, we can also write down the impact factor for a $\NoneV$ multiplet circulating in the loop,
\begin{align}
 c_{ggg^*}^{(1, \ \NoneV)}(p_2^{\lambda_2},p_3^{-\lambda_2},q)&=
  \left(\frac{\mu^2}{-s_{23}}\right)^\epsilon
  \left( 
    -\frac{2}{\epsilon^2}  
    -\frac{3}{2\epsilon}
    + \frac1{\epsilon} \log\left(\frac{\tau}{-s_{23}}\right)
	+ \frac{\pi^2}{2} 
\right)\,.
 \label{eq:c3_N1V}
\end{align}
Let us now consider the factorisation of one-loop primitive amplitudes with an external quark-antiquark pair, recalling the supersymmetric organisation \cref{eq:susy_q}.
We can immediately obtain the primitive amplitudes in $\NoneV$ from pure-gluon amplitudes by noting that in supersymmetric theories, the Ward identities such as \cref{eq:susy1,eq:susy2} hold to all orders. For MHV amplitudes, this in turn implies the equality of one-loop corrections
\begin{equation}
    a_5^{(1, \, \NoneV)}(2_\qb, 3_q, 4,1)= a_5^{(1,\, \NoneV )}(2, 3, 4,1)\,. \label{eq:susy_1loop}
\end{equation}
Thus, from the Regge factorisation of the amplitude, \cref{eq:Regge_CO_a4_loop_MHV},
%\begin{align}
%\begin{split}
%\real{ a_4^{(1, \ \NoneV)}(2_\qb,3_q,4,1)}\toRegge 
% &c_{\qb q g^*}^{(1, \ \NoneV)}(p_2, p_3,q)  
% +r_{g^*}^{(1, \ \NoneV)}(q;s_{12})\\
% +&c_{g g g^*}^{(1, \ \NoneV)}(p_4,p_1,-q)\,, \label{eq:aN1V}
% \end{split}
%\end{align}
we see the SUSY Ward identity \cref{eq:susy_1loop} immediately leads to the identification of vertices
\begin{equation}
 c_{\qb q g^*}^{(1, \ \NoneV)}(p_2^{\lambda_2},p_3^{-\lambda_2},q)=  c_{g g g^*}^{(1, \ \NoneV)}(p_2^{\lambda_2},p_3^{-\lambda_2},q)\,.\label{eq:c3q_N1V}
\end{equation}
Let us now study the non-supersymmetric amplitudes on the right-hand side of \cref{eq:susy_q}, which are given in ref.~\cite{Bern:1994fz}. We find that in the Regge limit, only the $g_L$ term has large-logarithmic dependence. The Regge decomposition of this amplitude,
\begin{align}
\begin{split}
 \real{a_4^{(1, \, g_L)}(2_\qb^{\lambda_2},3_q^{-\lambda_2},4^{-\lambda_1},1^{\lambda_1})}\toRegge 
 &c_{\qb q g^*}^{(1, \ g_L)}(p_2^{\lambda_2}, p_3^{-\lambda_2},q)  +r_{g^*}^{(1, \, g)}(q,s_{34})
 \\
 +&c_{g^* gg }^{(1, \ g)}(-q, p_4^{-\lambda_1}, p_1^{\lambda_1})\,,
\end{split}
\end{align}
defines the vertex
\begin{equation}
c_{\qb q g^*}^{(1, \ g_L)}(p_2^{\lambda_2}, p_3^{-\lambda_2},q)= 	
\left(\frac{\mu^2}{-s_{23}}\right)^\epsilon
  \left( 
    -\frac{1}{\epsilon^2}  
    +\frac{1}{\epsilon}
    + \frac1{\epsilon} \log\left(\frac{\tau}{-s_{23}}\right)
	+ \frac{\pi^2}{2} 
 +\frac{19}{18}
\right)\,.\label{eq:c3q_g}
\end{equation}
The fermion- and scalar-loop contributions are more subtle. Indeed, the amplitudes are zero for all helicity configurations, and for both $L$ and $R$ contributions~\cite{Bern:1994fz},
\begin{align}
 a_4^{(1, \, {\NoneX}_L)}(2_\qb,3_q,4,1)&=a_4^{(1, \, f_L)}(2_\qb,3_q,4,1) =a_4^{(1, \, s_L)}(2_\qb, 3_q,4,1) = 0\,, \label{eq:L}\\
 a_4^{(1, \, {\NoneX}_R)}(2_\qb,3_q,4,1)&=a_4^{(1, \, f_R)}(2_\qb,3_q,4,1) =a_4^{(1, \, s_R)}(2_\qb, 3_q,4,1) = 0\,.\label{eq:R}
\end{align}
Nevertheless, if these colour-ordered amplitudes obey the factorisation \cref{eq:Regge_CO_a4_loop_MHV}, we must take the quark-emission vertices to be equal and opposite that of the gluon-emission vertices, i.e.
\begin{align}
\label{eq:c3qf}
   c_{\qb q g^*}^{(1, \, f)}(p_2, p_3,q)&= -c_{ggg^*}^{(1, \, f)}(p_2, p_3,q)\,,\\
   c_{\qb q g^*}^{(1, \, s)}(p_2, p_3,q)&= -c_{ggg^*}^{(1, \, s)}(p_2, p_3,q)\,.\label{eq:c3qs}
\end{align}
%Point forward to 5pt section
We note the ampltiudes in \cref{eq:R} vanish, due to \cref{eq:tadpole}. We further note that \cref{eq:tadpole} holds for an arbitrary number of additional gluons. Considering the MRK limit of such amplitudes informs us the $f_R$ and $s_R$ amplitudes in particular cannot contribute to the peripheral emission vertex of a quark,
\begin{align}
\label{eq:c3fR}
   c_{\qb q g^*}^{(1, \, f_R)}(p_2, p_3,q)=c_{\qb q g^*}^{(1, \, s_R)}(p_2, p_3,q)&= 0\,.
\end{align}
It follows from \cref{eq:c3qf} and \cref{eq:c3qs} that
\begin{align}
\label{eq:c3fL}
   c_{\qb q g^*}^{(1, \, f_L)}(p_2, p_3,q)&= -c_{ggg^*}^{(1, \, f)}(p_2, p_3,q)\,,\\
   c_{\qb q g^*}^{(1, \, s_L)}(p_2, p_3,q)&= -c_{ggg^*}^{(1, \, s)}(p_2, p_3,q)\,.
   \label{eq:c3sL}
\end{align}
%The validity of this reasoning will be borne out in the next section, where we compare with the known colour-ordered results from the literature.
Returning to the supersymmetric decomposition \cref{eq:susy_q}, we can obtain the Regge limit of the $g_R$ primitive amplitudes from our knowledge of the amplitudes studied above,
\begin{equation}
 a_4^{(1, \, g_R)}(2_\qb,3_q,4,1) =  a_4^{(1, \, \NoneV)}(2_\qb,3_q,4,1) - a_4^{(1, \, g_L)}(2_\qb,3_q,4,1) - a_4^{(1, \, f_L)}(2_\qb,3_q,4,1) \,.
 \label{eq:aR1g_4}
\end{equation}
We find that the large logarithmic terms cancel, as do the $ggg^*$-vertices. This means that the $g_R$ amplitudes only contribute to the $\qb q g^*$ vertex, 
\begin{align}
\begin{split}
\real{ a_4^{(1, \ g_R)}(2_\qb,3_q,4,1)}\toRegge 
&c_{\qb q g^*}^{(1, \ \NoneV)}(p_2, p_3,q)
-c_{\qb q g^*}^{(1, \ g_L)}(p_2, p_3,q)
+c_{\qb q g^*}^{(1, \ f_L)}(p_2, p_3,q)
\\
\equiv
&c_{\qb q g^*}^{(1, \ g_R)}(p_2, p_3,q)\,.
 \label{eq:aR1g_4_v2}
\end{split}
\end{align}
Explicitly, this vertex is 
\begin{equation}
 c_{\qb q g^*}^{(1, \ g_R)}(p_2, p_3,q)= 	
\left(\frac{\mu^2}{-s_{23}}\right)^\epsilon
  \left( 
    -\frac{1}{\epsilon^2}  
    -\frac{3}{2\epsilon}
    -\frac{7}{2}
\right)\,.\label{eq:c3q_gR}
\end{equation}
%State pole factiorisation fails for R and L seperately
That the $g_R$ amplitude only contributes to the quark vertex makes intuitive sense, when considering the Feynman diagrams that contribute to this amplitude. 
The fact that \cref{eq:aR1g_4_v2} is not compatible with \cref{eq:Regge_CO_4} is of no concern, because the supersymmetric decompositions are only valid at one loop.
Further validation of \cref{eq:aR1g_4_v2} is provided by considering the Regge limit of colour-dressed amplitudes, which is the topic of the next section.
\subsection{Colour-dressed one-loop vertices for the peripheral emission of one parton}
\label{sec:4_dress}
In \cref{sec:colour-4} we reviewed the colour structure of four-parton one-loop amplitudes in the Regge limit, and in \cref{sec:A4loop_Regge} we reviewed the factorisation of the primitive amplitudes in this limit. In this section we will combine those results, which will allow us to re-derive the colour-dressed peripheral-emission vertices \cref{eq:IF_q}, \cref{eq:IF_g} and \cref{eq:IF_pp}.

Compatibility with the factorisation \cref{eq:Regge_NLL} means the one-loop amplitudes are given by the truncation to $\gs^4$, that is,
\begin{align}
\begin{split}
\disp{\mathcal{A}^{(1)[-]}_4
(f_2, f_3, f_4, f_1)}
\toRegge &2 s \ 
\mathcal{C}^{(1)}_{f_2 f_3 g^*}(p_2, p_3, q) \ 
\mathcal{R}^{(0)}_{g^*}(q) \ 
\mathcal{C}^{(0)}_{g^* f_4 f_1 }(-q, p_4, p_1)\\
+&
2 s \ 
\mathcal{C}^{(0)}_{f_2 f_3 g^*}(p_2, p_3, q) \ 
\mathcal{R}^{(1)}_{g^*}(q;s_{34}) \ 
\mathcal{C}^{(0)}_{g^* f_4 f_1 }(-q, p_4, p_1)
\\
+&
 2 s \ 
\mathcal{C}^{(0)}_{f_2 f_3 g^*}(p_2, p_3, q) \ 
\mathcal{R}^{(0)}_{g^*}(q) \ 
\mathcal{C}^{(1)}_{g^* f_4 f_1 }(-q, p_4, p_1)
\,.
\label{eq:Regge_4_NLO}
\end{split}
\end{align}
Let us consider \cref{eq:DDM_4g} for specific helicity configurations, and substitute the Regge limit of partial amplitudes obtained in \cref{sec:A4loop_Regge}. We begin with the trivial all-plus case, which we state for completeness:
\begin{align}
\begin{split}
\mathcal{A}_{4}^{(1)[\pm]}(g_2^{\oplus},g_3^{\oplus},g_4^{\oplus},g_1^{\oplus})&\toRegge0\,.
 \label{eq:DDM_4g_odd_pppp}
 \end{split}
\end{align}
For the single-minus case, the antisymmetry of the vertices lead to only the signature-odd amplitude contributing in the Regge limit 
\begin{align}
\begin{split}
\mathcal{A}_{4}^{(1)[-]}(g_2^{\oplus},g_3^{\oplus},g_4^{-\lambda_1},g_1^{\lambda_1})
\toRegge
  2s \, &\Big[ \gs^3 \,  c_{\Gamma} \, (N_c-N_f) \,  F^d_{a_2 a_3}
C^{(1,\, s)}_{ggg^*}(p_2^\oplus, p_3^\oplus, q) \Big] 
   \times R_{g^*}^{(0)}(q) \\
   \times
   &\Big[ \gs F^d_{a_4 a_1} C_{g^* gg }^{(0)}(-q, p_4^{-\lambda_1}, p_1^{\lambda_1}) \Big]\,.
 \label{eq:DDM_4g_pppm}
 \end{split}
\end{align}
Comparing with \cref{eq:Regge_4_NLO}, we see that \cref{eq:DDM_4g_pppm} is indeed compatible with simple Regge-pole factorisation and we obtain the vertex
\begin{equation}
    \mathcal{C}^{(1)}_{g g g^*}(p_2^{\oplus}, p_3^{\oplus}, q)
= \gs^3 \, F^c_{a_2 a_3}\,  (N_c-N_f) \, C^{(1,\, s)}_{ggg^*}(p_2^\oplus, p_3^\oplus, q)\,,
\label{eq:IF_pp_23_v2}
\end{equation}
in agreement with \cref{eq:IF_pp_23}.

For the two-minus case, we have
\begin{align}
\begin{split}
&\disp{\mathcal{A}_{4}^{(1)[-]}(g_2^{\lambda_2},g_3^{-\lambda_2},g_4^{-\lambda_1},g_1^{\lambda_1})}\toRegge
2s \, \left[\gs \, F^d_{a_2 a_3}C^{(0)}_{ggg^*}(p_2^{\lambda_2},p_3^{-\lambda_2},q) \right]\times R_{g^*}^{(0)}(q)
\\
&\mkern128mu \times c_\Gamma \gs^2
\Big\{ 
\left(N_c \ c_{ggg^*}^{(1, \, g)}(p_2^{\lambda_2},p_3^{-\lambda_2},q) + N_f \ c_{ggg^*}^{(1, \, f)}(p_2^{\lambda_2},p_3^{-\lambda_2},q)\right)
\\
& \mkern128mu
\hphantom{c_\Gamma \gs^2 \Bigg\{ } \
+
N_c \, r_{g^*}^{(1,\,g)}(q; s_{34})
\\
& \mkern128mu
\hphantom{ c_\Gamma \gs^2 \Bigg\{ } \
+
\left(N_c \ c_{g^* g g }^{(1, \, g)}(-q, p_4^{-\lambda_1}, p_1^{\lambda_1}) + N_f \ c_{g^* g g }^{(1, \, f)}(-q, p_4^{-\lambda_1}, p_1^{\lambda_1}) \right)
\Big\}
\\
 & \mkern128mu \times \left[\gs F^d_{a_4 a_1}C^{(0)}_{g^* g g }(-q, p_4^{-\lambda_1}, p_1^{\lambda_1}) \right]
 \label{eq:DDM_4g_odd_ppmm}
 \end{split}
\end{align}
Comparing these one-loop amplitudes with the Reggeisation ansatz \cref{eq:Regge_4_NLO} allows us to define the one-loop colour dressed vertex
\begin{align}
\begin{split}
\cC^{(1)}_{ggg^*}(p_2^{\lambda_2},p_3^{-\lambda_2},q) = &c_\Gamma \, \gs^3 F^d_{a_2 a_3} \, C^{(0)}_{ggg^*}(p_2^{\lambda_2},p_3^{-\lambda_2},q)
\\&\times
\left( N_c \ c_{ggg^*}^{(1, \, g)}(p_2^{\lambda_2},p_3^{-\lambda_2},q) + N_f \ c_{ggg^*}^{(1, \, f)}(p_2^{\lambda_2},p_3^{-\lambda_2},q)\right)\,
 \end{split}
\end{align}
in agreement with \cref{eq:IF_g}.
Similarly, \cref{eq:A4q-} becomes
\begin{align}
\begin{split}
&\disp{\mathcal{A}_4^{(1)[-]}
\left(\bar{q}_2^{\lambda_2}, q_3^{-\lambda_2}, g_4^{-\lambda_1}, g_1^{\lambda_1}\right)}
\toRegge  2s\, 
\left[\gs \, T^d_\ibi C^{(0)}_{\qb q g^*}(p_2^{\lambda_2},p_3^{-^{\lambda_2}\lambda_2},q) \right]\times R^{(0)}_{g^*}(q) 
\\
&
\qquad 
\times c_\Gamma \ \gs^2
\Bigg\{ 
\left(N_c \ c^{(1, \ g_L)}_{\qb q g^*}(p_2, p_3, q) 
- \frac{1}{N_c}c^{(1, \ g_R)}_{\qb q g^*}(p_2, p_3, q) 
+N_f \, c^{(1, \ f_L)}_{\qb q g^*}(p_2, p_3, q)\right)
\\
&
\qquad \qquad \qquad 
+
N_c \, r_{g^*}^{(1, \, g)}(q;s_{34})
\\
&
\qquad \qquad \qquad 
+
\left(N_c \ c^{(1, \, g)}_{g^* gg }(-q,p_4,p_1) + N_f \ c^{(1, \, f)}_{g^* gg }(-q,p_4,p_1)\right)
\Bigg\}
\\
&
\qquad 
\times
\left[\gs \, F^d_{a_4 a_1}C^{(0)}_{g^* g g}(-q,p_4^{-\lambda_1},p_1^{\lambda_1}) \right]\,,
 \label{eq:DDM_4q}
 \end{split}
\end{align}
from which we obtain
\begin{equation}
\begin{aligned}
&\mathcal{C}^{(1)}_{\qb q g^*}(p_2^{\oplus}, p_3^{\ominus}, q)
= \gs^3 \, c_{\Gamma} \, T^c_\ibi  \, C^{(0)}_{\qb q g^*}(p_2^{\oplus}, p_3^{\ominus}, q)
\\&\times
\left(N_c \ c^{(1, \ g_L)}_{\qb q g^*}(p_2, p_3, q) 
- \frac{1}{N_c}c^{(1, \ g_R)}_{\qb q g^*}(p_2, p_3, q) 
+N_f \, c^{(1, \ f_L)}_{\qb q g^*}(p_2, p_3, q)\right)\,,
\label{eq:IF_q_2}
\end{aligned}
\end{equation}
in agreement with \cref{eq:IF_q}.
In \cref{eq:DDM_4q} we see that the result \cref{eq:aR1g_4_v2} is responsible for producing a $1/N_c$ term in the $\qb q g^*$ vertex and not the $g^*gg$ vertex. These observations prove useful in \cref{sec:c_qqbg}, when studying the analogous primitive amplitudes in the five-parton case.
%Similarly, the prescriptions \cref{eq:c3fR,eq:c3fL} give rise to both a colour-dressed gluon-emission vertex in agreement with x, and the known colour-dressed quark-emission vertex .
\bibliography{refs.bib}
\end{document}